\documentclass[a4,11pt]{article}
\pdfoutput=1
\usepackage{jheppub}
\usepackage{amsmath}
\usepackage{epsfig}
\usepackage{amssymb}
\usepackage{mathrsfs}
\usepackage[utf8]{inputenc}
\usepackage{xspace}
\usepackage{ifthen}
\usepackage{url}

\def\etal{et al.}

\renewcommand{\eqref}[1]{eq.~(\ref{#1})\xspace}

\newcommand{\fig}[1]{\ref{#1}}
\newcommand{\figref}[1]{figure~\fig{#1}}

\newcommand{\sect}[1]{\ref{#1}}
\newcommand{\sectref}[1]{section~\sect{#1}}

\newcommand{\nonu}{\nonumber}

\newcommand{\tit}[1]{\textit{#1}}

\newcommand{\mrm}[1]{\mathrm{#1}}

\renewcommand{\d}{{\mathrm d}}
\newcommand{\e}{{\mathrm e}}

\newcommand{\g}{{\mathrm g}}

\newcommand{\n}{{\mathrm n}}
\newcommand{\p}{{\mathrm p}}
\newcommand{\q}{{\mathrm q}}

\newcommand{\Pb}{{\mathrm{Pb}}}
\newcommand{\A}{{\mathrm A}}

\newcommand{\qbar}{\overline{\mathrm q}}

\newcommand{\pT}{p_{\perp}}

\newcommand{\pTo}{p_{\perp 0}}

{\end{list}}
\newcounter{enumct}

\newcommand{\PY}{\textsc{Pythia}~8}
\newcommand{\DP}{\textsc{Dipsy}}

\newcommand*\smallbar[1]{\kern3pt\overline{\kern-3pt#1\kern-3pt}}

\title{Dipole evolution: perspectives for collectivity and $\gamma^*\A$
collisions} 
\author[a,b]{Christian Bierlich}
\author[a]{Christine O. Rasmussen}

\affiliation[a]{Dept.~of Astronomy and Theoretical Physics, Sölvegatan 14A, S-223 62  Lund, Sweden}
\affiliation[b]{Niels Bohr Institute, Blegdamsvej 17, 2100 København Ø, Denmark}

\emailAdd{bierlich@thep.lu.se}
\emailAdd{christine.rasmussen@thep.lu.se}

\abstract{The transverse, spatial structure of protons is an area
revealing fundamental properties of matter, and provides key input for
deeper understanding of emerging collective phenomena in high energy
collisions of protons, as well as collisions of heavy ions. In this
paper eccentricities and eccentricity fluctuations are predicted using the dipole
formulation of BFKL evolution. Furthermore, first steps are taken towards generation of 
fully exclusive final states of $\gamma^*\A$ collisions, by assessing the importance of 
colour fluctuations in the initial state. Such steps are
crucial for the preparation of event generators for a future electron-ion collider. 
Due to the connection between an impact parameter
picture of the proton structure, and cross sections of $\e\p$ and $\p\p$ collisions, 
the model parameters can be fully determined by fits to such quantities, 
leaving results as real predictions of the model.}

\preprint{LU TP 19-32, MCnet-19-17}
\begin{document}
\maketitle
\flushbottom
\section{Introduction}
In the research program at the Large Hadron Collider (LHC) and the 
Relativistic Heavy Ion Collider (RHIC), collisions of ultra 
relativistic heavy ions are hypothesized to result in the creation of a 
quark-gluon plasma (QGP) with partonic degrees of freedom. One of the 
main avenues for investigating and characterizing this plasma consists 
of measurements of azimuthal correlations between particle pairs 
separated in rapidity, connecting particle emission angles to the 
initial geometry of the collision. Non-trivial correlations reflecting 
collective properties were first observed in gold--gold and copper--copper 
collisions at RHIC \cite{Ackermann:2000tr}, but has since been investigated 
also in lead--lead (PbPb) collisions at the LHC 
\cite{Aamodt:2010pa,Chatrchyan:2011eka,ATLAS:2012at}. Such non-trivial 
azimuthal correlations had at that point already been hypothesized to be a 
signal for hydrodynamic behaviour \cite{Ollitrault:1992bk}, or, even
earlier, to involve microscopic dynamics of overlapping "quark tubes" or 
strings \cite{Abramovsky:1988zh}.

Similar results have been obtained in smaller collision systems such as 
proton--lead ($\p\Pb$) \cite{CMS:2012qk}, deuteron--gold \cite{Aidala:2017ajz}, 
and, perhaps most surprisingly, in proton--proton ($\p\p$) \cite{Khachatryan:2010gv}. 
Attempts to observe similar behaviour in even smaller collision systems, e.g.\ $e^+e^-$, has, while carrying 
interesting prospects, so far not produced positive results
\cite{Badea:2019vey}. Even though the discovery of collectivity in
$\p\p$ is almost ten years old, the origin of
such correlations in small collision systems is still highly debated
(see ref. \cite{Nagle:2018nvi} for a recent review), and its resolution
is among the top priorities for the future heavy ion program at LHC \cite{Citron:2018lsq}.
One possibility is that the correlations in these small collision systems are due 
to coherence effects \cite{Blok:2017pui} or initial state correlations 
\cite{Dusling:2012wy}. Another is a repetition of the argument
from heavy ion collisions, where the observed collective behaviour is a 
hydrodynamic response to the initial partonic spatial configuration 
\cite{Weller:2017tsr}. A picture where a hydrodynamic ``core'' coexists with 
a non-hydrodynamic ``corona'' has been shown by the EPOS model 
\cite{Werner:2007bf,Pierog:2013ria} to provide good descriptions of 
collectivity even in small collision systems.

The possibility of a hydrodynamic (or in fact any other) response to an initial
geometric configuration of partons, poses a challenge to the
traditional strategies for $\p\p$ event generators, such as 
\PY~\cite{Sjostrand:2014zea} or \textsc{Herwig}~7 \cite{Bellm:2015jjp},
both based on perturbative QCD (pQCD) with no obvious way to extract a 
spatial configuration for which to calculate a response. Attempts to 
calculate such a structure \cite{dEnterria:2010xip,Albacete:2016pmp,Bierlich:2017vhg} 
generally involve assuming a certain spatial distribution of partons 
in the proton and, using the eikonal approximation, then transferring this structure 
to a spatio-temporal structure of the multiple partonic interactions
(MPIs). The immediate drawbacks of such an approach are that (a) such 
models will in general contain parameters which need to be fitted to 
the same type of particle correlations as they wish to predict, and 
(b) assuming a spatial distribution of partons in a proton will 
generally contain many \textit{ad hoc} elements.

Even though the spatial distribution of partons in a proton cannot be 
assessed \textit{ab initio}, the evolution of said distribution can be 
calculated perturbatively in the formalism of Mueller 
\cite{Mueller:1993rr,Mueller:1994jq}. At high energies, average properties 
will retain little dependence on the initial configuration, i.e.\ be mostly 
dependent on the evolution. Since the transverse substructure of the 
colliding protons (or virtual photons) can be linked to total or semi-inclusive 
cross sections, any model parameters can be tuned to such quantities, and leave any 
further estimation of collective effects as real predictions of the model. Attempts to
predict the elliptic flow in $\p\p$ collisions using an implementation of 
Mueller's model was provided in 2011 \cite{Avsar:2010rf}, showing 
$v_{2,3}$ comparable to values found from $\Pb\Pb$ at RHIC and LHC energies.

This paper is concerned with presenting a new Monte Carlo implementation of 
Mueller's model, study its description of cross sections in pp and
$\gamma^*\p$ collisions, in order to provide estimates on parton level 
geometries in $\p\p$, proton--ion ($\p\A$) and ion--ion ($\A\A$) collisions, linked to collective phenomena. 
Mueller's model has been implemented as a Monte Carlo several times 
before, as it is not only useful for calculating spatial distributions 
of partons, but in fact has much wider applications due to the equivalence 
of the Mueller formalism with B-JIMWLK (Balitsky, Jalilian-Marian, Iancu, 
McLerran, Weigert, Leonidov and Kovner) \cite{Balitsky:1995ub,Balitsky:1998kc,
Balitsky:2001re,JalilianMarian:1997jx,JalilianMarian:1997gr,
Iancu:2001ad,Iancu:2000hn,Weigert:2000gi} evolution (see section \ref{sec:evol}). 
Such an implementation makes direct introduction of effects beyond the leading logarithmic approximation possible, 
e.g.\ conservation of energy and momentum without imposing kinematical constraints on the
splitting kernel \cite{Deak:2019wms}. This makes the implementation 
attractive for estimation of basic quantities dominated by small $x$ processes 
(e.g.\ cross sections) in cases where little guidance from data exists. In this 
paper (see section \ref{sec:results3}) we will also apply the formalism to 
extract Glauber--Gribov (GG) colour fluctuations in $\gamma^*\p$ collisions, in order 
to take the initial steps towards a generation of electron--ion ($\e\A$) 
collisions within the \textsc{Angantyr} framework 
\cite{Bierlich:2016smv,Bierlich:2018xfw} -- a possibility which is foreseen to 
aid the preparation of an $\e\A$ program currently being planned \cite{Accardi:2012qut}.

Earlier implementations of the Mueller dipole model include the public
\textsc{Oedipus} \cite{Salam:1996nb} and \DP~\cite{Avsar:2005iz,Flensburg:2011kk} 
codes, as well as a private implementation by Kovalenko \etal~
\cite{Kovalenko:2012ye}. All implementations treat only gluons in the
evolution, as will this work. The implementation in this paper is similar to the 
implementation in \DP~in some respects, but differs in other, while bearing 
less resemblance to the other two. The key differences between most of the 
used approaches, lies in the treatment of effects beyond leading order. 
Worth mentioning already here, is the treatment of sub-leading $N_c$ (number 
of colours) effects in the evolution, leading to saturation in the cascade.
In \DP, this is addressed through so-called swing mechanisms 
\cite{Avsar:2007xg,Avsar:2007xh}, which suppresses the contribution from large 
dipoles in dense environments by replacing them with small dipoles. In this paper
we consider only sub-leading $N_c$ effects in the collision frame, by including
multiple interactions in a way consistent with unitarity. Thus we make \emph{no attempt} 
at treating saturation in the cascade, as the 
focus is rather to study how well one can do with an approach that
includes only a minimal set of sub-leading corrections. Effects included in this
paper is energy-momentum conservation and recoil effects (which are beyond leading log)
and confinement (which is a non-perturbative effect). This also 
separates our approach from the \textsc{IP-Glasma} approach \cite{Schenke:2012wb},
which includes gluon saturation effects in the initial configuration
explicitly, and evolve using B-JIMWLK.

On a more technical note, the approach presented in this paper is
implemented within the larger framework of the \PY~Monte Carlo event
generator. This first of all means that the implementation will become publicly
available,\footnote{From a future version of Pythia, larger than version 8.300,
yet to be determined. See \texttt{https://home.thep.lu.se/Pythia} for up-to-date information.}
and to aid reproducibility and transparency, a large part of the manuscript, as
well as appendix \ref{sec:derivation}, are devoted to the details of the
implemented model. Our approach is simplistic in the sense that only a
minimal amount of corrections to Mueller's original model has been
added, and where ambiguities have arisen, the simplest possible choice
has been taken. 

The structure of the paper is as follows: After this introduction, the
pQCD model of Mueller is introduced. Then follows a description on how
observables are calculated within the Good-Walker framework as well as a
definition of the observables related to the substructure of protons.
The next section describes the overall features of the Monte Carlo
implementation, before we proceed to the results on cross sections,
eccentricities and colour fluctuations in processes with incoming
virtual photons. Lastly, a section is devoted to conclusions and
forthcoming work. 

\section{\label{sec:evol}Proton substructure evolution}

In this section we will outline the theoretical basis of the
initial state evolution approach used in subsequent sections, and 
briefly review its relation to other approaches. The theoretical basis is the well known 
dipole QCD model by Mueller \etal~ \cite{Mueller:1993rr,Mueller:1994jq}. 

\subsection{Dipole evolution in impact parameter space}
We consider in general a picture with a projectile with a dipole structure 
incident on a target. In the simplest case, the projectile is 
just a single dipole $r_{12}$, spanned between the coordinates 
$\vec{r}_1$ and $\vec{r}_2$, in impact parameter space. The probability at 
leading order for this dipole to branch when evolved in rapidity ($y$), is
\begin{equation}
	\label{eq:mueller}
	\frac{\d\mathcal{P}}{\d y} = \d^2\vec{r}_3~ \frac{N_c\alpha_s
  }{2\pi^2}\frac{r^2_{12}}{r^2_{13}r^2_{23}} \equiv \d^2\vec{r}_3~\kappa_3.
\end{equation}
Here $\vec{r}_3$ is the transverse coordinate of the emitted gluon and 
$\kappa_3$ is used as a short-hand for the splitting kernel. An 
observable $O$ known initially, will after an infinitesimal interval $\d y$ 
have the expectation value (denoted by a bar), assuming unitarity:
\begin{equation}
	\bar{O}(y + \d y) = \d y \int \d^2\vec{r}_3~\kappa_3 \left[O(r_{13})
  \otimes O(r_{23}) \right] + O(r_{12}) \left[ 1 - \d y\int \d^2\vec{r}_3~\kappa_3 \right],
\end{equation}
where $\otimes$ denotes the evaluation of the observable $O$ in the two dipole system $r_{13},r_{23}$. 
In the limit $\d y\rightarrow 0$ this becomes:
\begin{equation}
	\label{eq:oevolution}
	\frac{\partial \bar{O}}{\partial y} = \int \d^2\vec{r}_3~\kappa_3
  \left[O(r_{13}) \otimes O(r_{23}) - O(r_{12})\right].
\end{equation}

Remarkably, \eqref{eq:oevolution} allows for the evolution of any 
observable calculable in impact parameter space. In the case
of $S$-matrices in impact parameter space, the evaluation in the two dipole
system reduces to a normal product in the eikonal approximation. Thus $O(r_{13})\otimes O(r_{23}) 
\rightarrow S(r_{13})S(r_{23})$. Changing
to scattering amplitudes, $T$, by substituting $T \equiv 1 - S$, one obtains:
\begin{equation}\label{eq:b-jimwlk}
	\frac{\partial \smallbar{\langle T \rangle}}{\partial y} = \int \d^2
  \vec{r}_3~\kappa_3 \left[ \langle T_{13} \rangle + 
	\langle T_{23} \rangle - \langle T_{12} \rangle - \langle T_{13}T_{23} \rangle \right]. 
\end{equation}
This is the B-JIMWLK equation hierarchy \cite{Balitsky:1995ub,Balitsky:1998kc,
Balitsky:2001re,JalilianMarian:1997jx,JalilianMarian:1997gr,
Iancu:2001ad,Iancu:2000hn,Weigert:2000gi} in impact parameter space,
which, as shown already by Mueller \cite{Mueller:2001uk}, can be
generated directly from \eqref{eq:mueller}. Equation (\ref{eq:b-jimwlk}) includes a
non-linear term, $\langle T_{13}T_{23}\rangle$, and the treatment of this
term is defining for many of the various approaches dealing with initial
state evolution at low $x$.

Removal of the non-linear term yields the BFKL (Balitsky, Fadin, Kuraev and Lipatov) 
equation \cite{Kuraev:1977fs,Balitsky:1978ic}, 
which correctly  sums all of the leading logarithms in energy (or, more precisely in 
rapidity $(\alpha_s \cdot y)^n$) to all orders. Other than simply neglecting it, the
simplest treatment of the non-linear term is by a mean-field approach, where it factorizes as:
$\langle T_{13}T_{23}\rangle \rightarrow \langle T_{13}\rangle\langle T_{23}\rangle$. 
This approximation yields the BK (Balitsky and Kovchegov) equation \cite{Kovchegov:1999yj,Balitsky:1995ub}.

\subsection{The Mueller dipole model}
\label{sec:dipole-theory}
Several approaches have been proposed to utilize the simple, but powerful
evolution equation introduced in \eqref{eq:b-jimwlk}. Here we will focus on the 
Mueller dipole model that neglects the non-linear term completely, but is 
particularly suitable for calculation of geometric quantities.
Usually, \eqref{eq:b-jimwlk} is solved as an initial value problem: given
a scattering matrix at small initial rapidity ($y_0$), it determines the resulting
scattering matrix at any $y \geq y_0$. Note however, that \eqref{eq:oevolution} 
is applicable for any type of observable calculable in impact-parameter space, 
notably observables linked to the geometry of the partonic initial state. As an example, 
consider the average vertex coordinate position, $\langle z \rangle$, where $z$ 
is either the $x$ or $y$ coordinate of a dipole. For a single dipole 
$\langle z \rangle = (z_1 + z_2) / 2$,  for the two dipole system 
$\langle z \rangle = (z_1 + z_2 + z_3)/3$, where the two dipoles has a common point $z_3$, and directly:
\begin{equation}
	\frac{\partial \smallbar{\langle z \rangle}}{\partial y} = \int
  \d^2\vec{r}_3
  \kappa_3 \left(\frac{1}{3} z_3 -\frac{1}{6}(z_1 + z_2)\right).
\end{equation}

For more complicated geometric observables, such as eccentricity (see \sectref{sec:eccentricity}),
the analytic expressions become quite involved, and must be handled observable by observable. They are, 
however, quite easy to handle in a Monte Carlo, where any $O$ can be
evaluated event by event, and the expectation value extracted from a large statistics sample.

\begin{figure}[t]
\centering
\includegraphics[width=0.4\linewidth]{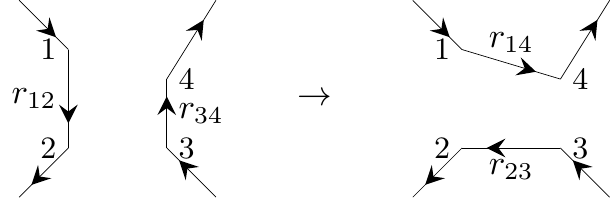}
\caption{\label{Fig:DipInt}
Schematic view of two colliding gluon dipoles. The initial dipoles denoted
$r_{12}$ and $r_{34}$ are allowed to interact via two-gluon exchange. This
results in the creation of two new dipoles, $r_{14}$ and $r_{23}$ and a
connection of the two dipole chains. The lines $r_{13}$ and $r_{24}$ are 
not drawn, but enters in \eqref{Eq:fij}.
}
\end{figure}

The starting point for the model, is the evolution of an Onium (or
$\gamma^* \rightarrow \q\qbar$)
state in transverse space and rapidity, following \eqref{eq:mueller}. Instead of
calculating average quantities directly from the evolution equation, Monte Carlo
events are generated, by performing a probabilistic evolution of a given
initial state, corresponding to a collision event performed by an experiment.
The calculational details of performing such an evolution are deferred to section 
\ref{sec:mc-implementation} and appendix \ref{sec:derivation}. It is, however,
important to note here the approximation of this evolution, namely that
all dipoles in the dipole-chain radiate independently, removing the non-linear
effect from the cascade itself.

After a full evolution in rapidity, a single dipole will have
evolved to a chain of dipoles, each of which are allowed to 
interact with dipoles from another evolved system through gluon 
exchange. The lowest order interaction between two dipoles, at amplitude 
level, is single gluon exchange, resulting in two gluon exchange at cross section level.
This cross section can be related to the elastic amplitude (cf.\ section
\ref{sec:good-walker}) through the
optical theorem. The dipole-dipole cross section depends on the distances between the interacting dipoles (the enumeration of dipoles follows figure \ref{Fig:DipInt}) as \cite{Kovchegov:2012mbw}:
\begin{align}\label{Eq:fij}
\frac{\d\sigma_{\mrm{dip}}}{\d^2\vec{b}} =& \frac{\alpha_s^2C_F}{N_c}\log^2
\left[\frac{r_{13}r_{24}}{r_{14}r_{23}}\right] \nonu\\
 \rightarrow& \frac{\alpha_s^2}{2}\log^2
 \left[\frac{r_{13}r_{24}}{r_{14}r_{23}}\right] \equiv f_{ij},
\end{align}
where the arrow indicates that the 't Hooft large-$N_c$
limit\footnote{The 't Hooft large-$N_c$ limit is the limit where factors 
of $\alpha_SN_c$ are kept fixed while factors of $1/N_c^2$ are
suppressed.} is taken to reach line 2 of \eqref{Eq:fij}, which then
defines $f_{ij}$. The 't Hooft large-$N_c$ limit is taken in order to
ensure consistency with the leading logarithmic approximation in the
(BFKL) evolution. The
distances $r_{ij}$ are indicated in figure \ref{Fig:DipInt}, except for 
$r_{13}$ and $r_{24}$, the distances between (anti-)colour-(anti-)colour
pairs (1,3) and (2,4). Where the dipole evolution comes with a factor of
$\alpha_sN_c$, the dipole-dipole cross section is proportional to
$\alpha_s^2$. Thus in the 't Hooft limit where $\alpha_S\sim1/N_c$, the 
dipole evolution is of order $N_c^0\sim1$, while the dipole-dipole 
interaction is of order $1/N_c^2$. This means the dipole-dipole
interaction is formally $N_c$-suppressed compared to the dipole evolution 
\cite{Kovchegov:2012mbw}.
 
A single collision can contain several dipole-dipole scatterings,
equivalent to MPIs in a standard parton language. Assuming that the
individual scatterings are uncorrelated, the contribution from each
scattering exponentiates, resulting in the unitarized scattering 
amplitude for a single event (see section \ref{sec:good-walker}):
\begin{align}\label{eq:unitamp}
T(\vec{b}) = 1 - \exp\left(-\sum_{ij}f_{ij}\right).
\end{align}
As each $f_{ij}$ comes with a factor of $1/N_c^2$,
the unitarized scattering amplitude correctly resums
$1/N_c^2$-suppressed terms in the interaction. In Regge terminology, 
each scattering $f_{ij}$ can be viewed as a Pomeron exchange. The first 
term in the expansion corresponds to single Pomeron exchange, and the 
latter terms to multi-Pomeron exchanges. The unitarized scattering 
amplitude can thus also be viewed as a resummation of all possible 
Pomeron exchanges in the collision frame. 

An expansion of the exponent in \eqref{eq:unitamp} into a power 
series results in factors of $(\sum_{ij}f_{ij})^n$. To second order 
this results in a term quadratic in $\sum_{ij}f_{ij}$, which corresponds 
to the mean-field approximation of the non-linear term from 
\eqref{eq:b-jimwlk}. As this non-linear term corresponds to 
saturation, we note that the dipole framework does not include 
saturation in the evolution, but, when using the unitarized 
scattering amplitude, non-linearity is included in the interaction frame,
and only there.

\subsection{Dipole evolution beyond leading order}
\label{sec:mueller-nlo}
Significant formal progress has been made in the pursuit of systematic
next-to-leading order (NLO) in $\alpha_s$ corrections to the BK equation \cite{Balitsky:2008zza}
and the full B-JIMWLK hierarchy \cite{Kovner:2013ona,Kovner:2014lca,Lublinsky:2016meo}.
Numerical studies of NLO BK \cite{Lappi:2015fma} have, however, shown that 
the equation becomes unstable for some values of the initial conditions, making it yet
unsuitable for a full Monte Carlo implementation. Recent work by Ducloué \etal~
\cite{Ducloue:2019ezk} have shown that, for a specific choice of the initial scattering matrix,
some problems of unphysical results can be overcome in the dilute-dense limit, by reformulating
the NLO evolution equation w.r.t.\ rapidity of the dense target.
This gives hope that a future improvement of the model, implemented as a Monte Carlo in this paper,
could include formal improvements beyond leading order, but at this point it is not deemed feasible.

An approach for going beyond leading colour in the cascade, which is
also suited for Monte Carlo implementation, is the so-called "swing"
mechanism, introduced by Avsar \etal~\cite{Avsar:2006jy,Avsar:2007xh}.
This can be understood as an extension of the identification of multiple
interactions in the collision frame with Pomeron loops, as presented in
the previous section. Since loops cannot be formed during the BFKL-like
evolution, only loops cut in the collision frame are included. The
problem is then posed as equivalent to forming $1/N^2_c$ suppressed
dipole configurations in the evolution, by allowing dipoles to reconnect in such a way
that the formalism becomes frame independent. This is another viable
path for future extensions beyond leading order. Further work on the
formalism is needed, however. Currently only a $2 \rightarrow 2$ dipole
swing has been thoroughly studied, which is not enough to make the
formalism fully frame independent. Going beyond $2\rightarrow 2$ is a
full study by itself, and not considered in the present paper.

In this paper we instead choose to include corrections beyond (formal) leading-log arising from
energy--momentum conservation. It is well known that the leading-log BFKL equation,
derived in the high-energy limit, will get sizable corrections at collider energies \cite{Beuf:2014uia}.
From studies of the full next-to-leading log BFKL
\cite{Fadin:1998py,Ciafaloni:1998gs}, it is shown that contributions
beyond leading log are very large, and a sizable amount are related to
energy-momentum conservation \cite{Salam:1999cn}. In a Monte Carlo such
corrections can be implemented directly, see details in appendix
\ref{sec:derivation}. Related are non-eikonal corrections. Non-eikonal
corrections arise due to the large but finite energy available during
the cascade. In the CGC approach this can be understood as sub-leading
effects to infinite Lorentz dilation of the projectile, which are
troublesome but manageable analytically \cite{Altinoluk:2014oxa}. In a
Monte Carlo implementation of the dipole model, the finite energy can be
treated as recoil effects in the dipole splittings.

A non-perturbative effect from confinement is also included in our
simulation. This must
be done both in the cascade, where large dipoles must be suppressed, and
in the interaction, where the range of the interaction must be limited
to take confinement into account. Following ref. \cite{Avsar:2007xg},
this is done by replacing $1/p^2_g$ in the Coulomb propagator implicitly
entering \eqref{Eq:fij} by $1/(p^2_g + M^2_g)$, where $M_g$ can be taken
as a confinement scale, or a fictitious gluon mass. This changes the expressions for the splitting kernel and the dipole--dipole interaction probability, the full expressions are written in section \ref{sec:mc-implementation}. 

\section{From substructure to observables}
The following section is dedicated to the introduction of the framework 
used for linking partonic substructure to physical observables, 
such as cross sections and flow coefficients. First, we describe the 
Good--Walker formalism for calculating cross sections for particles with an inner structure, from 
obtained scattering amplitudes, and secondly, the apparent scaling of 
flow coefficients with initial state eccentricity seen in heavy ion 
collisions is explained.

\subsection{The Good--Walker formalism and cross sections}
\label{sec:good-walker}
The Good-Walker formalism is a method of calculating cross sections of
particles with a well-defined wave function. It includes a normalised and
complete set of eigenstates $\{|\psi_i\rangle\}$ of the imaginary part of the 
scattering amplitude (neglecting the real part, which is vanishing at high energies), denoted
$\hat{T}(\vec{b})$ (related to the $\hat{S}$-matrix through $\hat{T} \equiv 1-\hat{S}$), with eigenvalues
$\hat{T}(\vec{b})|\psi_i\rangle=T_i(\vec{b})|\psi_i\rangle$. These
scattering states have equal quantum numbers, but differ in masses. 
The wave function of the incoming beams can thus be expressed in
terms of the above eigenstates, and written in short-hand notation as
\begin{align}
|\psi_I\rangle=|\psi_p,\psi_t\rangle=\sum_{p=1}^{N_p}\sum_{t=1}^{N_t}c_pc_t|\psi_p,\psi_t\rangle
\end{align}
with $|\psi_{p,t}\rangle$ denoting the projectile and target
wave functions, respectively, and $c_{p,t}$ the expansion coefficients. The
scattered wave function is found by operating with the transition
matrix on the incoming wave function, 
\begin{align}
|\psi_S\rangle = \hat{T}(\vec{b})|\psi_I\rangle= \sum_{p,t}c_pc_t~T_{p,t}(\vec{b})|\psi_p,\psi_t\rangle~, 
\end{align}
and from these definitions, the profile function for elastic scattering (at fixed Mandelstam $s$) 
can be defined:
\begin{align}
\label{eq:elprofile}
\Gamma_{\mrm{el}}(\vec{b}) =& \langle \psi_S|\psi_I\rangle 
  = \sum_{p,t}|c_p|^2|c_t|^2~T_{p,t}(\vec{b})\langle \psi_p,\psi_t|\psi_p,\psi_t\rangle \nonu\\
  =& \sum_{p,t}|c_p|^2|c_t|^2~T_{p,t}(\vec{b}) \equiv \langle T(\vec{b})\rangle_{p,t}~,
\end{align} 
where we have defined an average over projectile and target states in
the last equality and suppressed indices on $T$ inside the average
(which is done in all the following, unless specifically noted otherwise). 
Thus we obtain the cross sections and elastic slope in the eikonal 
approximation (again also at fixed Mandelstam $s$),
\begin{align}
\label{Eq:XS1}
\sigma_{\mrm{tot}}=&2\int\d^2\vec{b}\Gamma(\vec{b}) =
  2\int \d^2\vec{b}~\langle T(\vec{b})\rangle_{p,t},\\
\label{Eq:XS2}
\sigma_{\mrm{el}}=& \int \d^2\vec{b} |\Gamma(\vec{b})|^2 
  = \int \d^2\vec{b}~ \langle T(\vec{b})\rangle_{p,t}^2,\\
\label{Eq:XS3}
B_{\mrm{el}}=&\frac{\partial}{\partial
t}\log\left(\frac{\d\sigma_{\mrm{el}}}{\d
t}\right)\Big|_{t=0}=\frac{\int\d^2\vec{b}~ b^2/2~\langle
T(\vec{b})\rangle_{p,t}}{\int\d^2\vec{b}~\langle
T(\vec{b})\rangle_{p,t}}~.
\end{align}

In eqs.~(\ref{Eq:XS1}--\ref{Eq:XS3}) we have implicitly assumed a particle wave function
$\langle \psi|\psi\rangle=1$. In cases where the wave function is not
normalizable, one has to take into account the wave function in the above cross
sections. This includes processes with photons, where the wave function
is well-defined in pQCD for high virtualities. The total $\gamma^*\p$ cross section would thus require an
additional integration over wave function parameters: 
\begin{align}
\label{eq:gp-xsec}
\sigma^{\gamma^*\p}(s)=\int_0^1\d
z\int_0^{r_{\mrm{max}}}r\d
r\int_0^{2\pi}\d\phi\left(|\psi_L(z,r)|^2+|\psi_T(z,r)|^2\right)\sigma_{\mrm{tot}}(z,\vec{r}),
\end{align}
with $z$ the fractional momentum carried by the quark, $r$ the distance
between the quark and anti-quark, $\psi_{L,T}$ the longitudinal and
transverse parts of the photon wave function and $\sigma(z,\vec{r})$ the
dipole cross section calculated from the elastic profile function,
\eqref{Eq:XS1}. 
The photon wave function implemented in our approach is given 
in eqs.~(\ref{Eq:photonWf1}--\ref{Eq:photonWf2}) and the discussion for $\gamma^*\A$ is continued
in section \ref{sec:results3}.

\subsection{Eccentricity scaling of flow observables}
\label{sec:eccentricity}
Anisotropic flow is measured as momentum space anisotropies
and quantified in flow coefficients ($v_n$), obtained by a 
Fourier expansion of the azimuthal ($\phi$) spectrum:
\begin{equation}
	\frac{\d N}{\d\phi} \propto 1 + 2 \sum_n v_n \cos\left[n(\phi - \Psi_n)\right],
\end{equation}
with $\Psi_n$ the symmetry plane of the
$n$th harmonic. For a hydrodynamical expansion, it has been shown that
$v_2$ and $v_3$ are proportional to the initial state eccentricity 
in the corresponding harmonic, $v_n \propto \epsilon_n$, to a very 
good approximation \cite{Niemi:2012aj}, with the constant of 
proportionality depending on the properties of the QGP transporting 
the initial state anisotropy to the final state. A similar relation may
be expected when a pressure gradient is obtained without a thermalized or
hydrodynamized plasma \cite{Bierlich:2016vgw,Bierlich:2017vhg}. In the following, 
the eccentricities will therefore be taken as a proxy for flow 
observables, noting that the model imposed for the response may 
deviate from this linear scaling behaviour. In $\p\p$ and $\p\A$ 
collisions this type of behaviour becomes very apparent, due to the 
dominance of non-flow effects,\footnote{Including correlations from jets and due to particle decays.}
in particular at small event multiplicities. Non-flow mechanisms aside, 
it is clear that no matter what the actual response is,
measurable observables will be affected by large deviations in predicted eccentricities.

We follow the usual definition of the initial anisotropy or 
participant eccentricity \cite{Qiu:2011iv,Zhou:2016fvj}:
\begin{equation}
	\label{eq:epsilonn}
	\epsilon_n = \frac{\sqrt{\langle r^2 \cos(n \phi)\rangle^2 + \langle r^2 \sin(n \phi)\rangle^2}}{\langle r^2 \rangle}.
\end{equation}
Here $r$ and $\phi$ are usual polar coordinates, with the origin 
shifted to the center of the distribution. From \eqref{eq:epsilonn}, 
higher order cumulants can be calculated:
\begin{align}
	\epsilon^2_n\{2\} &= \langle \epsilon_n^2\rangle, \\
	\epsilon^4_n\{4\} &= 2\langle \epsilon_n^2\rangle^2 - \langle \epsilon_n^4 \rangle, \\
	4\epsilon^6_n\{6\} &= \langle \epsilon_n^6\rangle - 9 \langle \epsilon_n^4\rangle \langle 
	\epsilon_n^2\rangle + 12 \langle \epsilon_n^2\rangle^3, \\ 
	33\epsilon^8_n\{8\} &= 144\langle \epsilon_n^2\rangle^4 + 18\langle \epsilon_n^4\rangle^2 + 
	16\langle \epsilon_n^6\rangle \langle \epsilon_n^2\rangle - 144\langle\epsilon_n^4\rangle 
	\langle \epsilon_n^2 \rangle^2 - \langle \epsilon_n^8 \rangle.
\end{align}

In nuclear collisions, the normal participant nucleon eccentricity 
is used as as a baseline. The notion of ``participating'' is, however, 
a model dependent statement. We use the definition from \textsc{Angantyr} 
\cite{Bierlich:2018xfw,Bierlich:2016smv}, 
which defines participating nucleons as either ``inelastically'' or 
``absorptively'' (inelastic non-diffractively) wounded, see 
appendix \ref{sec:angantyr} for a brief review. For $\p\p$ 
collisions, and for fluctuations in nuclear collisions, we follow 
Avsar \etal~\cite{Avsar:2010rf}, and define a participant \tit{parton} 
eccentricity (though somewhat modified from the cited exploratory work). 
Assuming that the hydrodynamic evolution takes place at the end of the 
perturbative parton cascade, the participant parton eccentricity should 
be evaluated at this point in the evolution. In 
\sectref{sec:cms-pa-fluctuations} this participant parton eccentricity 
will be compared to a more purist initial state approach, where the 
final state parton cascade is not included. This is meant to inform a 
discussion about what the notion of an ``initial state'' really ought to 
entail. 

Parton level eccentricities are, however, not infrared safe. Consider the 
simple example of a soft gluon emission at the same impact parameter point 
as its mother. Such an emission will double count this spatial point at 
parton level, but disappear after hadronization, which will place two such 
partons inside the same hadron. To improve this, all contributions are 
weighted by a factor $\pT/(\pT + p_{\perp\mrm{min}})$, where
$p_{\perp\mrm{min}} = 0.1$ GeV ensures 
that considerably soft gluons will not double count.

Normalised symmetric cumulants will also be studied. Such quantities 
eliminate the dependence on the magnitude of the flow coefficients, and 
should thus remove the response factor between flow harmonics and 
eccentricities, and directly probe the substructure \cite{Albacete:2017ajt}. 
They are defined as:
\begin{align}
NSC(n,m)=&\frac{\langle v_n^2v_m^2\rangle-\langle v_n^2\rangle\langle
v_m^2\rangle}{\langle v_n^2\rangle\langle v_m^2\rangle}
\approx\frac{\langle \epsilon_n^2\epsilon_m^2\rangle-\langle \epsilon_n^2\rangle\langle
\epsilon_m^2\rangle}{\langle
\epsilon_n^2\rangle\langle\epsilon_m^2\rangle},
\end{align}
where the last approximate equality indicates the removal of the response.
Especially interesting for this study is $NSC(3,2)$, it being sensitive to 
initial-state fluctuations, namely the geometric correlation between 
$\epsilon_2$ and $\epsilon_3$, the elliptical and triangular parts of the 
Fourier expansion.

Finally it is noted that, since the model is implemented in a full event 
generator able to generate full final states for $\p\p$, $\p\A$ and $\A\A$ 
collisions, it is possible to investigate the event geometry as a function 
of final state multiplicity with the same acceptance as the experiment.

\section{Monte Carlo implementation}
\label{sec:mc-implementation}
In this section, the Monte Carlo implementation of Mueller's model is
briefly described. The full details are given in appendix
\ref{sec:derivation}. First, the details of the various initial states
are described, then some assumptions on the cascade and the interaction
are described, and lastly, some geometric properties of the evolution are
presented. 

\subsection{The initial states}

The new implementation is applicable for both virtual photon and 
proton beams. A photon state is represented by a single dipole, with a
wave function given as, 
\begin{align}\label{Eq:photonWf1}
|\psi_L(z,r)|^2=&\frac{6\alpha_{\mrm{em}}}{\pi^2}\sum_qe_q^2Q^2z^2(1-z)^2
  K_0^2\left(\sqrt{z(1-z)}Qr\right)\\
\label{Eq:photonWf2}|\psi_T(z,r)|^2=&\frac{3\alpha_{\mrm{em}}}{2\pi^2}\sum_qe_q^2Q^2\left[z^2+(1-z)^2\right]z(1-z)
  K_1^2\left(\sqrt{z(1-z)}Qr\right),
\end{align}
where we include the three lightest (massless) quarks. Here $z$ is the 
longitudinal momentum fraction carried by the quark, $(1-z)$ the longitudinal momentum fraction
carried by the anti-quark, $r$ is the distance 
between them, $Q^2$ the photon virtuality and $K_i$ the modified Bessel 
functions. For protons, the wave function is not known. Instead it is
represented by three dipoles in an equilateral triangle
configuration and normalized to unity, shown previously to give the best
description of data \cite{Avsar:2006jy}. The lengths of the initial dipoles 
are allowed to fluctuate on an event-by-event basis, chosen from a Gaussian 
distribution with mean $r_0$, and width $r_{\mrm{w}}$. 

\subsection{The dipole evolution}

To implement \eqref{eq:mueller} as a parton shower, it is
modified by a Sudakov factor:
\begin{align}
	\frac{\d\mathcal{P}}{\d y~\d^2\vec{r}_3 } &= \frac{N_c\alpha_s
  }{2\pi^2}\frac{r^2_{12}}{r^2_{13}r^2_{23}}
  \exp\left(-\int_{y_{\mrm{min}}}^{y}\d y\d^2\vec{r}_3~ \frac{N_c\alpha_s
  }{2\pi^2}\frac{r^2_{12}}{r^2_{13}r^2_{23}}\right)\nonu\\&\equiv\frac{N_c\alpha_s
  }{2\pi^2}\frac{r^2_{12}}{r^2_{13}r^2_{23}}\Delta(y_{\mrm{min}},y)~,
\end{align}
allowing for a trial emission from each dipole in the cascade. The
strategy of ``the winner takes it all'' is then employed, such that only
the trial emission with the lowest rapidity is chosen as a true
branching. This lowest rapidity then becomes the minimal rapidity in the next
(trial) emission. The process is reiterated until none of the trial emissions 
are below a maximal rapidity, governed by the energy of the collision,
\begin{align}\label{Eq:yMax1}
\p\p:\quad y_{\mrm{max}}^{\p}=&\log\left(\frac{\sqrt{s}}{m_p}\right),\\
\label{Eq:yMax2}\gamma^*\p:\quad y_{\mrm{max}}^{\p,
\gamma^*}=&\log\left(\frac{W}{m_0}\right),
\end{align}
with $m_p$ the proton mass, $m_0$ a reference scale set to 1 GeV,
and $W,\sqrt{s}$ the $\gamma^*\p$ and $\p\p$ center-of-mass (CM) energies,
respectively. Note that \eqref{Eq:yMax2} is an approximation to the
actual rapidity available for the dipole formed by a virtual photon. The
``true'' rapidity is not well-defined for virtual photons, as it depends 
not only on $W$, but also on $Q^2$ and momentum fractions carried by the 
quark and anti-quark ends of the dipole. This introduces different 
rapidity ranges available for either end of the dipole, complicating the 
evolution further. Equation~(\ref{Eq:yMax2}) was chosen as the simplest
possible rapidity range.

If confinement is taken into account (as described in section \ref{sec:mueller-nlo}), the evolution equation is modified
accordingly:
\begin{align}
	\label{eq:mod-split-kern}
\frac{\d P}{\d y~\d^2 \vec{r}} =&
  \frac{N_c\alpha_s}{2\pi^2} \frac{1}{r^2_{\mrm{max}}} \left[ \frac{\vec{r}_{13}}{|r_{13}|}K_1(|r_{13}|/r_{\mrm{max}}) -  \frac{\vec{r}_{23}}{|r_{23}|}K_1(|r_{23}|/r_{\mrm{max}})\right]^2\Delta(y_{\mrm{min}},y),
\end{align}
with $K_1$ the modified Bessel functions of the first kind and
$r_{\mrm{max}}$ a maximal radius of the initial dipole, left as a
tunable parameter.

\subsection{Geometric properties of the dipole evolution}

\begin{table}[t]
\centering
\begin{tabular}{l | c | c }
Parameter & Value & Meaning\\
\hline
$r_0$ [fm]	& 1. & Mean of normally distributed initial dipole sizes\\
$r_{\mrm{w}}$ [fm]	& 0. & Width of normally distributed initial dipole sizes\\
$r_{\mrm{max}}$ [fm]	& 1. & Maximally allowed dipole size (confined evolution only)\\
$\alpha_s$	& 0.21 & Value of fixed strong coupling\\
\end{tabular}
\caption{\label{Tab:PascalParams} The input parameters used in this
section.}
\end{table}

\begin{figure}[t]
\centering
\includegraphics[width=0.75\linewidth]{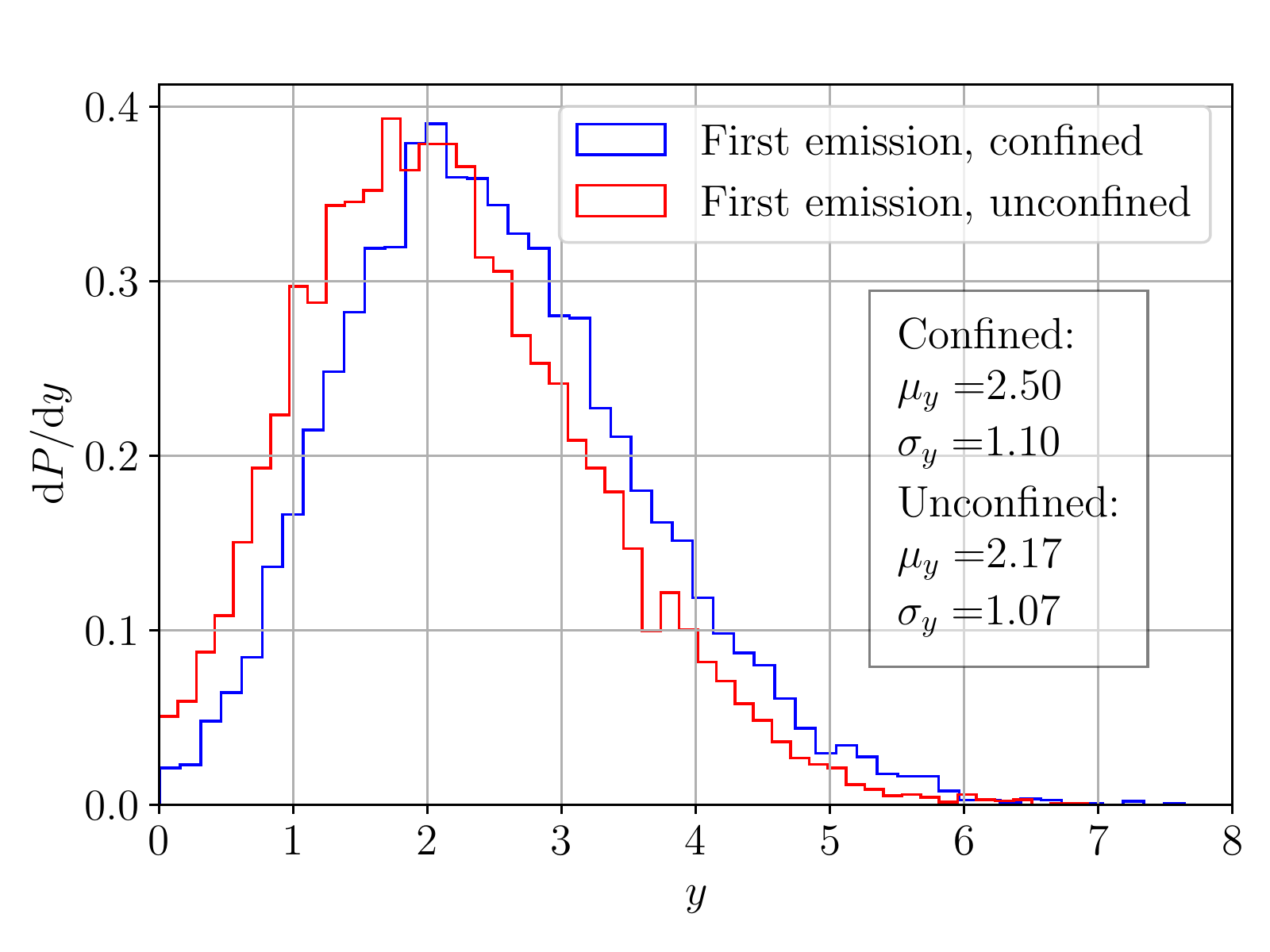}\\
\caption{\label{Fig:dPdY}The probability distribution in rapidity for unconfined (a) and
confined (b) dipole-evolution. The box shows the average and spread of
the distributions. 
}
\end{figure}

Given a specific parameter set, table \ref{Tab:PascalParams}, the 
probability distribution in rapidity for the first emission, 
$\d P/\d y$, is shown in figure \ref{Fig:dPdY}. This distribution has a 
mean at around two units of rapidity. Thus, on average, a new emission 
is assigned a rapidity of roughly two units larger than the mother 
(or emitter) dipole. It is worth noting that the inclusion of 
confinement effects slightly increases the mean as compared to the 
unconfined distribution. This is caused by the additional 
suppression of large dipoles, requiring large dipoles to be discarded 
in the evolution and an emission at a larger rapidity to be tried.

\begin{figure}[t]
\begin{minipage}[c]{0.475\linewidth}
\centering
\includegraphics[width=\linewidth]{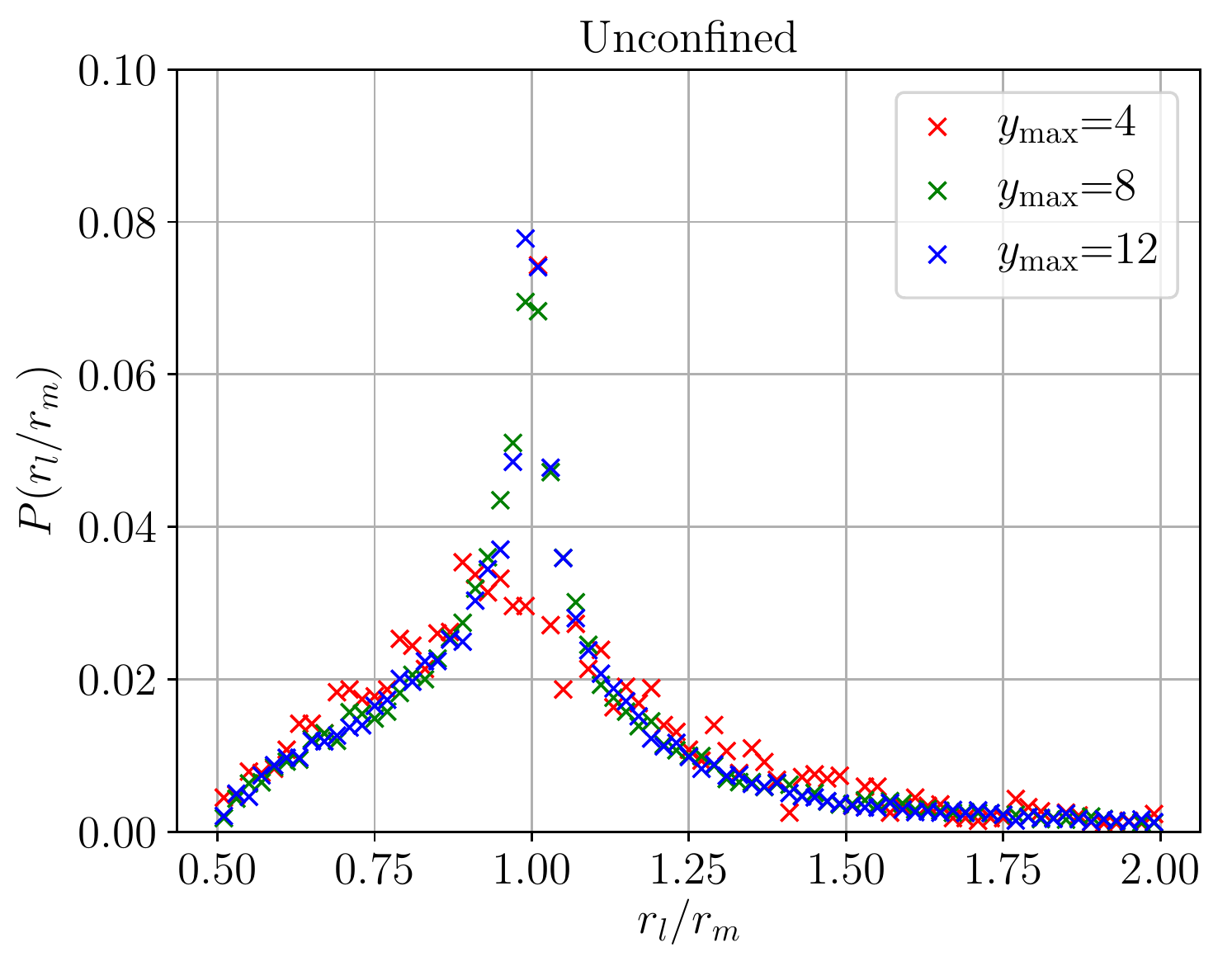}\\
(a)
\end{minipage}
\hfill
\begin{minipage}[c]{0.475\linewidth}
\centering
\includegraphics[width=\linewidth]{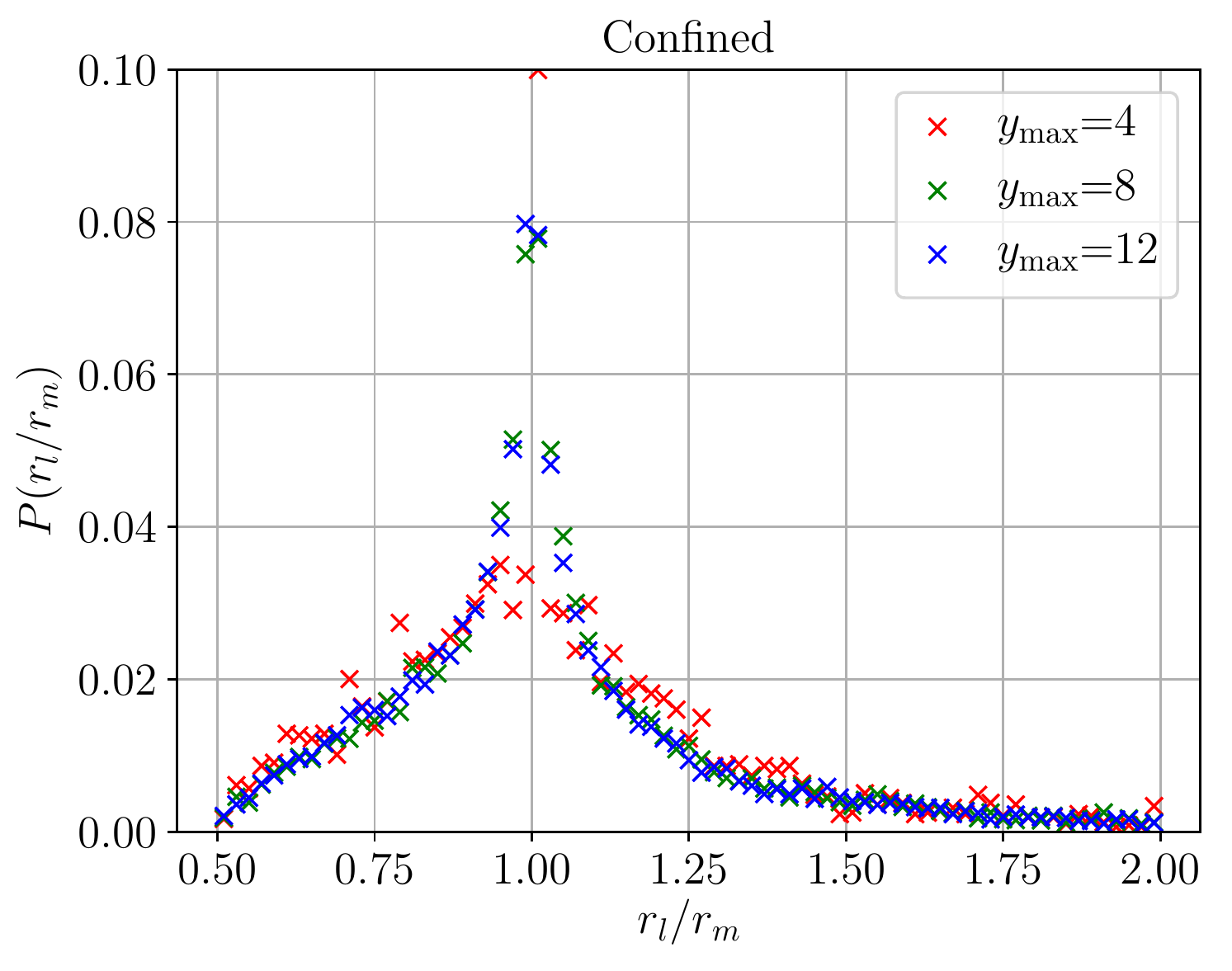}\\
(b)
\end{minipage}
\begin{minipage}[c]{0.475\linewidth}
\centering
\includegraphics[width=\linewidth]{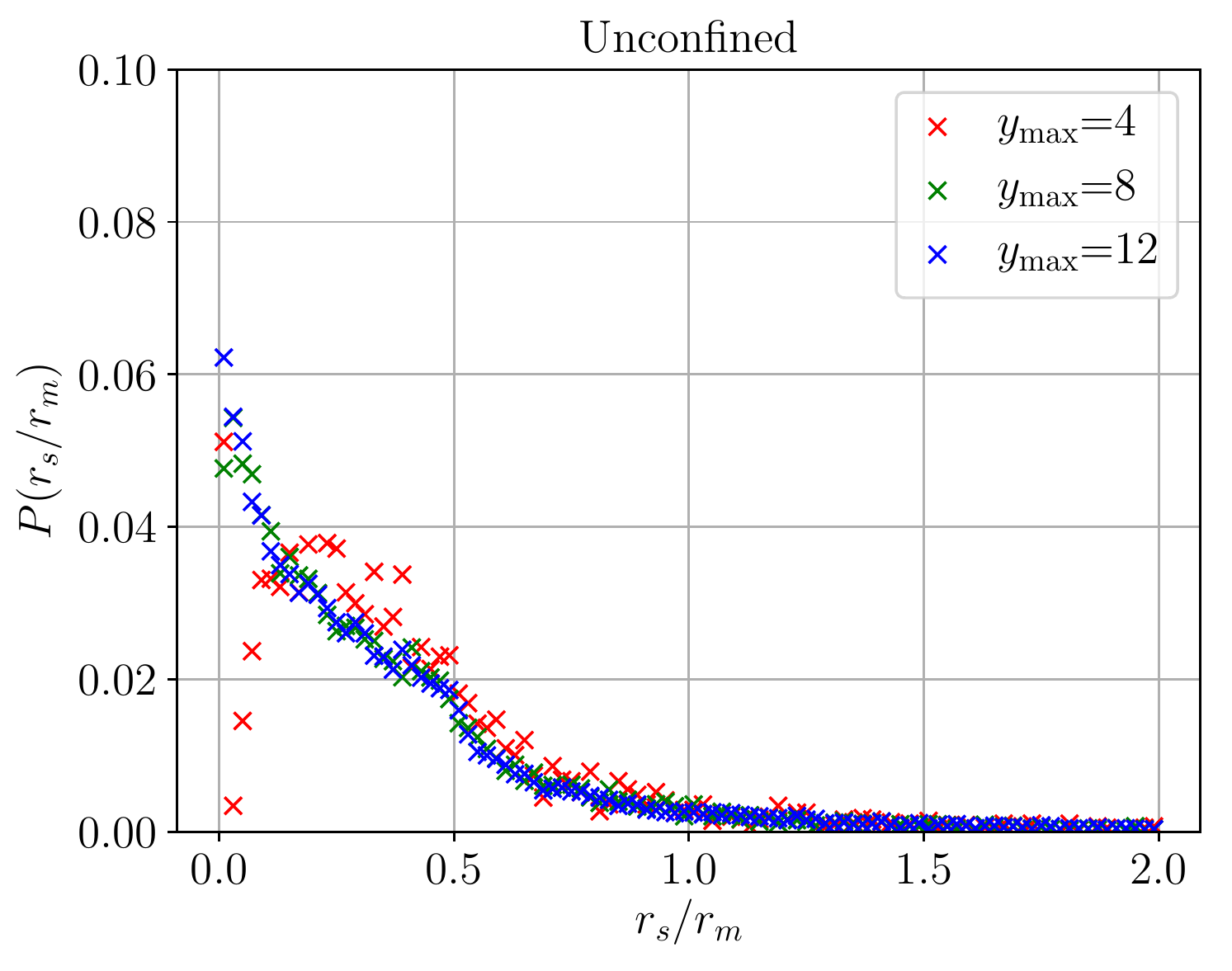}\\
(c)
\end{minipage}
\hfill
\begin{minipage}[c]{0.475\linewidth}
\centering
\includegraphics[width=\linewidth]{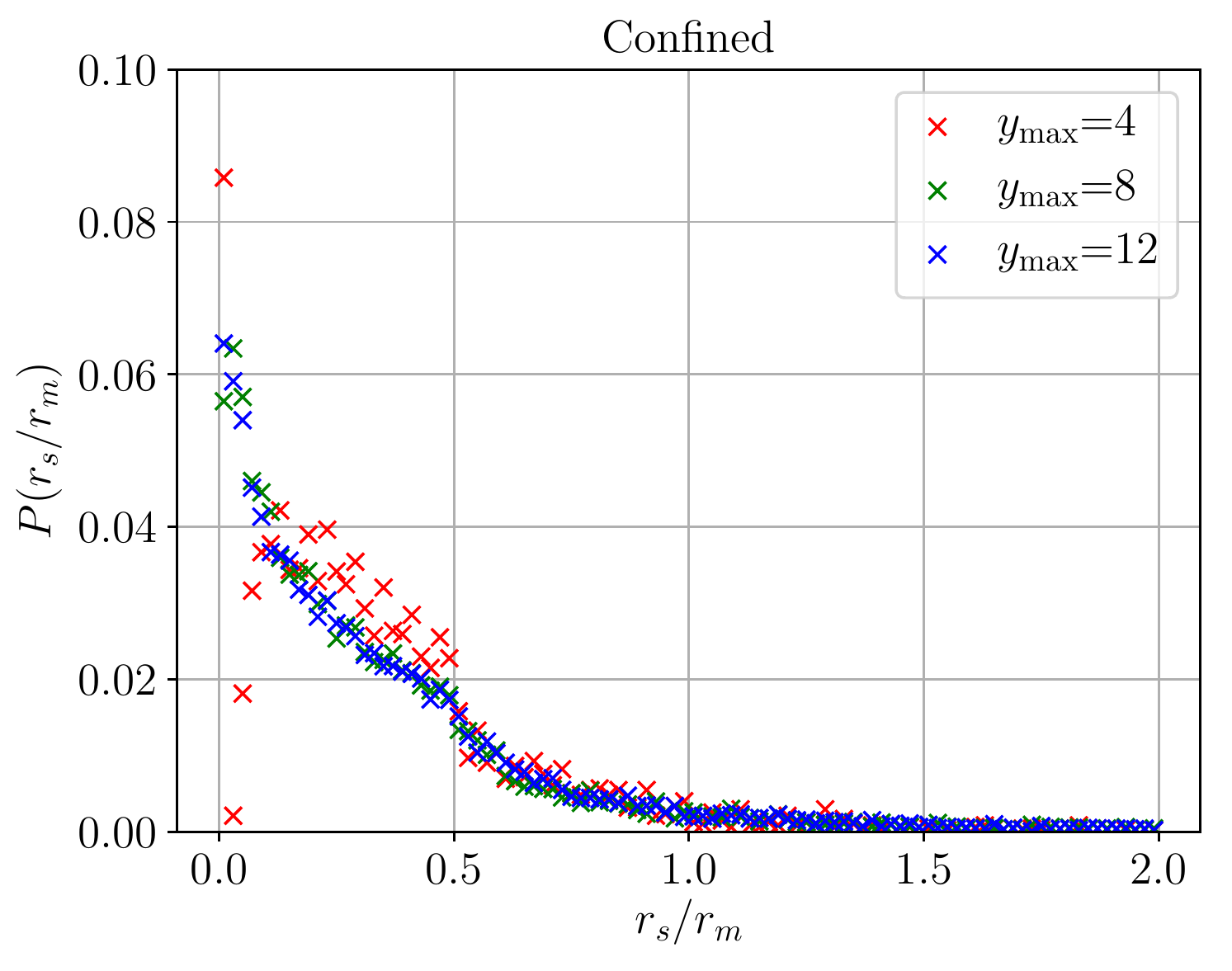}\\
(d)
\end{minipage}
\caption{\label{Fig:Rratios}The scaled lengths of the daughter dipoles w.r.t.\ the mother
dipole as a function of maximal rapidity for unconfined (a,c) and
confined (b,d) dipole-evolution. Figures (a,b) shows the larger of the
daughter dipoles, while figures (c,d) shows the smaller of the daughter
dipoles. The parameters used in the dipole
evolution are the same as presented in table \ref{Tab:PascalParams}. 
}
\end{figure}

In each step of the dipole evolution a mother dipole is
split into two daughters. Figure \ref{Fig:Rratios} shows the distribution in sizes of
the smaller and larger dipole, scaled w.r.t.\ their mothers' size for
three different evolutions ($y_{\mrm{max}}=4,8,12$). Here, it is evident
that on average the larger dipole retains the size of the mother, while
the distribution of the smaller is much broader. At lower $y_{\mrm{max}}$
there is a bump in the distribution at around 30-40\% of the mothers
size, while this bump is less pronounced at larger maximal rapidity.

\begin{figure}[t]
\begin{minipage}[c]{0.475\linewidth}
\centering
\includegraphics[width=\linewidth]{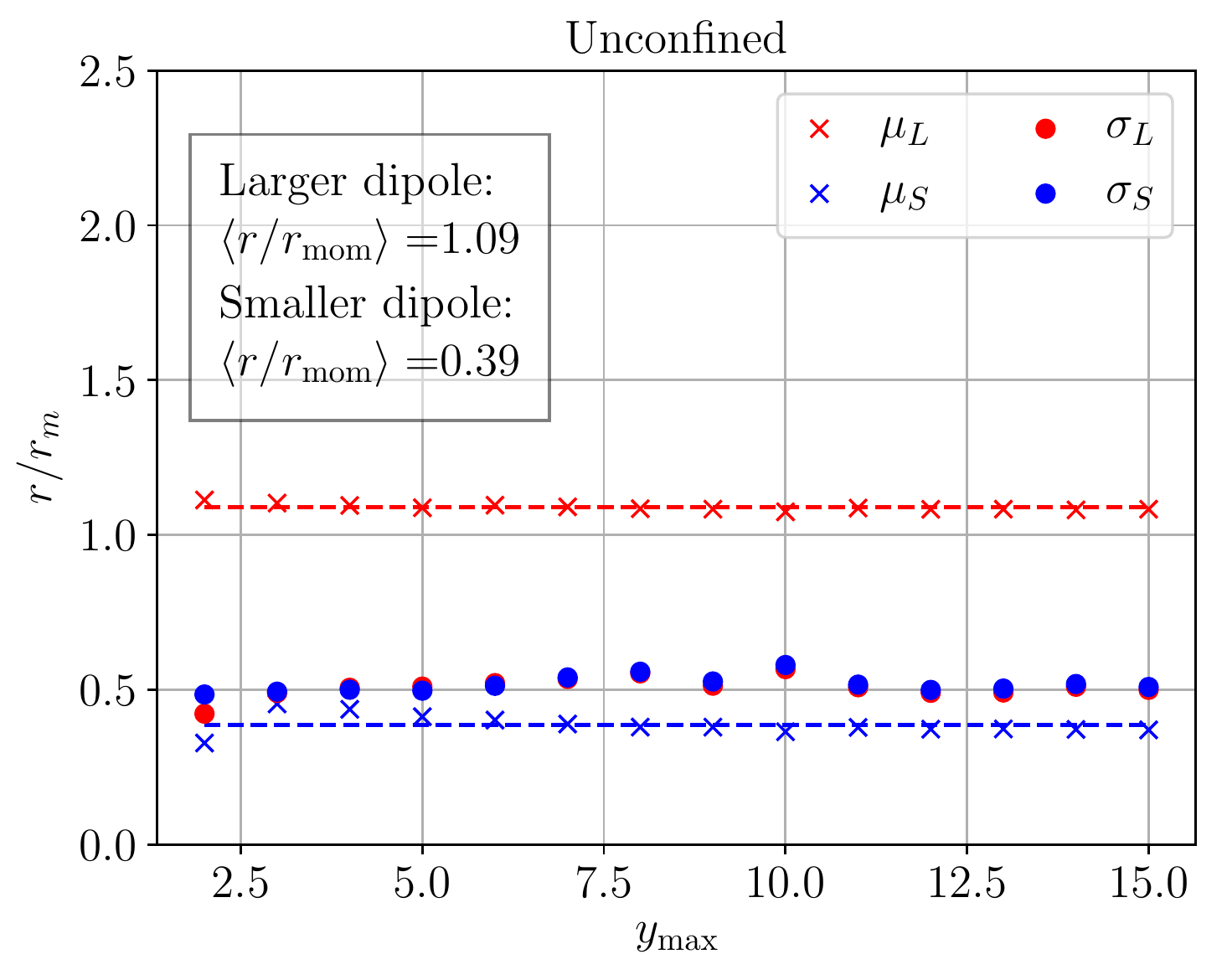}\\
(a)
\end{minipage}
\hfill
\begin{minipage}[c]{0.475\linewidth}
\centering
\includegraphics[width=\linewidth]{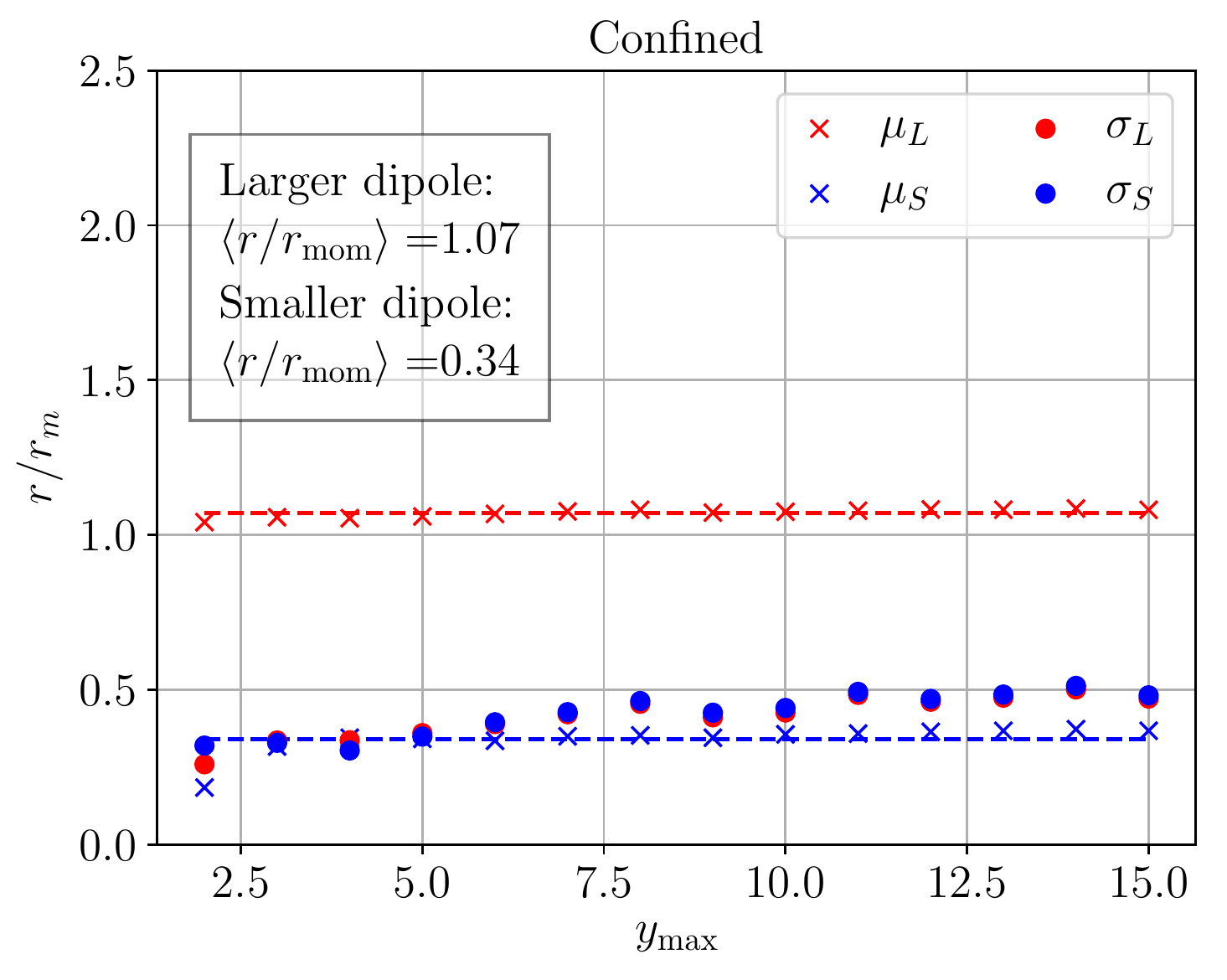}\\
(b)
\end{minipage}
\caption{\label{Fig:RratiosMeanAndSpread}The scaled lengths of the daughter dipoles w.r.t.\ the mother
dipole as a function of maximal rapidity for unconfined (a) and
confined (b) dipole-evolution. The parameters used in the dipole
evolution are the same as presented in table \ref{Tab:PascalParams}. 
}
\end{figure}

Figure \ref{Fig:RratiosMeanAndSpread} shows the corresponding average and standard deviation in
the lengths of all the daughter dipoles scaled w.r.t.\ the length of their 
mothers', as a function of maximal rapidity of the evolution. As stated above, it is clear
that while the larger of the two daughter dipoles can be taken to be identical to the mother dipole,
the size of the smaller dipole has larger fluctuations. The average size of the smaller dipole
is, however, fixed at roughly a third of the mother dipole for all $y_{\mrm{max}}$.

\begin{figure}[t]
\centering
\begin{minipage}[c]{0.5\linewidth}
\centering
\includegraphics[width=0.8\linewidth]{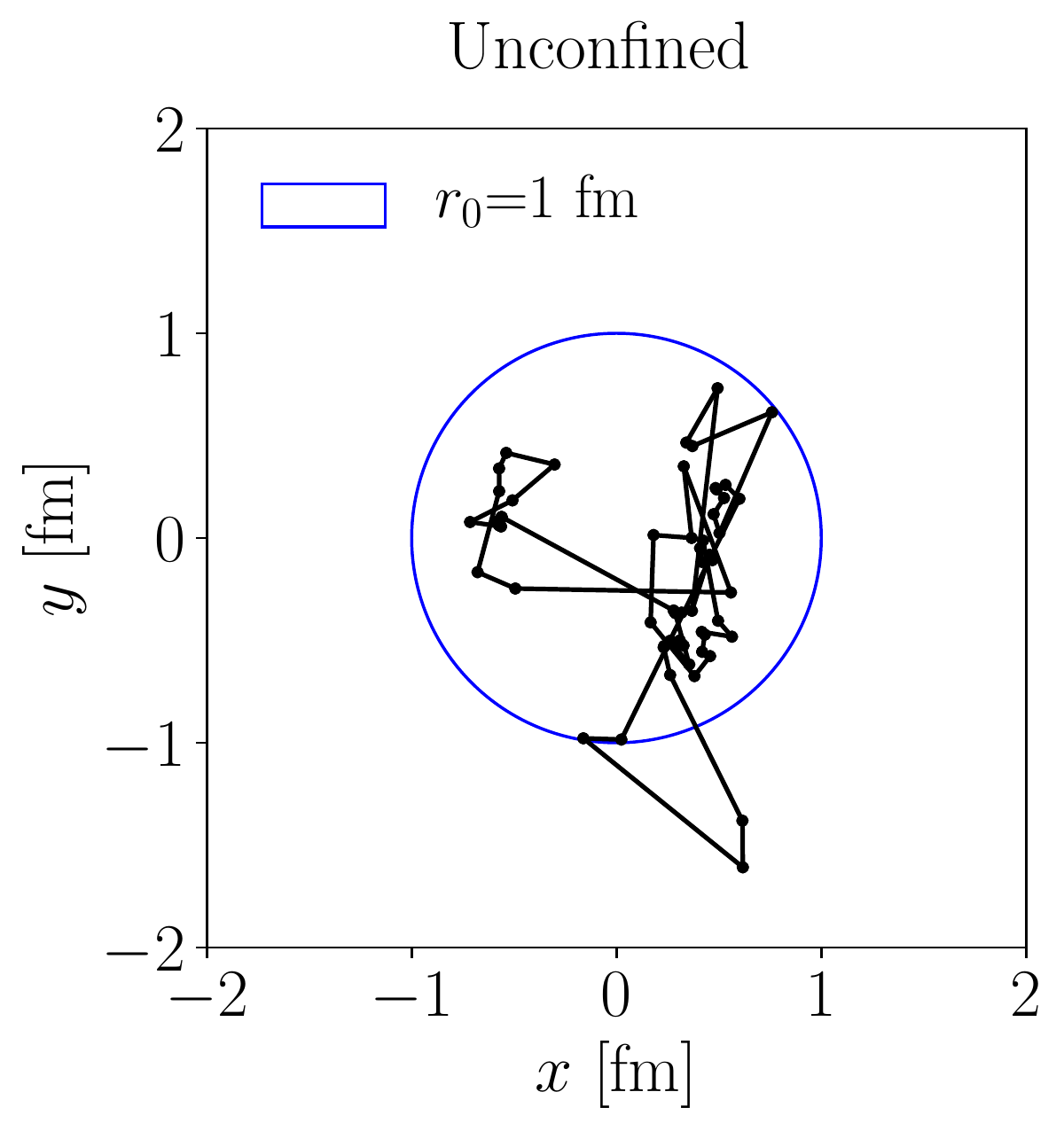}\\
(a)
\end{minipage}%
\begin{minipage}[c]{0.5\linewidth}
\centering
\includegraphics[width=0.8\linewidth]{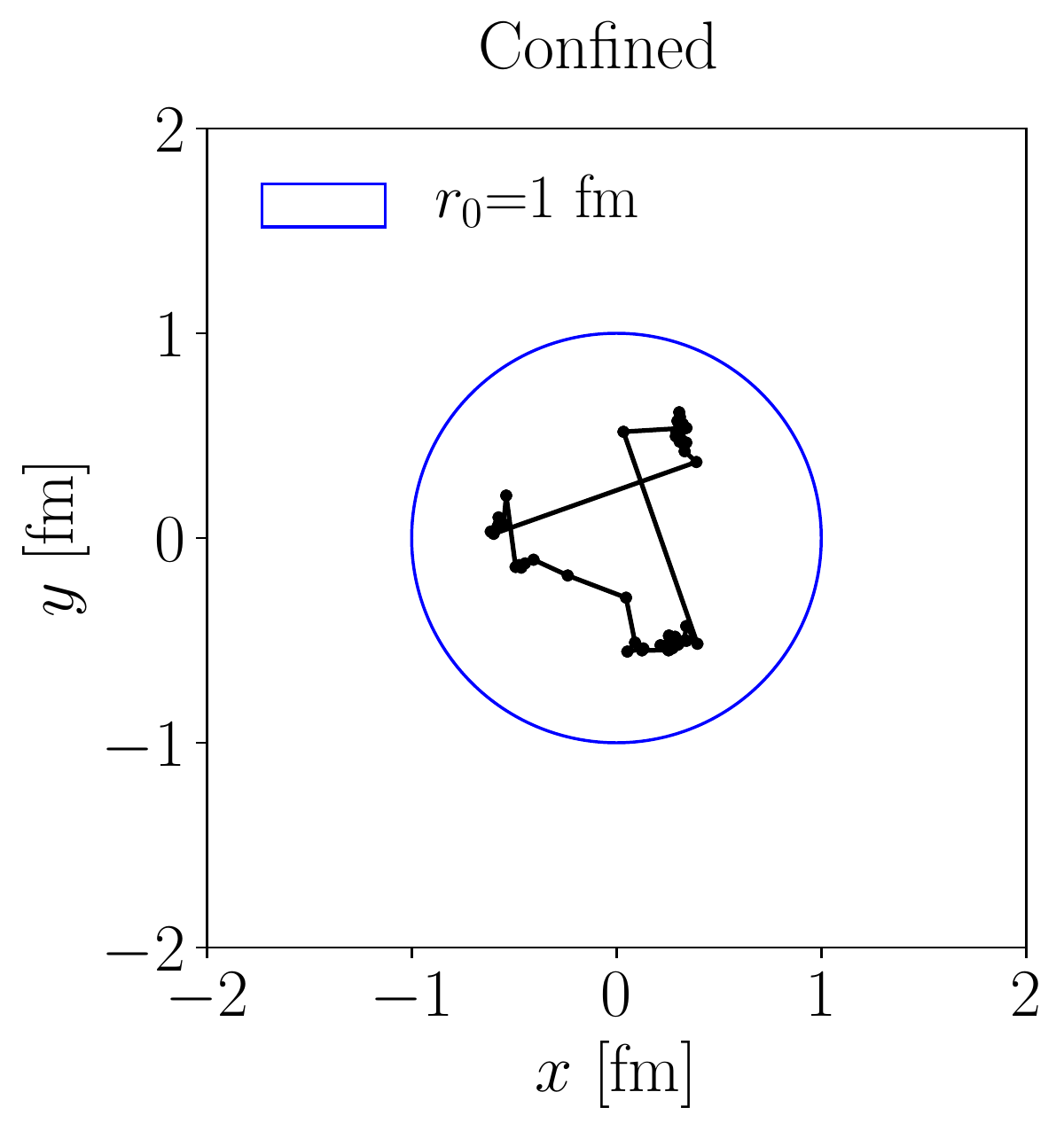}\\
(b)
\end{minipage}
\caption{\label{Fig:EvolvedProton}
An initial state proton consisting of three dipoles in an
equilateral triangle configuration after a full evolution at 7 TeV 
(corresponding to $y_{max}=8.86$). (a) has been evolved without
confinement, while (b) has been evolved with confinement. The 
parameters of table \ref{Tab:PascalParams} have also been applied in this 
evolution.
}
\end{figure}

After a full evolution, an initial proton consisting of three dipoles
will have evolved to a larger set of dipoles of mostly smaller sizes
than the initial dipoles, cf.\ figure \ref{Fig:EvolvedProton}. From these
two figures it is evident that the effect of confinement plays a large
role in the evolution, effectively reducing the number of large dipoles
in the final configuration. Thus, as $\sigma_{\mrm{dip}}\sim r^2$, 
confinement is expected to play a large role when evaluating the cross 
sections. Confinement also introduces more activity -- or hot spots -- 
around the endpoints of the dipoles.   

\subsection{Pascal approximation for dipole evolution}
\label{sec:pascal}
The full dipole evolution can be approximated based on the 
geometric observation above. On average one dipole splitting happens 
per two units of rapidity, and the lengths of the two resulting dipoles
after the splitting, are approximately equal to and one third the size
of the mother dipole respectively. This behaviour is tabulated 
in table \ref{Tab:Pascal} for four generations of evolution. Similar
results have been observed within the \DP~framework, although their
dipole swing slightly increases the size of the smaller dipole in a
branching \cite{Avsar:2007ht} to half of the size of the mother dipole.

\begin{table}[t]
\centering
\begin{tabular}{l | c | c | c | c | c}
& y = 0 & y = 2 & y = 4 & y = 6 & y = 8 \\
\hline
$r$    & 1 & 1 & 1 & 1 & 1 \\
$r/3$  & 0 & 1 & 2 & 3 & 4 \\
$r/9$  & 0 & 0 & 1 & 3 & 6 \\
$r/27$ & 0 & 0 & 0 & 1 & 4 \\
$r/81$ & 0 & 0 & 0 & 0 & 1 \\
\hline 
$N_{\mrm{dip}}$ & 1 & 2 & 4 & 8 & 16 \\
\end{tabular}
\caption{\label{Tab:Pascal} Approximate behaviour of dipole evolution
for four generations
	of dipoles. The number of dipoles in column $n$ row $k$ is equal to the binomial
	coefficient ${n}\choose{k}$.}
\end{table}

The number of dipoles in table \ref{Tab:Pascal} follows the coefficients of the binomial theorem,
with the number in column $n$ row $k$ being equal to ${n}\choose{k}$, and can thus be 
arranged to form Pascal's triangle. The total number of dipoles after
a given number of generation, as well as the number of dipoles of a
certain size,
can be quickly approximated this way. Knowing the positions of the initial dipoles
and the emitted dipole sizes, positions of all dipoles can also be inferred.

To exploit further these simple relations in the dipole evolution, we
have created an alternative toy-model denoted the Pascal approximation.
Here, the step size in rapidity ($\Delta y$) and the size of the smaller
dipole ($r_s=fr_m$) in a branching are implemented as tunable parameters, 
with $r_m$ the size of the mother dipole and $f$ a tunable fraction. 
The number of steps taken in total is calculated from the step size, 
$N_{\mrm{steps}}=y_{\mrm{max}}^{b}/\Delta y$ with $b=\p,\gamma^*$ and
$y_{\mrm{max}}$ given in eqs.~(\ref{Eq:yMax1}--\ref{Eq:yMax2}). 
To mimic the recoil effects in the full dipole evolution, a Gaussian
smearing of the daughter lengths is introduced with mean $\mu=r_m, r_s$ 
for the larger and smaller daughters, respectively. Knowing the lengths
of the mother and daughter dipoles, they are placed in transverse space by 
calculating the angles of the triangle spanned by the endpoints of the 
three (connected) dipoles.

This simple approximation is useful for introducing toy-models for 
sub-leading effects, such as confinement and saturation, as basic quantities like 
total number of dipoles after a given evolution, can be calculated
analytically. A crude model for confinement
is introduced by requiring the length of the emitted dipoles to not exceed a tunable
maximally allowed cutoff, $r_{\mrm{max}}$. If this occurs, the branching
is discarded and the next step in rapidity is tried. Once the full
evolution has occurred, each of the dipoles are allowed to interact using
the dipole-dipole scattering amplitude given in \eqref{Eq:fij} 
(or \eqref{Eq:fijConf} for the confined version).

\begin{figure}[t]
\centering
\begin{minipage}[c]{0.5\linewidth}
\centering
\includegraphics[width=1.0\linewidth]{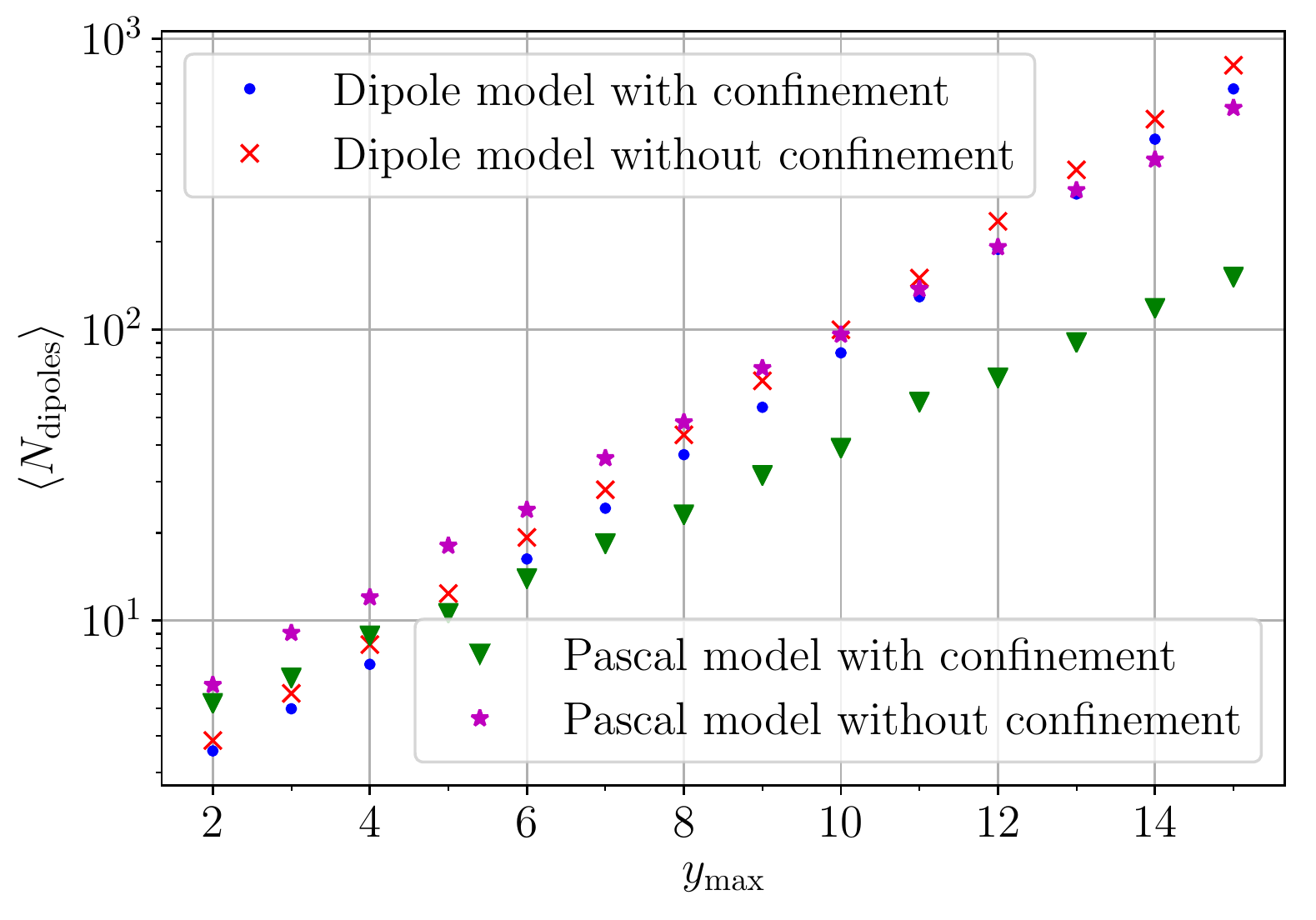}\\
(a)
\end{minipage}%
\begin{minipage}[c]{0.5\linewidth}
\centering
\includegraphics[width=0.8\linewidth]{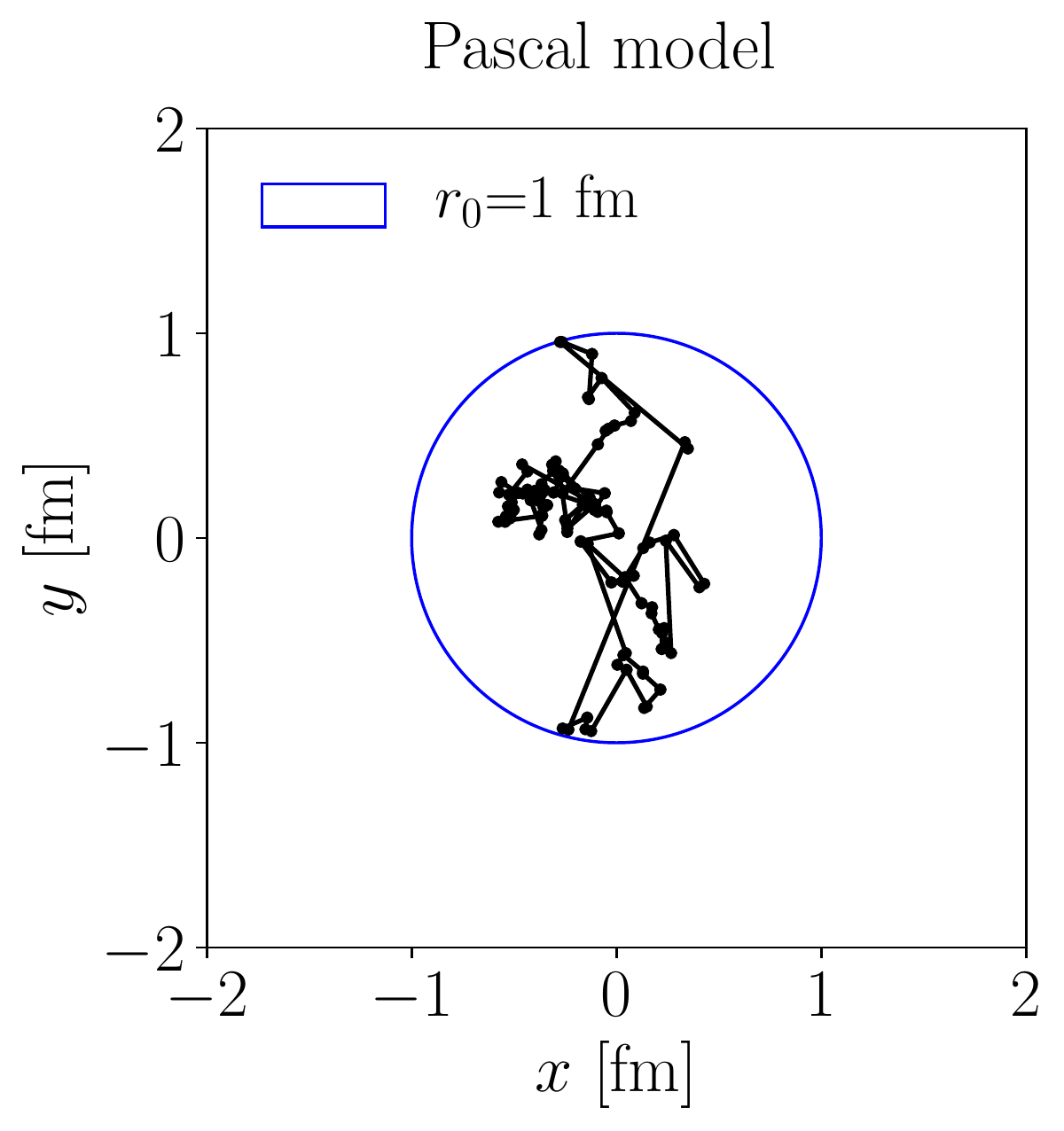}\\
(b)
\end{minipage}
\caption{\label{Fig:Pascal}
(a) The average number of dipoles inside a proton after a full evolution
to maximal rapidity $y$. (b) An initial state proton consisting of three dipoles in an
equilateral triangle configuration after a full evolution with the Pascal
model at 7 TeV (corresponding to $y_{max}=8.86$). 
}
\end{figure}

In fig \ref{Fig:Pascal}(a) the average number of dipoles in a
single proton after a full evolution to maximal rapidity $y_{\mrm{max}}$ is shown. The 
same parameters as given in table \ref{Tab:PascalParams} are used, while
the parameters $f,\Delta y$ are extracted from figures \ref{Fig:dPdY}
and \ref{Fig:Rratios}. The unconfined Pascal model follow the same
trend as the full dipole model, albeit having a slightly larger average
number of dipoles at small maximal rapidity. It is evident that the 
confined Pascal model has a different slope than the full dipole model 
and the unconfined Pascal model, with the effect of the crude model for 
confinement clearly seen at large $y_{\mrm{max}}$. Here, the confined 
Pascal model results in a smaller average number of dipoles as compared 
to the full dipole model. 

In figure \ref{Fig:Pascal} (b) the dipole configuration of an evolved
proton for $y_{\mrm{max}}=8.86$ is shown. This has more features in common with full, unconfined dipole evolution than full,
confined dipole evolution, with dipoles being more randomly distributed, than 
focused around hot spots. 

A practical advantage of the Pascal approximation, besides being a toy
model for testing models for sub-leading effects, is its computational speed.
For simple cascade-quantities like numbers of dipoles, results can be calculated
analytically. With inclusion of geometry, event-by-event results can 
be generated approximately a factor of 1000 faster, for large maximal rapidity
($y_{\mrm{max}}\geq10$). It thus serves as a decent replacement for the full dipole evolution 
model for calculation of cascade properties with limited computational 
resources. For full calculations of amplitudes and cross sections, the efficiency gain is
not nearly as large, as in that case the bottleneck is the calculation
of all $f_{ij}$ in \eqref{Eq:fij}.

\subsection{Dipole-dipole interactions}
The dipole-dipole interactions are defined to occur at rapidity zero and
given by \eqref{Eq:fij}. If confinement is introduced in the 
splitting kernel (\eqref{eq:mod-split-kern}), one also has to change the interaction probability in 
order to make the event generation consistent. This modifies 
\eqref{Eq:fij} to
\begin{align}\label{Eq:fijConf}
f_{ij}=&\frac{\alpha_s^2}{2}\left[K_0\left(
  \frac{r_{13}}{r_{\mrm{max}}}\right) 
  + K_0 \left(\frac{r_{24}}{r_{\mrm{max}}}\right) 
  -K_0\left(\frac{r_{14}}{r_{\mrm{max}}}\right) 
  -K_0\left(\frac{r_{23}}{r_{\mrm{max}}}\right)\right]^2,
\end{align}
where $K_0$ is the modified Bessel function and $r_{\mrm{max}}$ the
maximally allowed size for a dipole in the evolution. 

The choice of collision frame, however, is not trivial. Obviously, no
observables should depend on the frame-choice of the collision. In
practice, the choice does matter, as no sub-leading color corrections
are included in the dipole evolution. Previous studies have shown that
for symmetric systems, e.g.\ $\p\p$ collisions, the optimal frame choice
is the center-of-mass (CM) frame \cite{Mueller:1996te}. This is also utilized
in our approach, cf.\ \eqref{Eq:yMax1}, where both beams are
evolved the same distance in rapidity. In asymmetric systems, such as 
$\gamma^*\p$ or $\p\A$ systems, the CM frame lies more towards the heavier 
of the two objects, and it has been previously argued that the optimal 
frame here would be the rest frame of the heavier beams \cite{Mueller:1996te}. 
This, however, is not the choice we have taken. The maximal rapidity 
chosen in \eqref{Eq:yMax2} is found in what we call the 
center-of-rapidity frame. Here, both beams are evolved the 
same distance in rapidity, similarly to what is chosen in symmetric
collision systems. As already stated, this work does not attempt to include sub-leading colour
effects in the evolution, thus frame-independence is not possible to obtain.
Hence the simplest choice has been made to use the same frame for all 
systems, i.e.\ the center-of-rapidity frame given in eqs.~(\ref{Eq:yMax1}--\ref{Eq:yMax2}). 

\subsection{\label{sec:mpi}Assigning spatial vertices to MPIs}
In order to utilize the formalism developed so far in real $\p\p$, $\p\A$ 
and $\A\A$ events, the dipole cascade is matched to the \PY~MPI model
\cite{Sjostrand:1987su}. This allows for evaluation of geometric initial
state quantities, such as eccentricities (see section
\ref{sec:eccentricity}),
at fixed number of charged hadrons in the final state, using a similar 
definition of charged particles as the experiments.
The \PY~MPI model considers $\p\p$ collisions, treating all partonic 
sub-collisions as separate $2 \rightarrow 2$ QCD scatterings, which are 
uncorrelated up to momentum conservation. Other factors present in the
MPI model is a rescaling of the parton density between each scattering,
preservation of valence quark content and a sophisticated treatment of
beam remnants \cite{Sjostrand:2004pf}.

In the MPI framework, the sub-collisions and their kinematics are 
selected using the normal $2\rightarrow2$ QCD cross section. But since 
this cross section diverges at low $p_\perp$, the expression is regulated 
using a parameter, $p_{\perp 0}$:
\begin{equation}
  \label{eq:mpixsec}
  \frac{\d\sigma_{2\rightarrow 2}}{\d\pT^2} \propto
  \frac{\alpha^2_s(\pT^2)}{\pT^4} \rightarrow
  \frac{\alpha^2_s(\pT^2 + \pTo^2)}{(\pT^2 + \pTo^2)^2}.
\end{equation}
For matching of vertices to each individual partonic sub-collision, it is also
useful to note that MPIs are generated in decreasing order of $p_\perp$, starting 
from a (process-dependent) maximal scale. 
This decreasing order is generated from a Sudakov-like expression of the form:
\begin{equation}
	\label{eq:pythiaMpi}
	\frac{\d\mathcal{P}}{\d p_{\perp i}} =
	\frac{1}{\sigma_{\mrm{ND}}}\frac{\d\sigma}{\d p_{\perp i}}
        \exp\left[-\int_{p_{\perp i}}^{p_\perp i-1}
	  \frac{1}{\sigma_{\mrm{ND}}}
          \frac{\d\sigma}{\d p_\perp'}\d p_\perp' \right],
\end{equation}
with $\d\sigma/\d p_{\perp i}$ given by \eqref{eq:mpixsec}. The
impact-parameter of the collision is also taken into account in the
evolution by connecting the average number of MPIs to the overlap
$\mathcal{O}(b)$ of the two colliding protons. This introduces additional
factors of $\mathcal{O}(b)/\langle \mathcal{O}(b)\rangle$ in \eqref{eq:pythiaMpi} 
along with the need to select the impact parameter
consistently.\footnote{Here $\langle \mathcal{O}\rangle \equiv \int
\d^2 \vec{b} \mathcal{O}(b) / \int \d^2 \vec{b} \left[1 - \exp(-k
\mathcal{O}(b))\right]$, where $k$ is constrained by the ratio of the
dampened $2\rightarrow 2$ QCD cross section in equation
(\ref{eq:mpixsec}) to the total non-diffractive cross section.}
Furthermore, 
new partons are generated by initial- and final-state radiation.

Recently, a method of assigning space-time information 
to the MPIs in \PY~was introduced
\cite{Bierlich:2017vhg}. Here, the transverse 
coordinates are sampled from a two-dimensional Gaussian distribution 
defined by the overlap of the mass distributions of the two colliding 
protons. The width of the Gaussian is a free parameter (which should not be
too far from the proton radius) and a mean equal to 
the impact parameter chosen in the MPI framework. Initial- and final-state 
radiation are then treated as small displacements of the selected anchor 
points of the MPIs. This introduces and additional smearing of an MPI 
vertex whenever a parton is radiated off from the partons involved in
the MPI. The smearing is done using another Gaussian with a width of 
$\sigma_\perp/\p_\perp$, where $\sigma_\perp$ is a parameter with default 
value $0.1$ GeV$\cdot$fm.

In this work, we utilize the dipole framework to generate the space-time 
vertices, instead of the default two-dimensional Gaussian distribution.
We currently do not use the dipole model to generate the $\pT$-spectrum
for the MPIs, but retain the $\pT$-distribution obtained internally with 
\PY. This means that the dipole model is only used to obtain
information on the spatial location of the MPIs. Using the dipole 
framework to generate space-time vertices requires (as with
the Gaussian model) some assumptions, as this matching can not be derived from
first principles. In order to obtain a reasonable matching, the following is noted:
\begin{itemize}
	\item Each branch of the (projectile) dipole cascade can be identified as a
	virtual emission, which goes on shell if, and only if, it collides with a 
	corresponding virtual emission from the target.
	\item Each proton--proton collision has many \emph{potential} sub-collisions
	between all combinations of virtual emissions. We order the sub-collisions
	in terms of contribution to total cross section, thus the MPI with
  largest $\pT$ is identified with the dipole-dipole scattering with the largest $f_{ij}$.
\end{itemize}

The concrete matching is done by first generating two dipole cascades,
and allowing them to collide with the same impact parameter used in the generation of the MPIs.
This produces a list of possible dipole--dipole collisions, each with an interaction
strength $f_{ij}$. As the MPIs are generated (from hardest to softest), they are
each assigned a vertex sampled from this list with a weight equal to  $f_{ij}$, 
normalised to the summed dipole-dipole interaction strength (and \tit{not} the unitarized
interaction). The vertex is simply given as the mean of the transverse
coordinates of the dipoles in the interaction. Once a set of dipoles have been assigned to an MPI, they
are both flagged as used, and cannot initially be re-used to ensure a reasonable spread.
In cases where the list runs out of interactions containing only unused dipoles, the dipoles 
are allowed to re-interact, though not with the same dipole as initially. 

As opposed to the default model, vertices are now selected from a
distribution which event-by-event is asymmetric, and contains "hot
spots" with large activity, as shown in \figref{Fig:EvolvedProton} (b)
for the full evolution including confinement effects. 

The matching of largest $f_{ij}$ to hardest MPI requires further discussion, as
one could argue the opposite. The dipole-dipole scattering amplitude is driven 
by the distances between the endpoints of the interacting dipoles, as indicated
in figure \ref{Fig:DipInt}. One can argue that small $f_{ij}$  
corresponds to small distances, which in turn corresponds to large $\pT$ 
of the gluons emitted in the interaction. Hard MPIs would with this
reasoning correspond to small $f_{ij}$. This is indeed the choice made in 
the \DP~event generator for exclusive final states \cite{Flensburg:2011kk},
but opposite to the choice made above. We do, however, also
note that the exclusive final states generated by \DP~describes $p_\perp$ 
spectra of charged particles poorly, in particular the high-$p_\perp$ part
of the distributions vastly overshoots data. We therefore currently refrain 
from associating the dipole sizes directly to the $p_\perp$ of emerging partons, 
but rather give larger attention to the cross section. We note that large 
$f_{ij}$ interactions dominate the cross sections. A guiding principle is 
therefore to ensure that such interactions are \tit{always} identified with 
an MPI, by assigning it first.

There are several possible future improvements of the matching technique. 
A small improvement of the existing technique, could include
to also identify initial state radiation with emissions going on shell,
and assign them vertices from the cascade as such. 
Going beyond improvement of matching techniques, would be a full re-evaluation of 
the MPI model, with the dipole cascade and interactions as a starting
point. The consequences of such an approach could be studied in a
toy-model where the $\pT$ obtained with the dipole formulation could be 
utilized instead of the $\pT$'s obtained within the \PY~model. It would
then be possible to study if this method gives rise to similar
problems as the \DP~MPI description has in the high~$\pT$ tail.
 
Instead of creating a completely new model like \DP, it should be possible
to use the dipole model to improve the existing \PY~model.
A first step would be to replace 
$\sigma_{\mrm{ND}}$ in \eqref{eq:pythiaMpi} with a dynamically calculated
cross section, event-by-event. Secondly, the parameter $\pTo$ in \eqref{eq:mpixsec}
could be re-evaluated in terms of the dipole model. The physical interpretation
of $\pTo$ in the MPI model, is that of a colour screening scale. The perturbative
treatment of \eqref{eq:pythiaMpi} would naively break down at some minimal
scale $\sim \hbar/r_p \sim \Lambda_{QCD}$, where $r_p$ is the (colour screening)
size of a proton, left as a free parameter. In the dipole model, this colour screening
length could be identified as either the transverse size of the cascade after the evolution,
or the length of the largest colour connected dipole chain. In that way
the energy dependence
of $\pTo$ would also come for free, instead of having to assume a power-law 
dependence, as is the default assumption in \PY.

\subsubsection{Heavy ion collisions}

The method described above can be directly applied to heavy ion collisions as they are modelled
in the \textsc{Angantyr} framework for heavy ion collisions in \PY~(see appendix \ref{sec:angantyr} for
a brief review, and refs. \cite{Bierlich:2016smv,Bierlich:2018xfw} for a full description). In
the \textsc{Angantyr} model, sub-collisions are chosen using a Glauber-like approach. Sub-collisions are in
turn associated with one out of several types of pp events, depending on the properties of the 
sub-collision. Since all these events are generated using the MPI model described above, the generalization
is only a matter of generating vertices for each sub-collision in its local coordinate system, and then
moving them to the global coordinate system defined by the Glauber calculation. 

\section{Results I -- comparing cross sections}

In this section we present results on integrated cross sections for
$\p\p$ and $\gamma^*\p$ collisions. For $\p\p$ we present results for
both the dipole evolution model and for the Pascal model, while for
$\gamma^*\p$ we focus our attention on the dipole model. The main
purpose of this section is tuning: the model parameters have to be
estimated by comparisons with data, preferably data that we do not aim
to make predictions for in later sections. 

It is thus not the aim of this section to be able to describe the cross
sections perfectly -- but more generally, to get an overall agreement
between model and data, especially at LHC energies, where we aim to make
predictions for the substructure observables.

More dedicated models are available to describe the cross sections at
all energies, from the GeV range to the TeV range, results of which
are shown alongside results from the dipole model in the $\p\p$ section. The most
widespread model is based on the 1992 total cross section fit by Donnachie and Landshoff (DL) 
\cite{Donnachie:1992ny} and the models for elastic and diffractive cross 
sections by Schuler and Sj\"ostrand (SaS) \cite{Schuler:1993wr}. Another,
more recent model by Appleby \etal~ (ABMST) \cite{Appleby:2016ask} is more
complex than SaS, and able to describe latest LHC data better.
The models are both implemented in \PY, with some additions to the original
models \cite{Rasmussen:2018dgo}. In this paper we compare to the original
models, and not those adapted to \PY.

\subsection{Results for $\gamma^*\p$}

We begin with the results on photon-proton total cross sections. Here,
we compare the dipole evolution model to data obtained from H1 \cite{Adloff:2000qk}
at different energies and virtualities. We note that the photon
wave function implemented only includes the three lightest quarks, and
none of the vector meson states present at low $Q^2$. Thus we
expect the results to be less precise at low virtualities, where the
probability for the photon to fluctuate into a hadronic state becomes
non-negligible. Similarly, the masses of the quarks should
be taken into account if the argument of the Bessel functions become
close to the squared quark masses, i.e.\ if 
\begin{align}
z(1-z)Q^2\simeq& m_q^2\,
\end{align}
occurring in the limits $z\rightarrow 0,1$ or if $Q^2$ small. 
The contribution from $c$-quarks are neglected for simplicity, the uncertainty
arising from this approximation is discussed at the end of the section.

The H1 data presents results on the proton structure function
$F_2(x,Q^2)$ at a large range of virtualities and energies. This is
translated into a photon-proton total cross section as follows:
\begin{align}
	\sigma_{\mrm{tot}}^{\gamma^*\p}=\frac{4\pi^2\alpha_{\mrm{em}}\hbar^2c^2}{Q^2}F_2(x,Q^2)
\end{align}
with the CM energy given as $W^2=Q^2(1-x)/x$ and $\hbar c$ a unit conversion
factor.

\begin{figure}[t]
\centering
\begin{minipage}[c]{0.5\linewidth}
\centering
\includegraphics[width=1.0\linewidth]{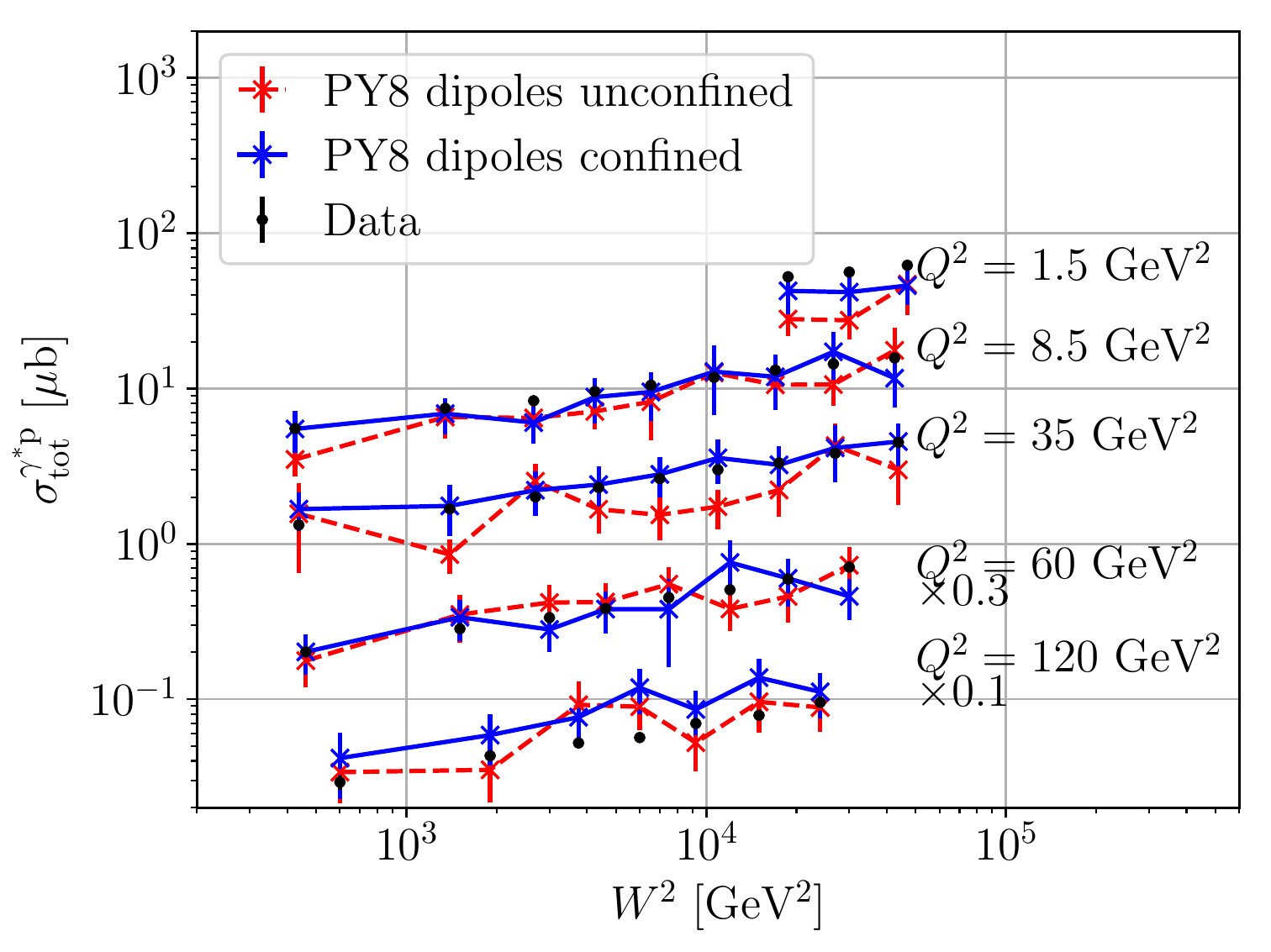}\\
(a)
\end{minipage}%
\begin{minipage}[c]{0.5\linewidth}
\centering
\includegraphics[width=1.0\linewidth]{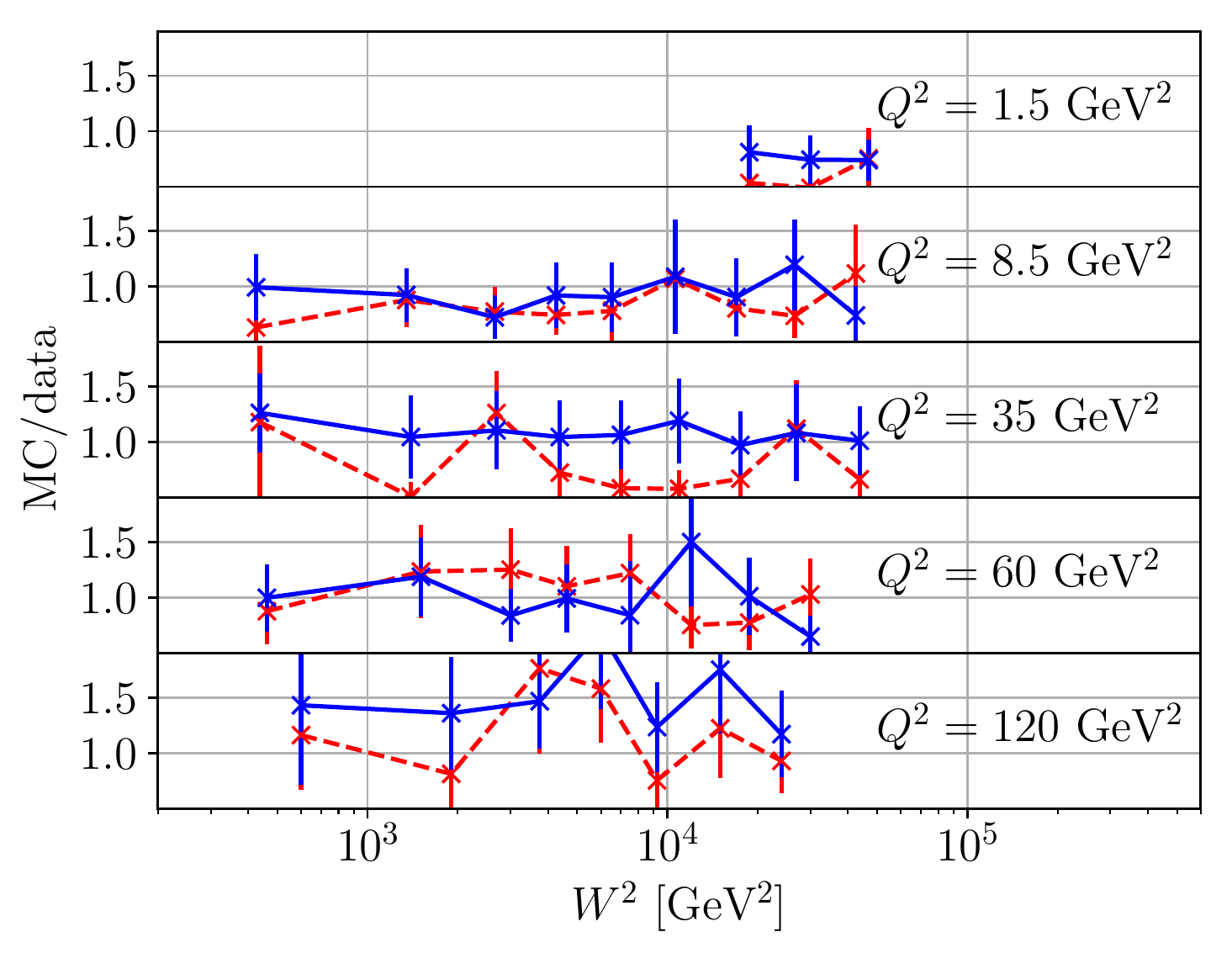}\\
(b)
\end{minipage}
\caption{\label{Fig:epTune} The total photon-proton cross section,
$\sigma_{\mrm{tot}}^{\gamma^*\p}$, as a function of squared 
photon-proton center-of-mass energy, $W^2$, for several virtualities
(a). Note that the distributions for the two highest virtualities
($Q^2=60,120$ GeV$^2$) have been scaled with a factor of $0.3,0.1$,
respectively, for better visibility. (b) shows the ratio MC/data as 
a function of squared center-of-mass energy, $W^2$, for the five
different virtualities.
}
\end{figure}

It is evident from figure \ref{Fig:epTune} that we undershoot data at low
$Q^2$. At intermediate virtualities the model does a fairly good job,
while at the highest virtuality probed the prediction overshoots
data with roughly 50\%. In order to quantify the performance of the
models, a $\chi^2$ test has been performed, taking into account the
errors of the measurement:
\begin{align}
\chi^2=&\sum_{i=W^2}\frac{(D[W^2]-M[W^2])^2}{\sigma_{D[W^2]}^2+\sigma_{M[W^2]}^2}
\end{align} 
with $D$ denoting the cross section measured in data at a given squared 
energy $W^2$, $M$ the model prediction for that squared energy and
$\sigma_{D,M}^2$ the variance of the data and model predictions,
respectively.

The model has been tuned with the \textsc{Professor2} framework 
\cite{Buckley:2009bj}, and the parameters are shown along with the
$\chi^2/N_{\mrm{dof}}$ in table \ref{Tab:gampTune}. The parameters of 
the tune are reasonable, giving a initial dipole size roughly of order 1
fm with a width of the Gaussian fluctuations at around $0.1$ fm. Adding 
confinement allows for a slightly larger initial dipole size, as the 
largest dipoles in the evolution will be suppressed as compared to the 
unconfined model, while also the upper integration limit on the 
photon is allowed to increase when turning on confinement. The width of the 
fluctuations and the strong coupling appear not to be affected by the 
confinement effect. Taking the full H1 data set into account, the
confined model gives a reasonable $\chi^2/N_{\mrm{dof}}$, and performs
slightly better than the unconfined model.

\begin{table}[t]
\centering
\begin{tabular}{l | c | c }
& \multicolumn{2}{| c }{$\gamma^*\p$}\\
Parameter & unconfined & confined \\
\hline
$r_0$ [fm]	& 1.08 & 1.15 \\
$r_{\mrm{max}}$ [fm]	& - & 3.50 \\
$r_{\mrm{w}}$ [fm]	& 0.10 & 0.10 \\
$r_{\mrm{max}}^{\gamma^*}$ [fm]	& 2.07 & 2.56 \\
$\alpha_s$	& 0.21 & 0.22 \\
\hline
$\chi^2/N_{\mrm{dof}}$ (shown $Q^2$ values) & 2.41 & 0.57 \\
$\chi^2/N_{\mrm{dof}}$ (full H1 data set) & 2.99 & 1.98 \\
\end{tabular}
\caption{\label{Tab:gampTune} The parameter values obtained when tuning
to the $\sigma_{\mrm{tot}}^{\gamma^*\p}$ data set and the $\chi^2$
obtained for the two models.}
\end{table}

Since the charm contribution to the $\gamma^*\p$ cross section has been neglected, an assessment of the uncertainty arising from this approximation should be made.
Adding massless charm quarks shifts the total $\gamma^*\p$ cross section upwards by 67\%, estimated by the ratio:
\begin{equation}
	\frac{e^2_u + e^2_d + e^2_s + e^2_c}{e^2_u + e^2_d + e^2_s} - 1 = 4/6.
\end{equation}
This rise in cross section can be tuned away in a way similar to the procedure described above. Adding quark masses (lighter quark masses neglected) reduces the contribution. 
The reduction is larger for smaller $Q^2$. The quantitative effect of adding masses was studied in ref. \cite{Avsar:2007ht}. For small $Q^2$ the decrease compared to the massless charm case is $\sim 15\%$ and for large $Q^2$ the decrease is $\sim 40\%$. Both represent un-tuned values. A conservative, un-tuned estimate of the uncertainty in figure \ref{Fig:epTune} from neglecting (massive) charm quarks is therefore up to $\sim 25\%$.
Retuning will allow for shifting the cross section upwards in the low $Q^2$ region where the values in figure \ref{Fig:epTune} undershoots, improving the overall agreement.

\subsection{Results for $\p\p$}
\label{sec:pp-xsec}
\begin{figure}[t]
\centering
\begin{minipage}[c]{0.5\linewidth}
\centering
\includegraphics[width=1.0\linewidth]{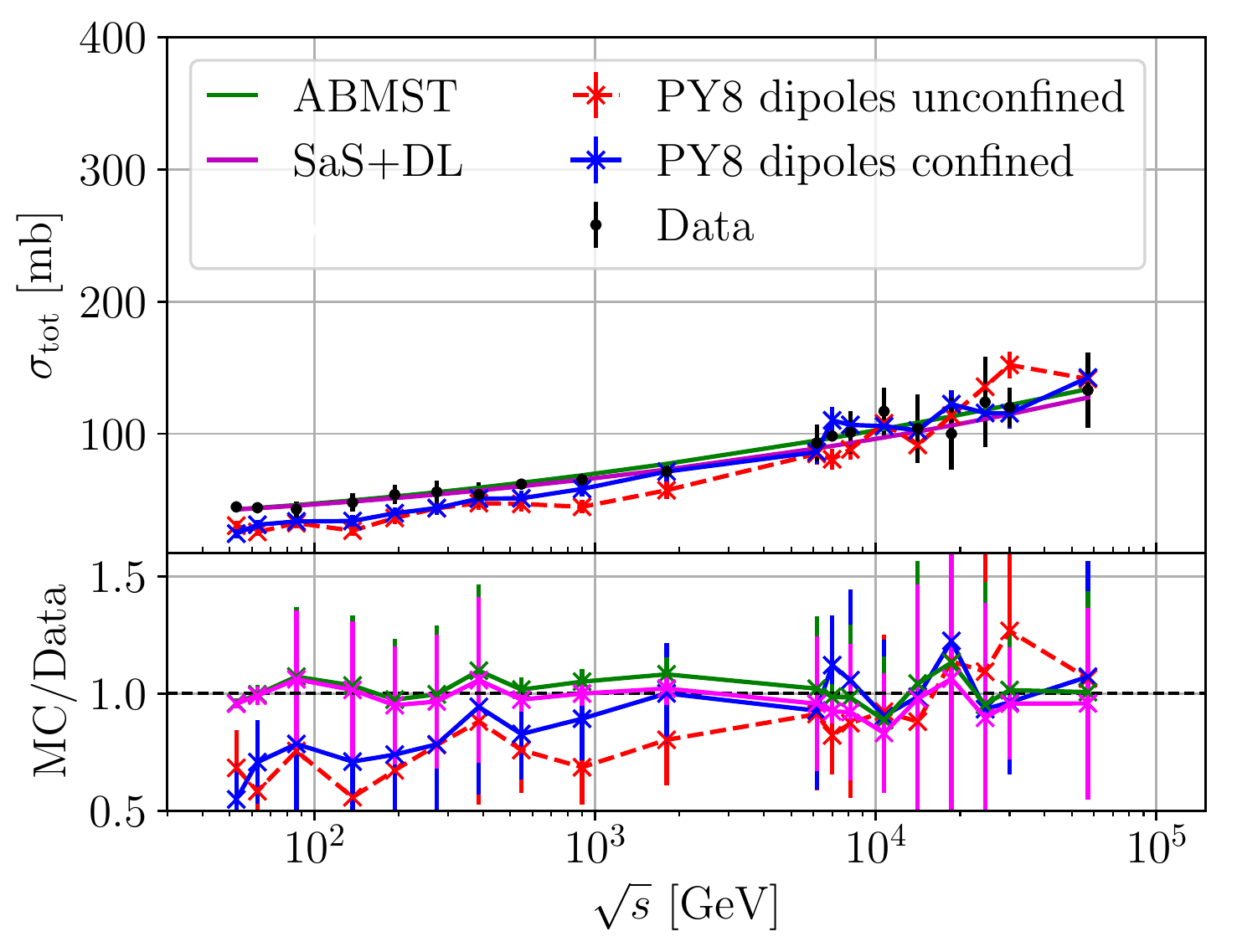}\\
(a)
\end{minipage}%
\begin{minipage}[c]{0.5\linewidth}
\centering
\includegraphics[width=1.0\linewidth]{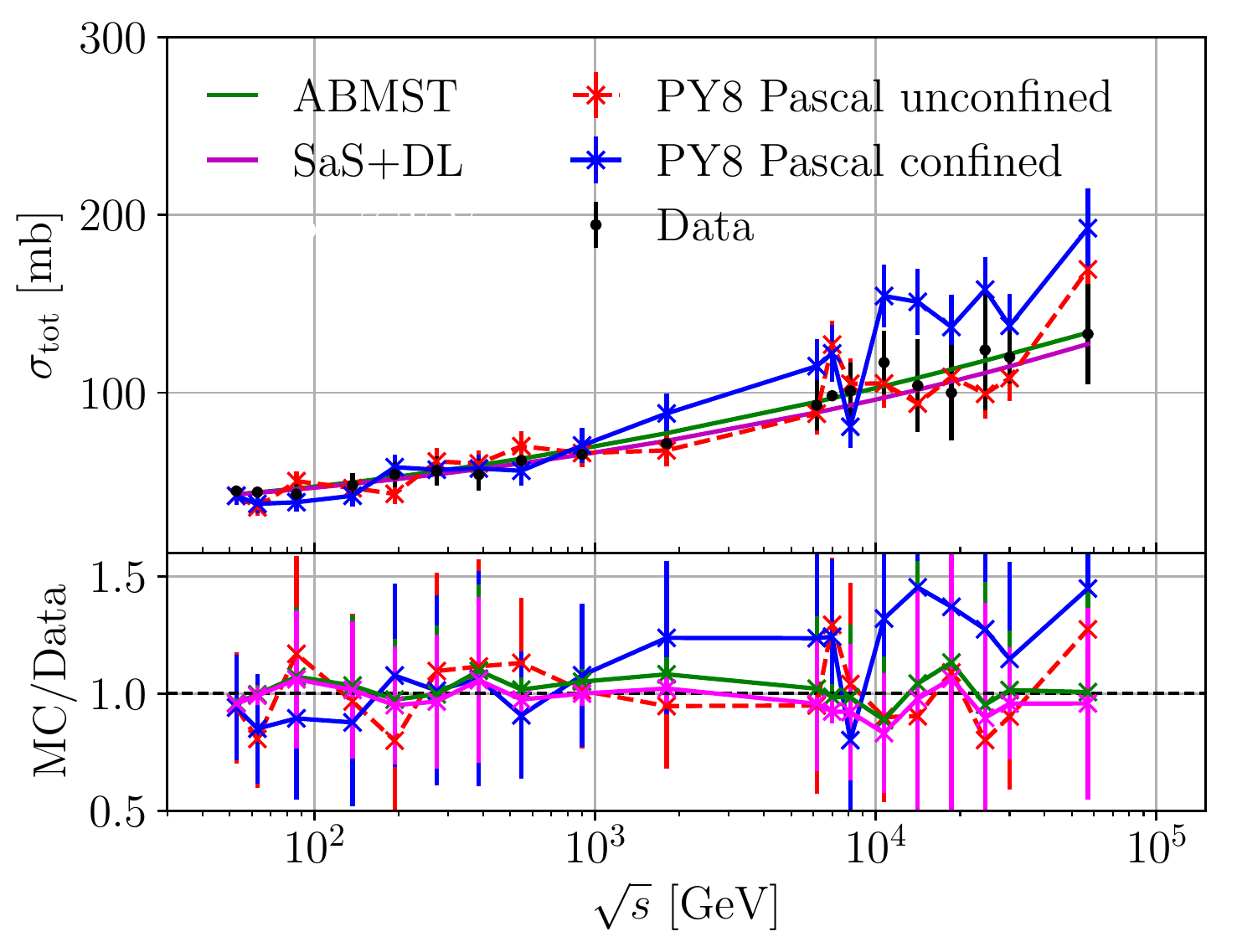}\\
(b)
\end{minipage}
\caption{\label{Fig:ppSigTot} The total $\p\p$ cross section as a
function of $\sqrt{s}$ for the dipole (a) and Pascal (b) models. Both
show the confined and unconfined versions in solid blue and dashed red
lines, respectively. Both figures show the ABMST (solid green lines) and
SaS+DL (solid magenta lines) for comparison.}
\end{figure}

In figure \ref{Fig:ppSigTot} we show the total cross section as a function 
of CM collision energy for both the full dipole model (a) and the
Pascal model (b). Both figures show results using the confined (solid blue 
lines) and unconfined (dashed red lines) models as well as the ABMST
(solid green lines) and SaS+DL model (solid magenta lines). It is evident 
that the full dipole model undershoots data at low $\sqrt{s}$, whereas it agrees
with data at roughly $\sqrt{s}\geq10^2$ GeV, with the confined model having a
smaller $\chi^2/N_{\mrm{dof}}$ (cf.\ table \ref{Tab:ppTune}) than the unconfined 
model using only this data set. The Pascal model, figure \ref{Fig:ppSigTot} (b), 
shows an overall shift towards higher cross sections as compared to the full 
dipole model, thus describing the lower energies well while slightly 
overshooting the higher energies. With only this data set, both Pascal
models have a lower $\chi^2/N_{\mrm{dof}}$ than the dipole models. As explained 
in section \ref{sec:pascal}, the key difference between the models, is the 
treatment of confinement as a hard cutoff.
In both figures, it is evident that both the
SaS+DL and ABMST models perform better, not surprising as these
models have been created to reproduce (a subset of) this data.
 
\begin{figure}[t]
\centering
\begin{minipage}[c]{0.5\linewidth}
\centering
\includegraphics[width=1.0\linewidth]{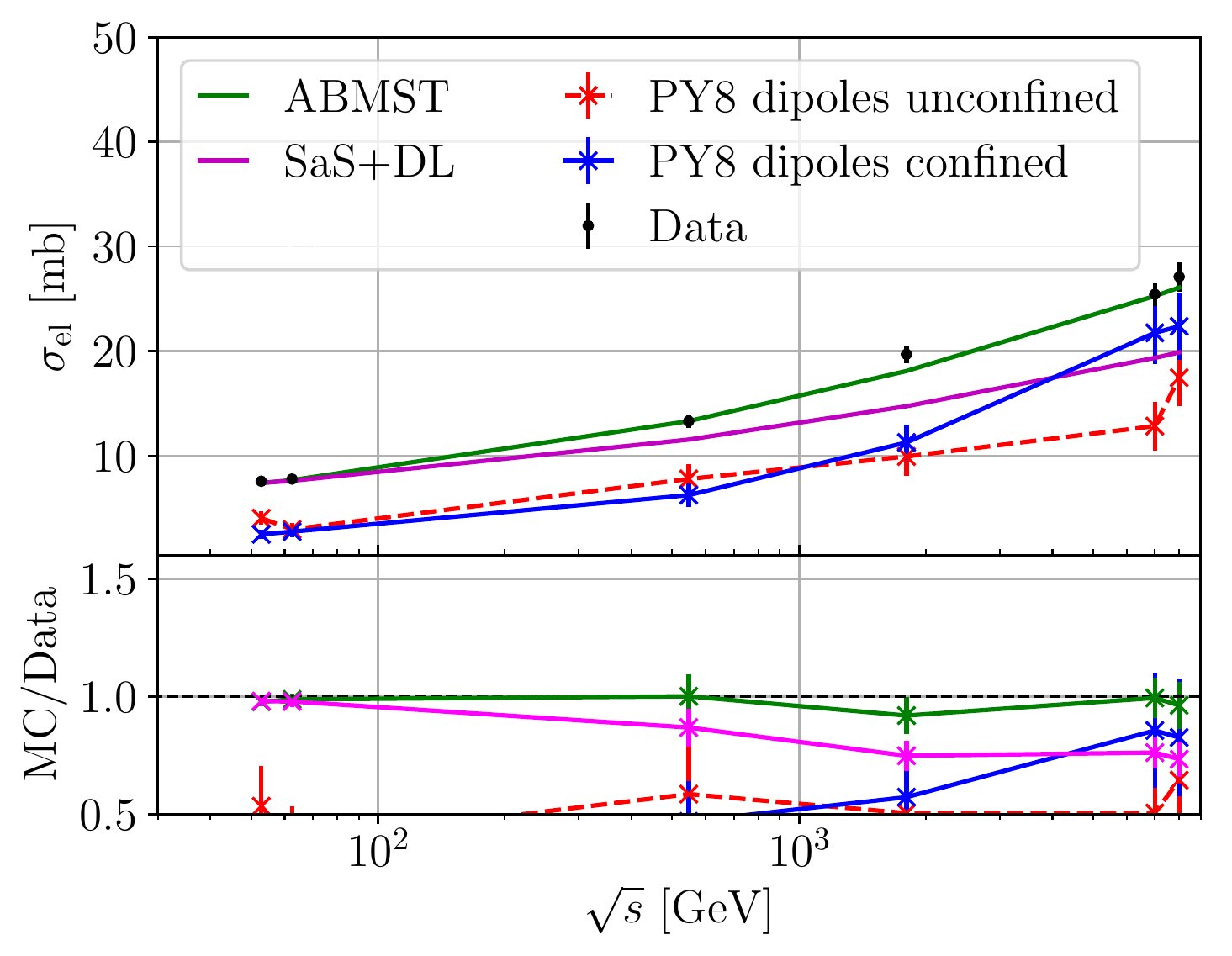}\\
(a)
\end{minipage}%
\begin{minipage}[c]{0.5\linewidth}
\centering
\includegraphics[width=1.0\linewidth]{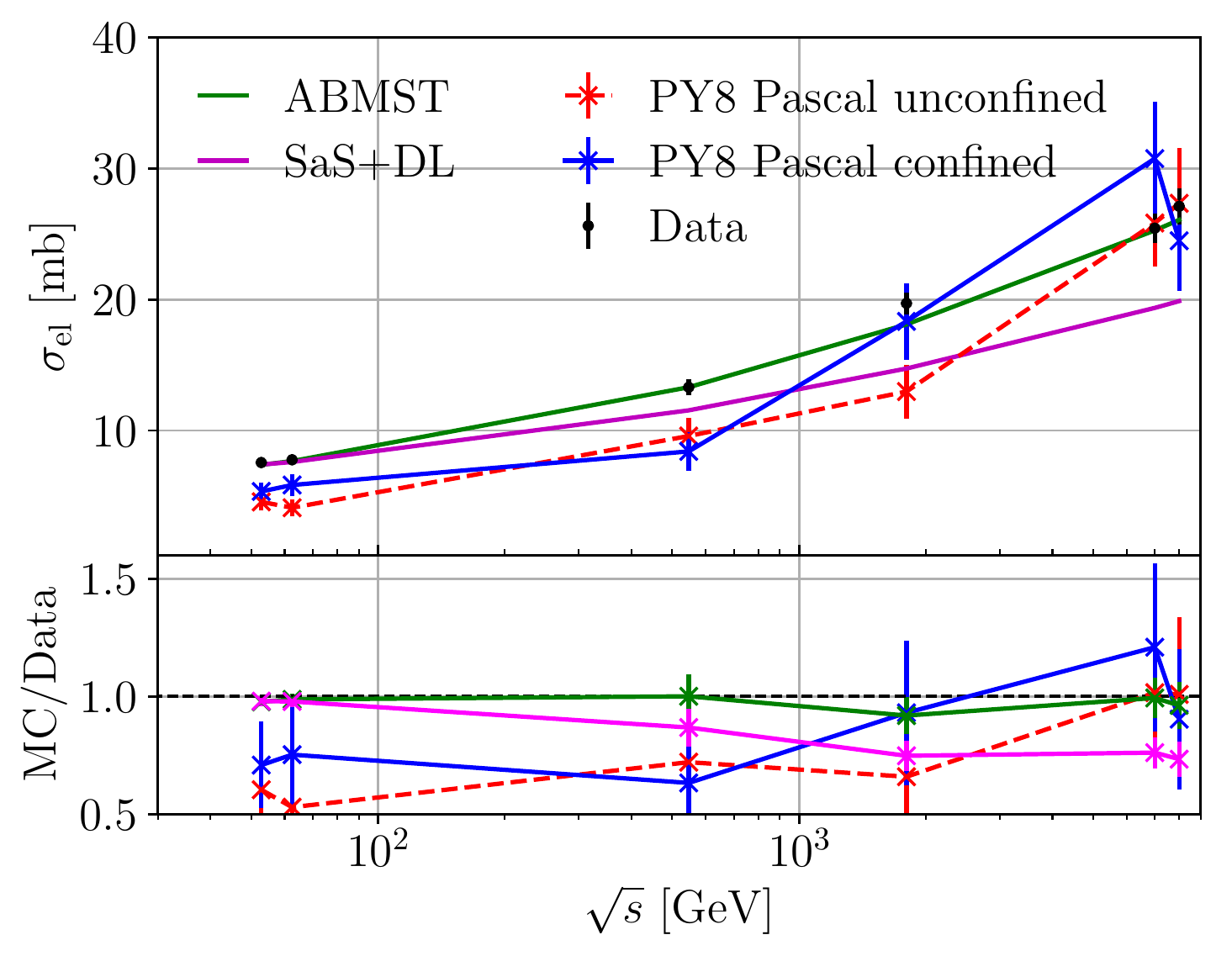}\\
(b)
\end{minipage}
\caption{\label{Fig:ppSigEl} The elastic $\p\p$ cross section as a
function of $\sqrt{s}$ for the dipole (a) and Pascal (b) models. Both
show the confined and unconfined versions in solid blue and dashed red
lines, respectively. Both figures show the ABMST (solid green lines) and
SaS+DL (solid magenta lines) for comparison.
}
\end{figure}

In figure \ref{Fig:ppSigEl} we show the elastic $\p\p$ cross section 
for the full dipole model (a) and the Pascal model (b). Neither of the
dipole models are able to describe this cross section, being roughly
50\% below data in the entire energy range, except for the very last
bins, i.e.\ at LHC energies. The Pascal model, however, agrees with data
at lower energies better than the full model.
Also here, the two dedicated models describe the elastic data better than the dipole and
Pascal models, with the SaS+DL model deviating from the data at LHC
energies, while ABMST describes data in the entire energy range. This is
partly due to a modification of the elastic slope in the SaS model, and 
partly due to the additional trajectories included in the ABMST model: where
SaS+DL only contains a single Pomeron in the description of the elastic
cross section, ABMST has two -- along with additional terms not
dominating at these energies. This of course introduces more freedom to
the model, thus a better agreement with data at high energies.

\begin{figure}[t]
\centering
\begin{minipage}[c]{0.5\linewidth}
\centering
\includegraphics[width=1.0\linewidth]{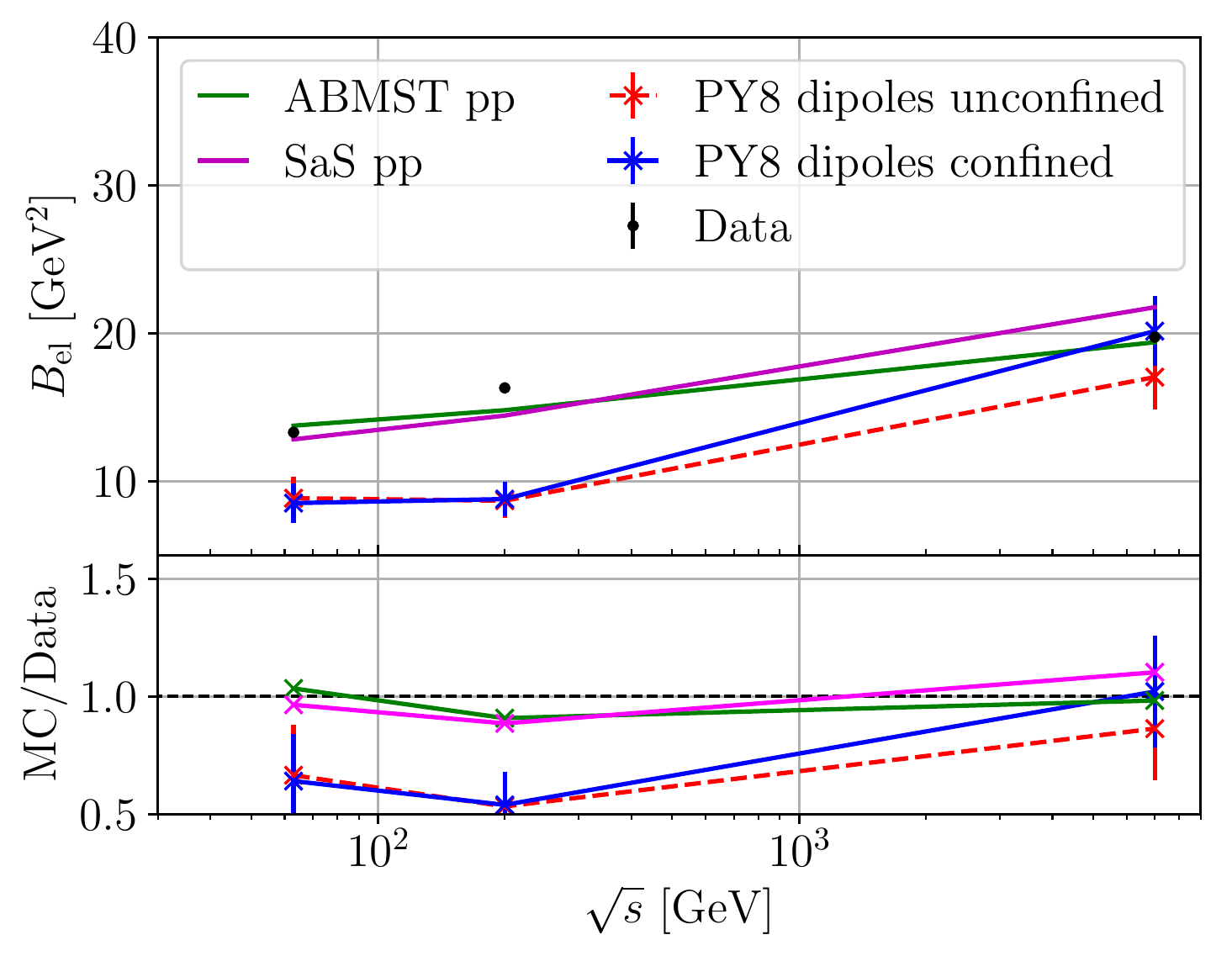}\\
(a)
\end{minipage}%
\begin{minipage}[c]{0.5\linewidth}
\centering
\includegraphics[width=1.0\linewidth]{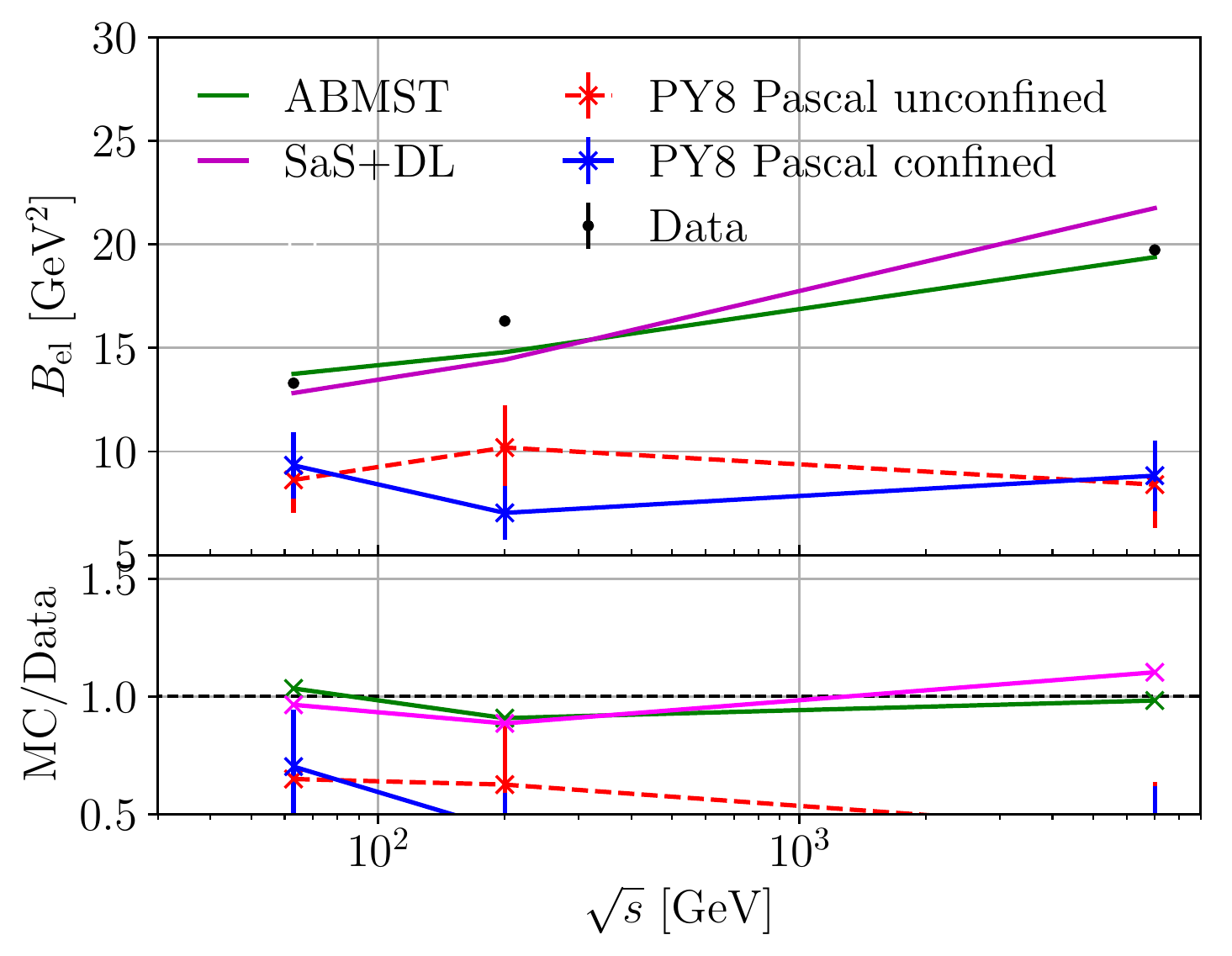}\\
(b)
\end{minipage}
\caption{\label{Fig:ppSlope} The elastic slope for $\p\p$ collisions as a
function of $\sqrt{s}$ for the dipole (a) and Pascal (b) models. Both
show the confined and unconfined versions in solid blue and dashed red
lines, respectively. Both figures show the ABMST (solid green lines) and
SaS+DL (solid magenta lines) for comparison.
}
\end{figure}

The last result is the elastic slope at $t=0$, shown in figure
\ref{Fig:ppSlope}. Second to the total cross section, this is the 
most important distribution for us to describe, as this is sensitive 
to the internal structure of the proton, i.e.\ the actual impact 
parameter value used in the calculation, while e.g.\ the elastic 
cross section is only sensitive to the average impact parameter. 
Figure \ref{Fig:ppSlope} again shows the results for the full dipole 
model (a) and the Pascal model (b). Here, both models are
undershooting the data by roughly 50\% in the entire energy range,
except that the dipole model is able to describe data in the very last
bin. Also here, the ABMST and SaS+DL models predictions are closer to the data
than the dipole and Pascal models.

We expect that the introduction of a running strong coupling will aid in the description
of the data. This introduction appears in two places: in the
dipole evolution and in the dipole-dipole scattering. A larger strong
coupling in the evolution decreases the average step size in rapidity and
increases the average size of the emitted dipoles, thus allowing for a 
larger number of larger dipoles at the end of the evolution. 
This, along with the increased dipole-dipole scattering cross section
with increased strong coupling, would essentially increase all the cross 
sections, and thus also the elastic slope. The scale choice in such a 
running coupling would not be obvious, however, and we thus postpone 
the inclusion of a running coupling to future work.

The combined results on $\sigma_{\mrm{el}}$, $\sigma_{\mrm{tot}}$ and
the elastic slope deserves a further comment. From the optical theorem the differential elastic cross section is:
\begin{equation}
	\frac{\d\sigma_{\mrm{el}}}{\d t} = \frac{\sigma^2_{\mrm{tot}}}{16 \pi}(1+\rho^2).
\end{equation}
Neglecting the real part of the amplitude puts $\rho = 0$. The left hand
side is often approximated by an exponential: $\d \sigma_{\mrm{el}}/\d t
= \exp \left(B_{\mrm{el}}\cdot t \right)$, giving $\sigma_{\mrm{el}} =
\sigma^2_{\mrm{tot}}/(16 \pi B_{\mrm{el}})$ when integrated over $t$.
The results in figures \ref{Fig:ppSigTot}, \ref{Fig:ppSigEl} and
\ref{Fig:ppSlope} are not in agreement with this simple proposition.
This can either mean that the exponential approximation is not a good
one (which is manifestly true for large $|t|$), or that the shape of
$T(b)$ in our model simply is too narrow, such that the elastic cross
section and hence $B_{\mrm{el}}$ is not well described.
If the latter is the case, a solution to the problem could be to include
more dipoles than three in the initial proton state. This would increase
the total and elastic cross sections at low $\sqrt{s}$, while the effect 
at higher $\sqrt{s}$ could be tuned away by a slightly smaller value of
$r_{\mrm{max}}$ or $\alpha_s$. Other studies of proton substructure \cite{Albacete:2017ajt}
has indicated that the number of hot-spots required for a satisfactory
qualitative description of LHC data is larger than three, but at the same time,
previous studies of the Mueller dipole formalism in the DIPSY event generator \cite{Avsar:2006jy}
indicated that a triangle configuration of initial dipoles is the most suitable
choice for a good description of the cross sections. A study of this sort will
therefore likely have to rely on attemps of simultaneous description
of both sub-structure and cross sections, and will be referred to a future
study.

Table \ref{Tab:ppTune} shows the parameters obtained when tuning to all three
observables ($\sigma_{\mrm{tot}}, \sigma_{\mrm{el}}, B_{\mrm{el}}$)
using \textsc{Professor2}. We also show the $\chi^2/N_{\mrm{dof}}$ for
three data sets of various sizes. It is striking that the inclusion of
the elastic cross section to the $\chi^2$-calculation swaps the behaviour 
of the full dipole model -- without the elastic data set, the confined
model has a lower $\chi^2/N_{\mrm{dof}}$ than the unconfined model,
while the opposite is true with the inclusion of the elastic cross 
section. This swap is caused by the first two data points in the elastic 
cross section, where the unconfined version of the full dipole describes data 
slightly better than the confined version. 
The parameters of the dipole model obtained with the tunes show the
behaviour also observed in $\gamma^*\p$: adding confinement allows for an 
increased initial dipole size and slightly larger fluctuations around this size. The
initial dipole size seems reasonable for both the confined dipole model
and the unconfined Pascal model, giving sizes of the order
$r_0\sim0.7-1.$ fm also confirmed by \DP~($r_0\sim0.7$ fm) and
proton charge radii measurements (giving roughly $r_0\sim0.9$ fm). 

\begin{table}[t]
\centering
\begin{tabular}{l | c | c | c | c | c | c }
& \multicolumn{2}{| c }{Full dipole model} & \multicolumn{2}{| c}{Pascal model} & \multicolumn{1}{| c}{ABMST} & \multicolumn{1}{| c}{SaS+DL}\\
Parameter & unconfined & confined & unconfined & confined & & \\
\hline
$r_0$ [fm] & 0.53 & 0.70 & 1.20 & 1.10 & &  \\
$r_{\mrm{max}}$ [fm]& - & 3.00 & - & 2.05 & &  \\
$r_{\mrm{w}}$ [fm]& 0.17 & 0.27 & 0.10 & 0.10 & &  \\
$\alpha_s$& 0.24 & 0.22 & 0.15 & 0.16 & &  \\
$f_r$& - & - & 0.25 & 0.40 & &  \\
$\Delta y$ & - & - & 2.20 & 2.45 & &  \\
\hline
$\chi^2/N_{\mrm{dof}}$ : $\sigma_{\mrm{tot}}$ & 5.22 & 3.34 & 0.75 & 1.04 & 0.28 & 0.41 \\
$\chi^2/N_{\mrm{dof}}$ : $\sigma_{\mrm{tot}}$, $B_{\mrm{el}}$ & 6.89 & 5.40 & 2.80 & 5.34 & 0.25 & 0.34 \\
$\chi^2/N_{\mrm{Dof}}$ : $\sigma_{\mrm{tot}}$, $B_{\mrm{el}}$, $\sigma_{\mrm{el}}$ & 11.21 & 13.67 & 4.63 & 5.14 & 0.20 & 0.46 \\
\end{tabular}
\caption{\label{Tab:ppTune} The parameter values obtained when tuning
to the $\sigma_{\mrm{tot}},\sigma_{\mrm{el}},B_{\mrm{el}}$ data set and the $\chi^2$
obtained for the different models.}
\end{table}

As already stated, the inclusion of a running strong coupling is
expected to improve results for both the dipole and Pascal model.
As we currently are aiming to describe proton substructure at the TeV 
scale, we can, however, ignore the small deviations from 
$\sigma_{\mrm{tot}}$ and $B_{\mrm{el}}$ at lower energies for the moment. 

\section{Results II -- eccentricities in small and large systems}

In this section we turn our attention to predictions related to the
geometry of an event. The parton-level eccentricities of
both small and large systems are examined using the matching between the dipole model
and the MPI framework described as in section \ref{sec:mpi}. Results from the
dipole model\footnote{We do not show eccentricities calculated using the Pascal approximation, 
as it is, at this point, mainly intended as a toy model. If the Pascal approximation should be used for studies of eccentricities,
we should point out that the large spread in daughter sizes as seen in \figref{Fig:RratiosMeanAndSpread} 
must be incorporated, in order to provide reasonable estimates for flow fluctuations.} 
are shown along with the default models of \PY: in $\p\p$ 
collisions, the default scheme of \PY~is a transverse placement according 
to a 2D Gaussian, while for larger systems two default \PY~methods are
available -- the usual Glauber approach and the 2D Gaussian $\p\p$ model 
extended to larger systems. In order to compare to data, all events are
hadronized with \PY~after the parton-level eccentricities are
calculated. Results are presented and in a single case compared to data from ALICE
\cite{Acharya:2019vdf}. Parton level eccentricities were calculated 
with a $\pT$ weighting, cf.\ section \ref{sec:eccentricity}, and events 
accepted if they passed the ALICE high-multiplicity trigger.
Eccentricities and normalized symmetric cumulants are presented as a
function of average central multiplicity ($|\eta|<0.8$).

Recall from section \ref{sec:mpi} that \PY~includes a $\pT$-dependent Gaussian smearing of 
parton vertices in the initial- and final-state shower. It is not 
clear from first principles whether such effects should be included in
the calculation of geometric quantities or not. Consider, on one hand, creation of 
a QGP at early times, right after the collision. Here a parton shower will not
be able to influence the geometry of the event, before a hydrodynamic response should
be taken into account. On the other hand, one can imagine a system with large gradients
(such as a small collision system) which will take time to hydrodynamize, and will therefore
be influenced by geometric fluctuations from the final state shower as
well. It is important to note, however, that no QGP is assumed in any results
presented below as no QGP is assumed in neither \PY~nor
\textsc{Angantyr}. 

\begin{figure}[t]
\centering
\includegraphics[width=0.5\linewidth]{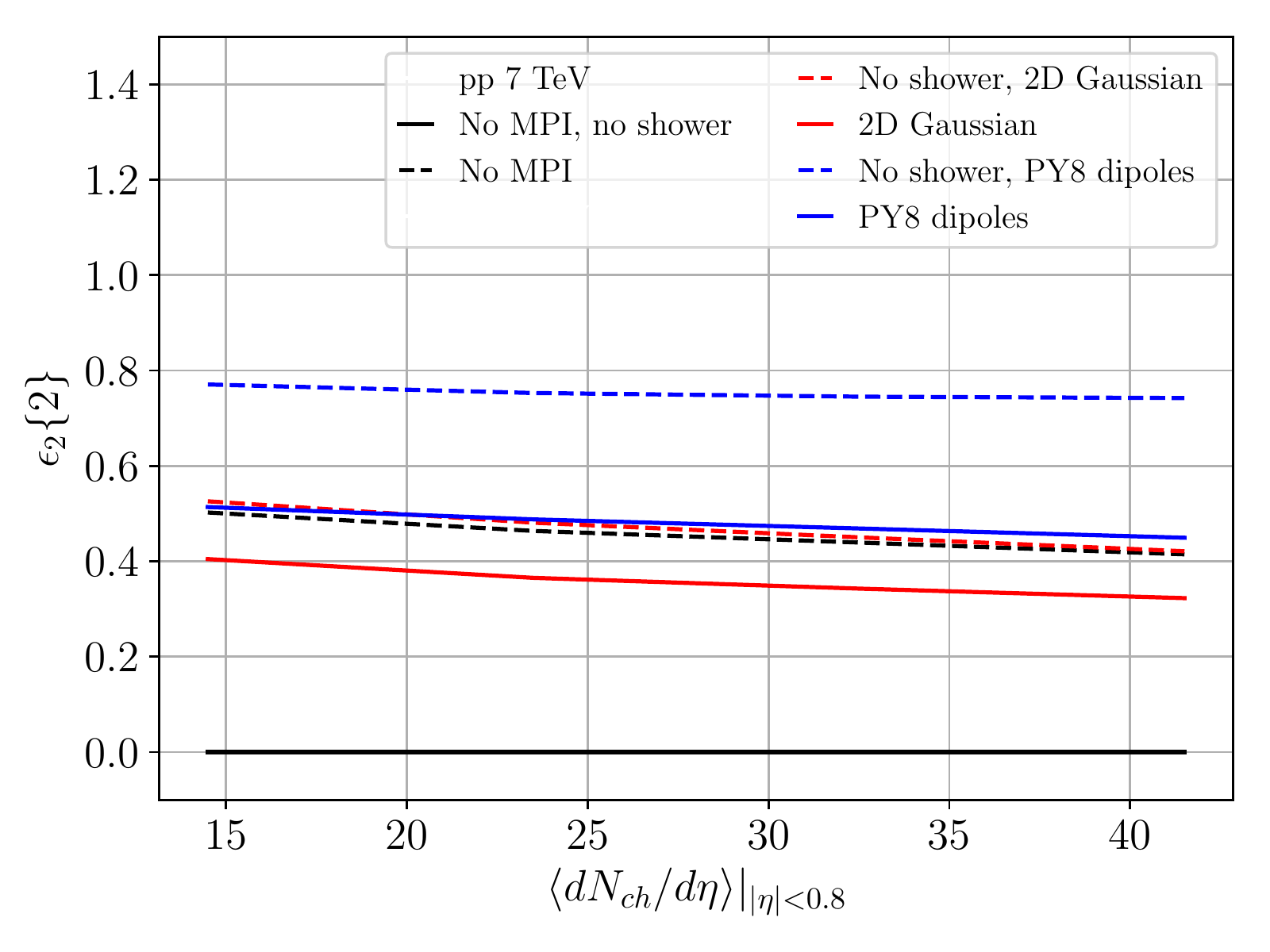}\\
\caption{\label{Fig:ecc-zero1} The eccentricities in $\p\p$ collisions
obtained with the several different options: No MPI vertex assignment
and no shower smearing (solid black lines), no MPI vertex assignment,
with shower smearing (dashed black lines), the 2D Gaussian MPI vertex
with/without shower smearing (solid/dashed red lines, respectively) and 
the dipole model MPI vertex assignment with/without shower smearing
(solid/dashed blue lines).
}
\end{figure}

Opening up for a discussion, we show results in figure \ref{Fig:ecc-zero1} with and 
without shower. It is evident that the eccentricities are vastly
affected by the models. First consider the simplest case, i.e.\ placing all MPIs
in the proton center and not introducing any shower smearing. This gives no
eccentricity as expected, cf.\ solid black line in figure
\ref{Fig:ecc-zero1}. Symmetric distributions, such as the 2D Gaussian
shower smearing and the MPI vertex assignment, should in principle give
no eccentricity. But, as we are sampling only a finite number of MPIs
from such symmetric distributions, an eccentricity does appear for
these models, cf.\ dashed red and dashed black lines of figure 
\ref{Fig:ecc-zero1}. The two methods overlap, thus the exact same effect 
can be introduced with either (a) no MPI vertex assignment with a Gaussian smearing
from the shower or (b) a Gaussian MPI vertex assignment and no shower smearing. 
That the two overlap is not so surprising
as both are Gaussian smearings, and applying such a smearing during the
shower or assigning it to the MPIs should make no difference: both
methods give rise to a more lenticular overlap region of the two
colliding protons. 

Applying Gaussian smearing twice, i.e.\ both in the MPI vertex assignment and
during the shower smears the lenticular shape from the MPI assignment
slightly, thus causing the eccentricity to drop, cf.\ the solid red line
in figure
\ref{Fig:ecc-zero1}. The largest effect on the eccentricity is seen when
purely considering MPI vertex assignment with the dipole model, cf.\ the
dashed blue line in figure \ref{Fig:ecc-zero1}. The eccentricity with
the dipole model is approximately twice as large as with the Gaussian
model, thus indicating that event-by-event asymmetries in the initial
state gives rise to larger fluctuations and thus larger eccentricities.
Adding the Gaussian shower smearing on top of the dipole model, solid
blue line of figure \ref{Fig:ecc-zero1}, washes out some of these
features -- i.e.\ makes the almond shape of the overlap region rounder.

\begin{figure}[t]
\centering
\begin{minipage}[c]{0.5\linewidth}
\centering
\includegraphics[width=1.0\linewidth]{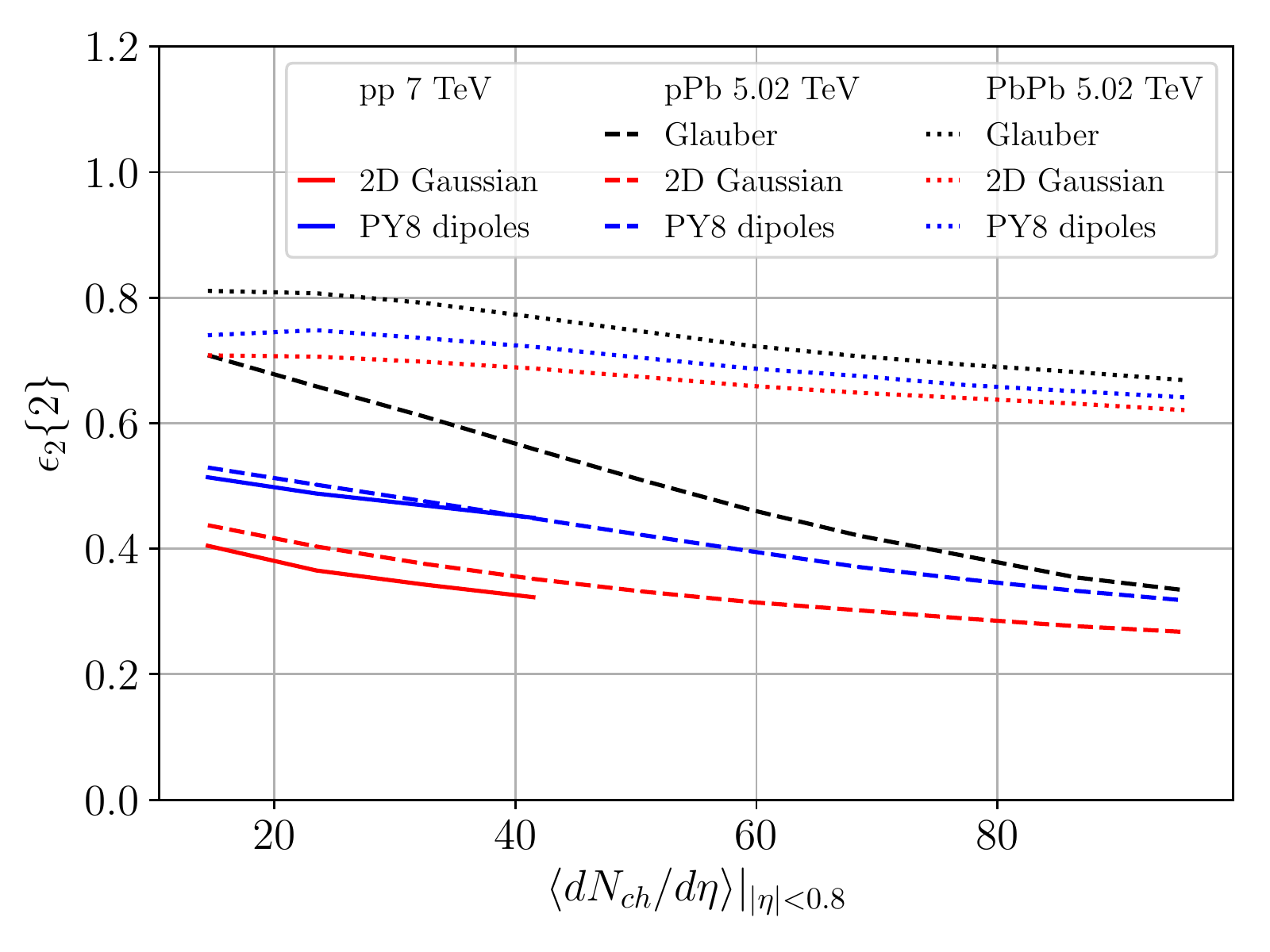}\\
(a)
\end{minipage}%
\begin{minipage}[c]{0.5\linewidth}
\centering
\includegraphics[width=1.0\linewidth]{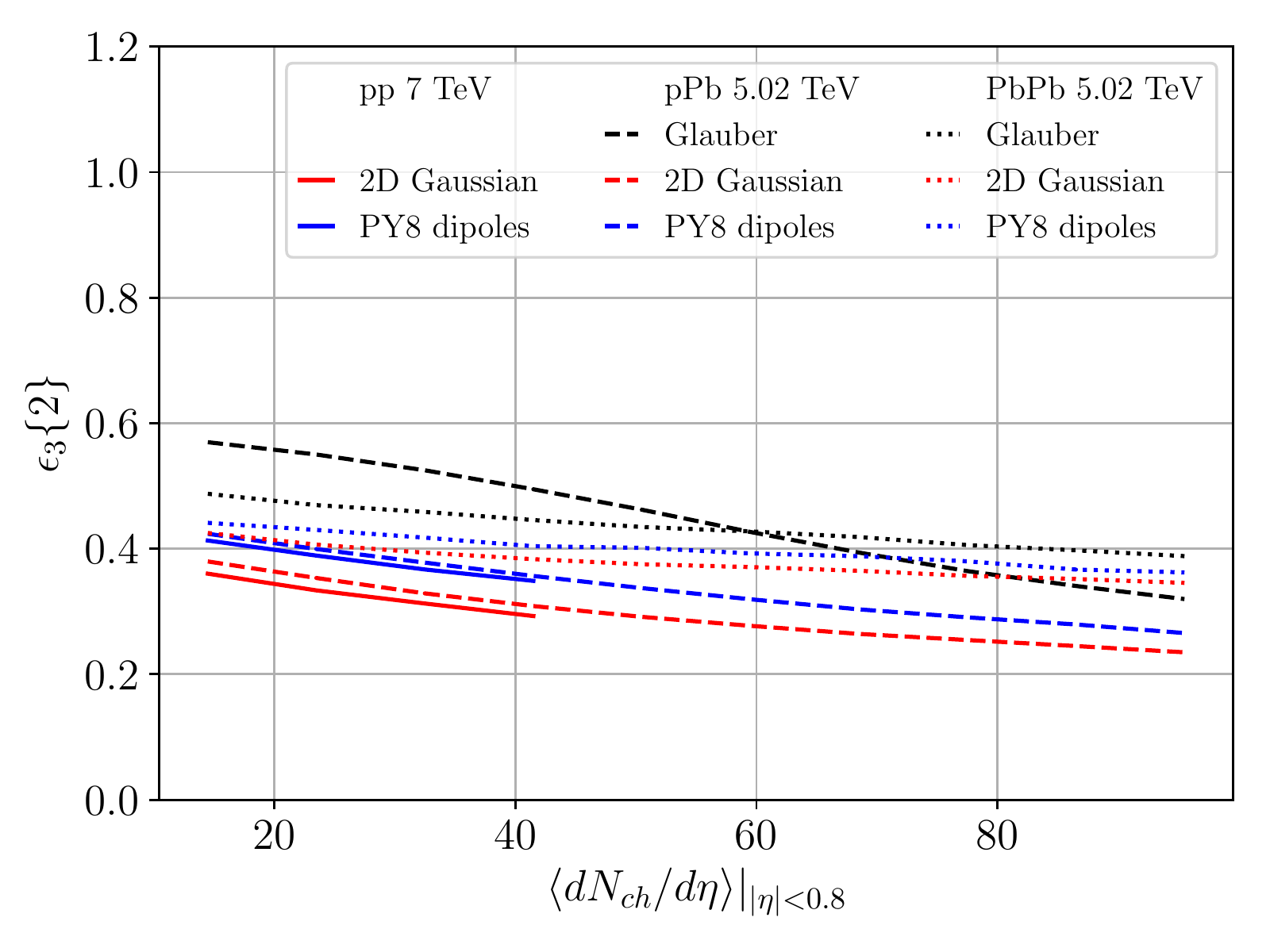}\\
(b)
\end{minipage}
\caption{\label{Fig:eccn2} The second (b) and third (c) order 
eccentricities using two-particle cumulants for $\p\p,\p\Pb,\Pb\Pb$ 
collisions (solid, dashed, dotted, respectively) using the Glauber 
(black), Gaussian (red) and dipole (blue) models.
}
\end{figure}

Figure \ref{Fig:eccn2} shows the eccentricities
$\epsilon_{2,3}\{2\}$ in three different systems. Beginning with
$\epsilon_2$ we observe in $\p\p$ that the dipole evolution gives rise to
a larger eccentricity than the 2D-Gaussian. In the dipole evolution, the asymmetry 
is built in at the cascade level, where in the 2D-Gaussian, where MPIs are sampled from
a symmetric distribution, asymmetry only arises due to the sample size.
Proceeding to larger systems, $\p\Pb$ and $\Pb\Pb$, it is evident that the
same trend is seen: the dipole model gives rise to larger $\epsilon_2$
than the symmetric model. The Glauber model, however, is consistently
larger than the other two models at low multiplicity, while all three 
models appear to approach the same eccentricity at higher multiplicities. 
Thus it becomes evident that the proton initial state does have an effect 
on eccentricities, and that it is especially evident in low-multiplicity
events, e.g.\ peripheral $\Pb\Pb$ collisions. 

Unfortunately, the low-multiplicity events are often marred by large
non-flow effects. Measuring the eccentricities with higher-order 
cumulants can remove some of the contributions from non-flow, thus 
making it easier to compare data to models. We present results for
$\epsilon_{2}$ with higher-order cumulants in appendix
\ref{sec:eccAppendix}, as the results are similar in shape as figure
\ref{Fig:eccn2}, but differ in normalisation. Figure \ref{Fig:eccn2} (b) show
$\epsilon_3\{2\}$ for all three systems. Here, it becomes more difficult
to distinguish between the models in symmetric systems, while a large
discrepancy between the Glauber approach and the other two is seen in
$\p\Pb$. 

\begin{figure}[t]
\centering
\begin{minipage}[c]{0.5\linewidth}
\centering
\includegraphics[width=1.0\linewidth]{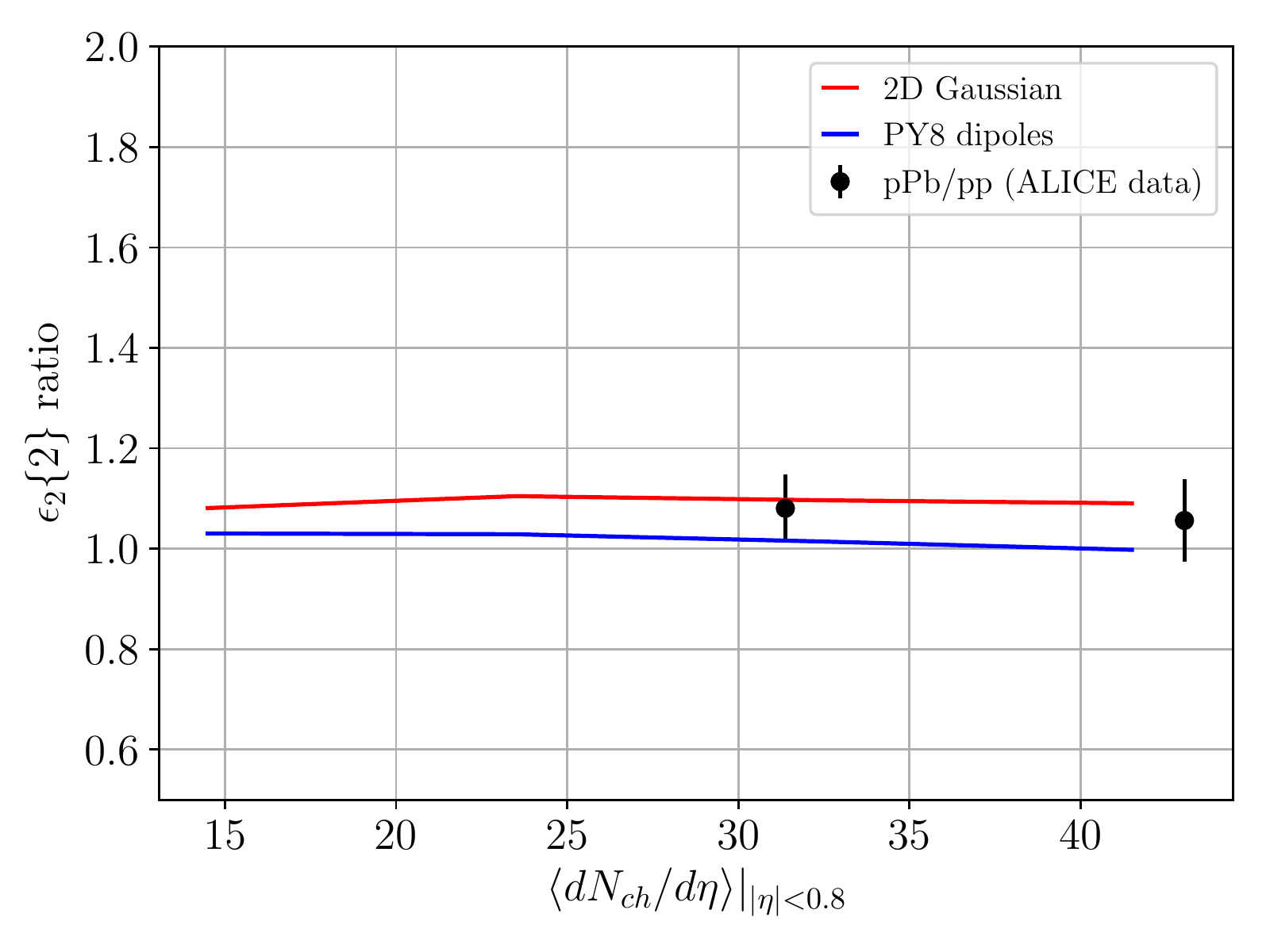}\\
(a)
\end{minipage}%
\begin{minipage}[c]{0.5\linewidth}
\centering
\includegraphics[width=1.0\linewidth]{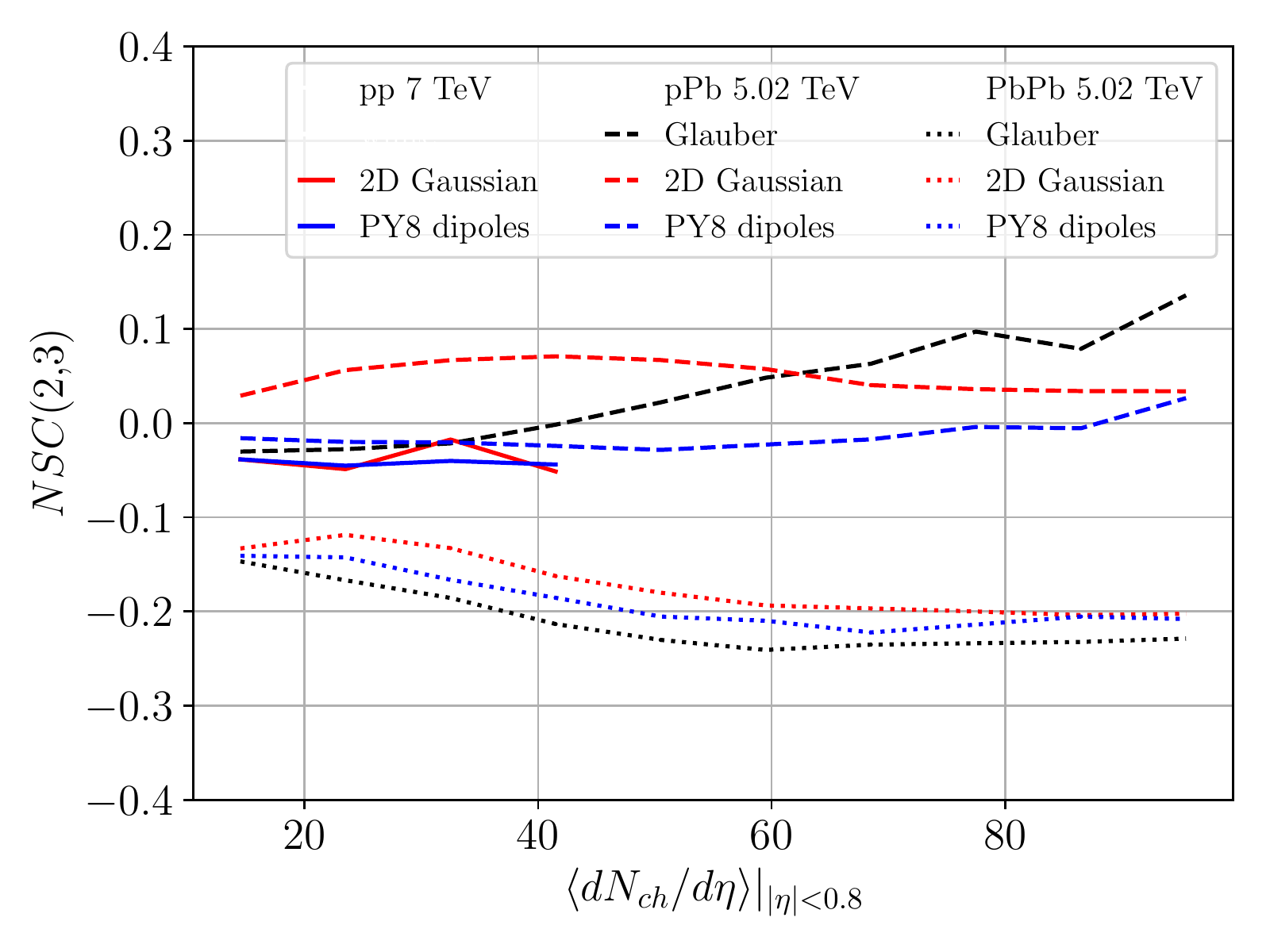}\\
(b)
\end{minipage}
\caption{\label{Fig:eccn2ratio} (a) The ratio of second order eccentricities 
obtained in $\p\Pb$ w.r.t.\ the baseline $\p\p$ sample using the Gaussian 
(red) and dipole (blue) models. Data points calculated from ALICE
figures \cite{Acharya:2019vdf}. (b) The normalized symmetric cumulants $NSC(3,2)$
for $\p\p,\p\Pb,\Pb\Pb$ collisions (solid, dashed, 
dotted, respectively) using the Glauber (black), Gaussian (red)
and dipole (blue) models.
}
\end{figure}

Another feature seen in figure \ref{Fig:eccn2} is that the dipole 
model gives roughly the same results for $\epsilon_{2,3}$ in both 
$\p\p$ and $\p\Pb$ collisions. If one assumes that the response functions
are the same for the two systems (however one may have obtained these
response functions, QGP or by string-string interactions), the ratio of $\p\Pb$ to $\p\p$
eccentricities should thus be comparable to the ratio of flow
coefficients measured with the ALICE detector. This ratio is shown 
in figure \ref{Fig:eccn2ratio} for the second-order eccentricity. Both
the Gaussian and dipole models are compatible with the ALICE data,
however, so we cannot presently discriminate between the two.
Additional measurements of the flow coefficients in low-multiplicity 
events are thus required in order to discriminate between models in
this observable.

Figure \ref{Fig:eccn2ratio} (b) shows the normalized symmetric cumulant,
$NSC(3,2)$. This has been constructed to study the correlations 
between the eccentricities and normalized to the uncorrelated 
eccentricities in order to remove the effects of the response 
function. ALICE reports that all three systems have the same
$NSC(3,2)$ at the same average multiplicity, indicating that the
correlation between the flow coefficients are the same in different
collision systems. We observe no such effect. Focusing on the dipole
model, the correlations appear equal in magnitude for $\p\p$ and $\p\Pb$,
but $\Pb\Pb$ results are consistently below the smaller systems. Results
for the Gaussian model shows no similarities at all between systems, as
the $\p\Pb$ $NSC(3,2)$ is positive, while $\p\p$ and $\Pb\Pb$ are negative.
Thus the normalized symmetric cumulants for $\p\Pb$ systems would be an 
ideal place to discriminate between the symmetric and asymmetric initial 
state. $\Pb\Pb$ results for all three models are in agreement with 
\textsc{IP-Glasma} predictions presented in the ALICE paper. The main
difference between the dipole model and the \textsc{IP-Glasma} approach
is the inclusion of saturation in the cascade of the latter. As the two
approaches give similar results, we do not find that saturation plays a
large role in this observable.
 
\subsection{\label{sec:cms-pa-fluctuations}Flow fluctuations in pPb collisions}

\begin{figure}[t]
\centering
\begin{minipage}[c]{0.5\linewidth}
\centering
\includegraphics[width=1.0\linewidth]{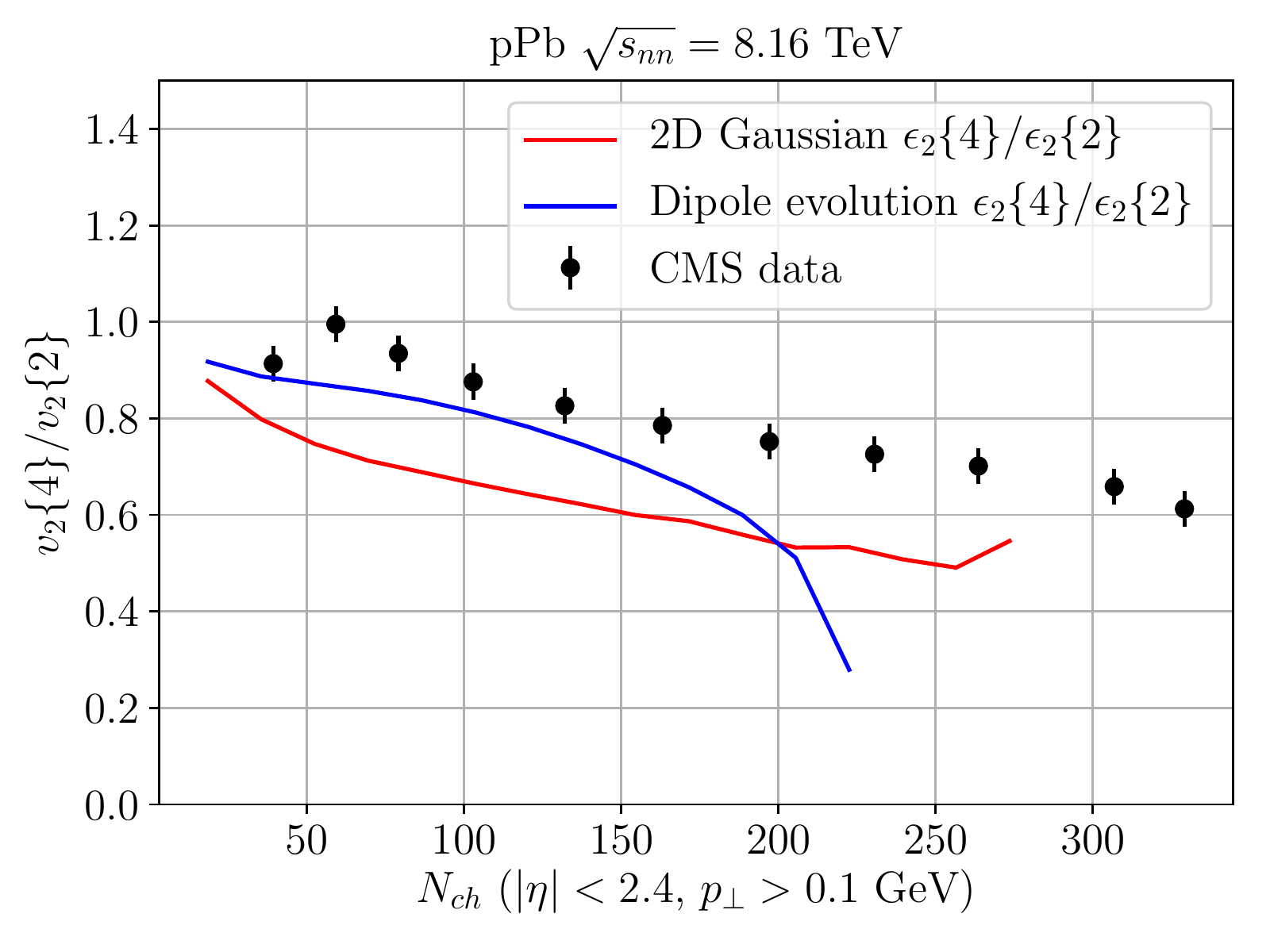}\\
(a)
\end{minipage}%
\begin{minipage}[c]{0.5\linewidth}
\centering
\includegraphics[width=1.0\linewidth]{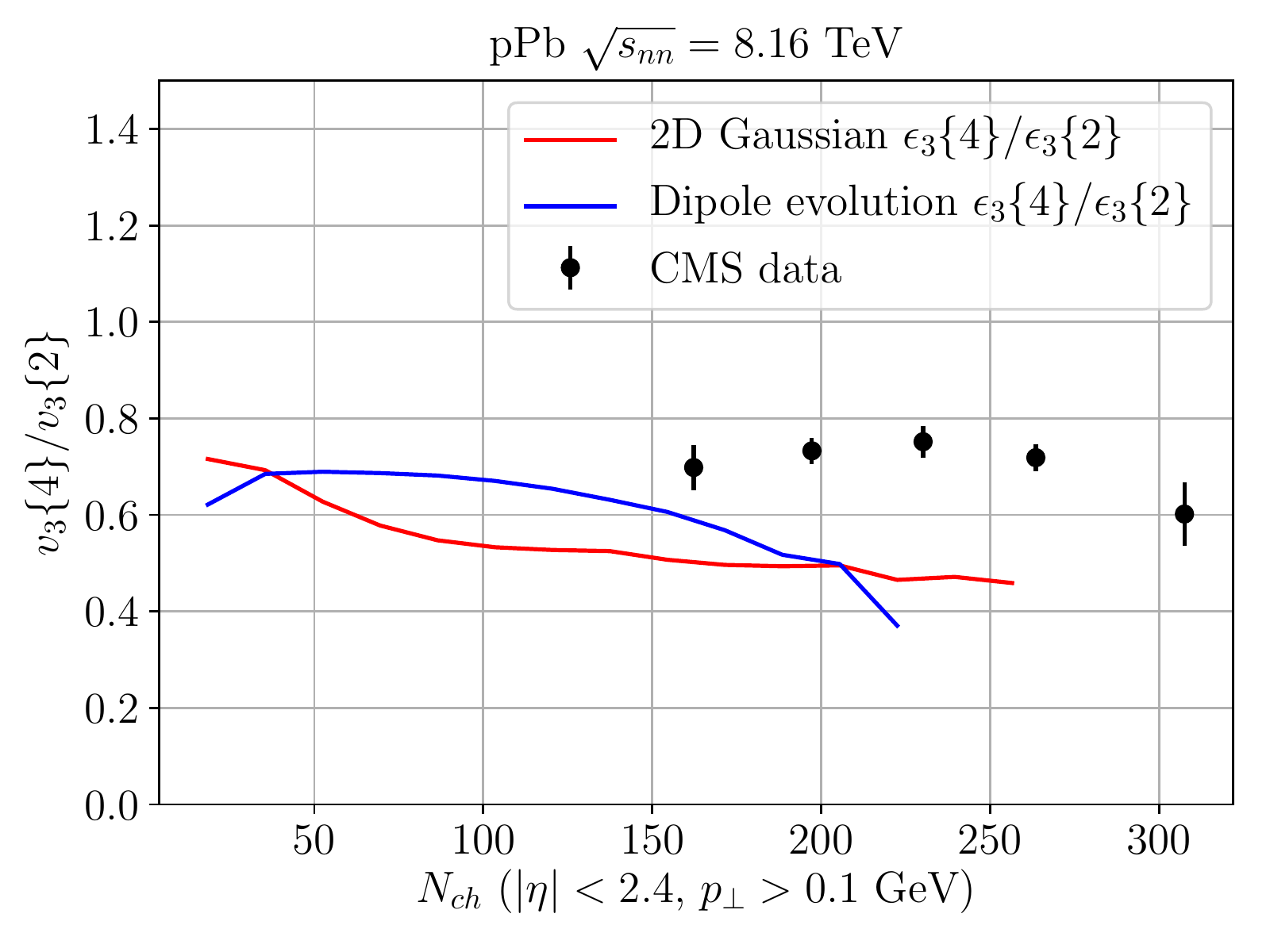}\\
(b)
\end{minipage}
	\caption{\label{Fig:cms-nch} Ratios of $v_n\{4\}/v_n\{2\}$ with $n=2$ (a) and $n=3$ (b) as measured by CMS as function of multiplicity in pPb collisions, compared to eccentricity ratios calculated with the Gaussian and the dipole models. 
}
\end{figure}

Recently, CMS presented results on multi-particle correlations using
higher-order particle cumulants in $\p\Pb$ collisions
\cite{Sirunyan:2019pbr}. Ratios of the flow-coefficients based on these
cumulants were presented, including the first measurements of the ratio
of $v_3\{4\}/v_3\{2\}$ in $\p\Pb$. In figure \ref{Fig:cms-nch} we show
the predictions for the ratios with the confined dipole model and the default 
Gaussian model as a function of multiplicity. Both models reasonably 
reproduce the shape seen in the elliptical ratio, figure
\ref{Fig:cms-nch} (a) showing $v_2\{4\}/v_2\{2\}$,
while the normalisation of the dipole model is slightly better than with
the Gaussian model. For the triangular ratio, figure \ref{Fig:cms-nch}
(b), both models appear to undershoot data at high multiplicities, where
data is available. As opposed to model predictions presented in the CMS
paper \cite{Giacalone:2017uqx,Bernhard:2016tnd}, our predictions have 
\tit{not} been applied a 10\% \textit{ad hoc} increase in
normalisation. And where the model predictions presented in
\cite{Giacalone:2017uqx,Bernhard:2016tnd} predicts roughly the same ratio for both
$\epsilon_2$ and $\epsilon_3$, neither the dipole nor the Gaussian model
predicts the same normalisation for the two ratios, cf.\ the
height of figure \ref{Fig:cms-nch} (a) and (b) differs.  

\begin{figure}[t]
\centering
\begin{minipage}[c]{0.5\linewidth}
\centering
\includegraphics[width=1.0\linewidth]{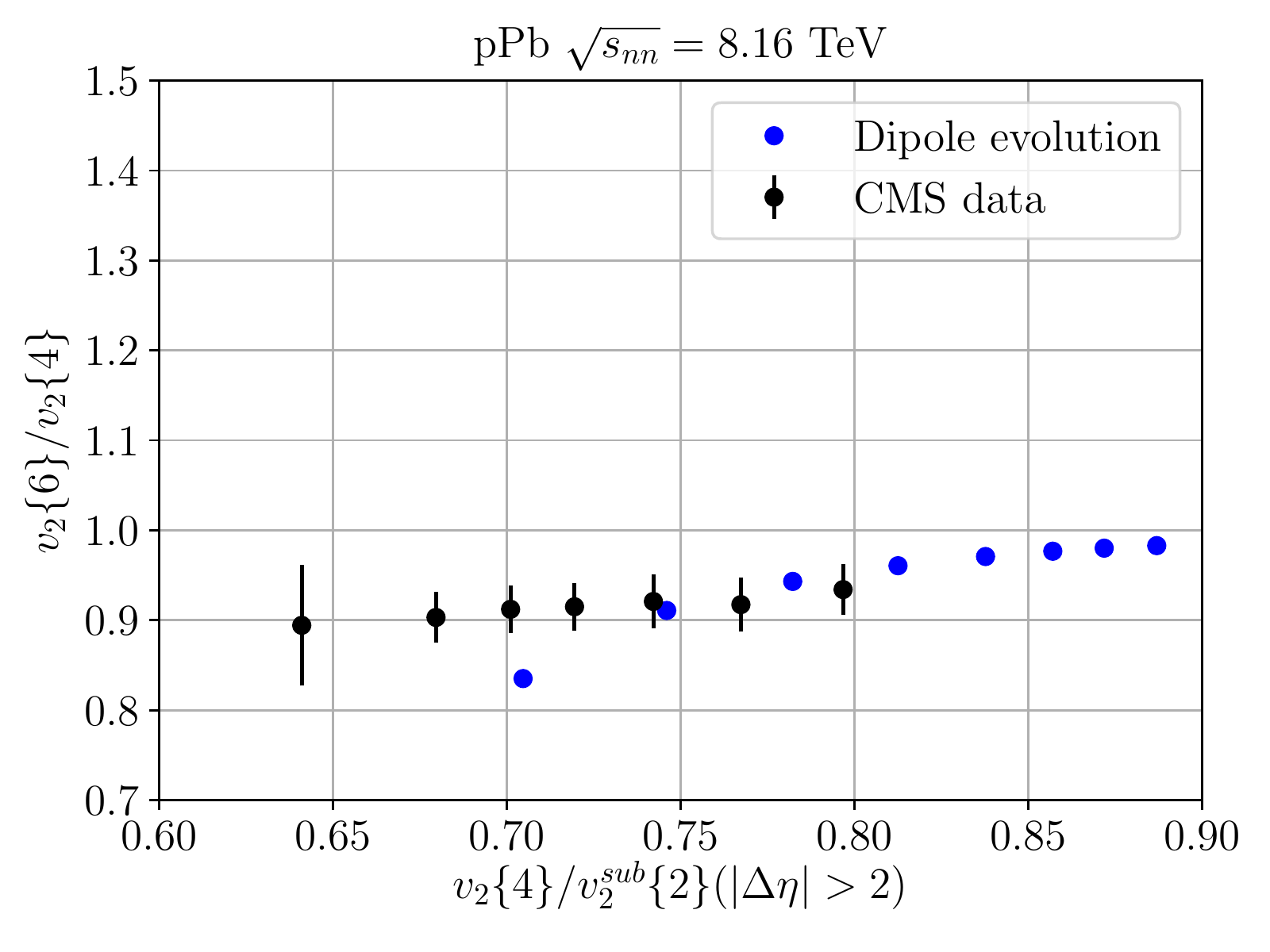}\\
(a)
\end{minipage}%
\begin{minipage}[c]{0.5\linewidth}
\centering
\includegraphics[width=1.0\linewidth]{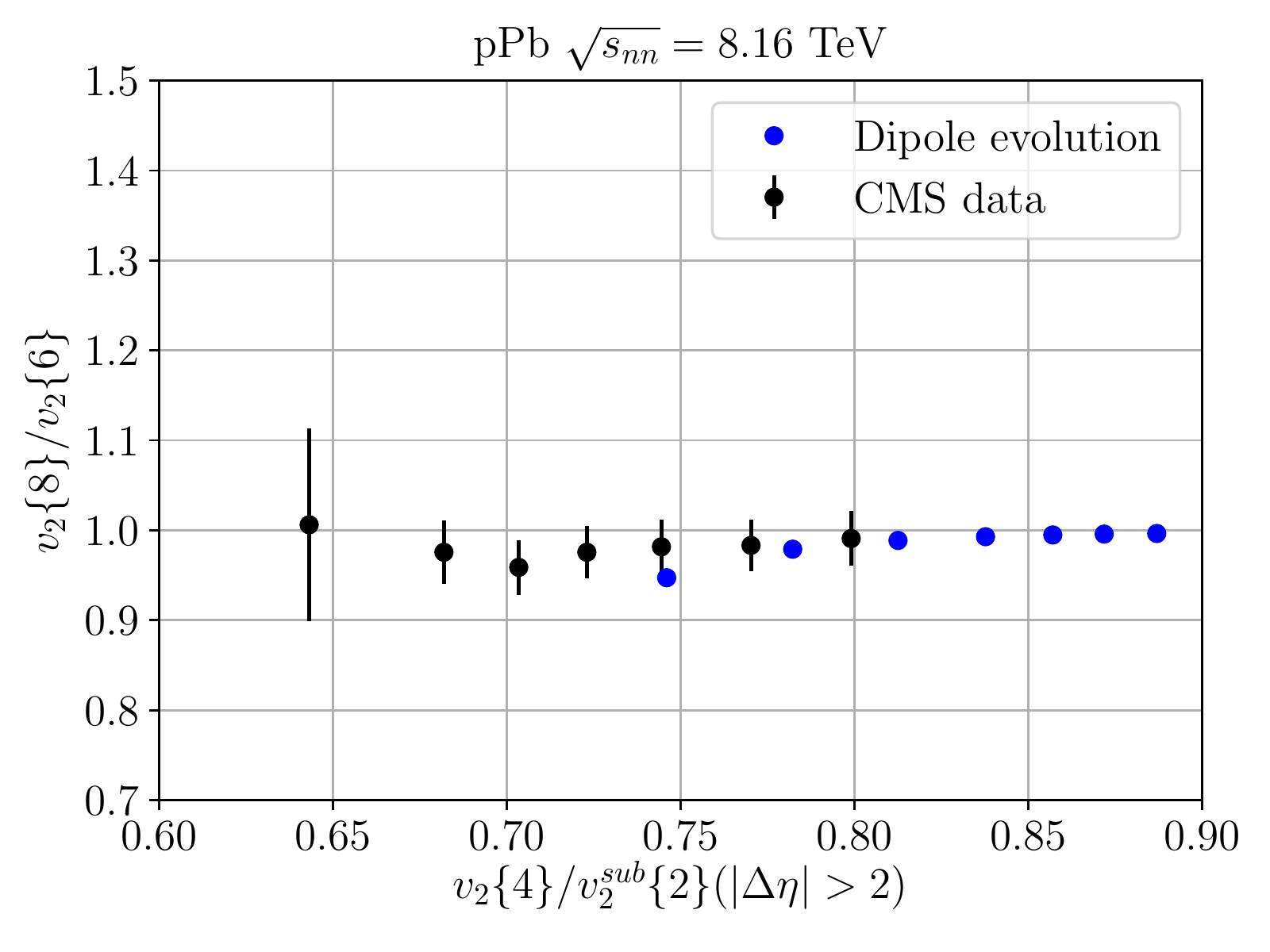}\\
(b)
\end{minipage}
\caption{\label{Fig:cms-dbl-ratios} Correlations between higher order flow harmonics as measured by the CMS experiment, compared to correlations between higher order eccentricity ratios calculated in the dipole model.
}
\end{figure}

Figure \ref{Fig:cms-dbl-ratios} shows the higher-order cumulant ratios for
elliptic flow as a function of the lower-order ratio presented in figure
\ref{Fig:cms-nch} (a). For higher order cumulants, the Gaussian model predicts 
purely imaginary values for even powers of the cumulants, hence it has been left out of the figures.
The dipole model, however, is able to describe data reasonably well. The
dipole predictions decrease with decreasing $v_2\{4\}/v_2\{2\}$ ratio in
figure \ref{Fig:cms-dbl-ratios} (a), while being roughly constant at unity
in figure \ref{Fig:cms-dbl-ratios} (b). This is in accordance with the
model predictions presented by CMS \cite{Yan:2013laa}, assuming a
non-Gaussian model for the initial state. We note that the
eccentricities presented with the dipole model here are 
(a) based on a pQCD model, and (b) related to final state multiplicities
calculated in the same acceptance as the experiment.

\section{\label{sec:results3}Results III -- dynamic colour fluctuations in Glauber calculations}
A general feature of several models describing both collisions of protons and of nuclei, is 
the notion of interacting nucleons and nuclear sub-collisions, calculated in the formalism
of Glauber \cite{Glauber:1955qq,Miller:2007ri}. The basic formalism is mainly concerned with
calculating the full $\A\A$ scattering matrix or amplitude from knowledge of the nucleon-nucleon 
amplitude and spatial positions. Multiple interactions between nucleons factorize
in transverse coordinates, so in the eikonal limit the $S$-matrix for scattering between two nuclei
$A$ and $B$ becomes:
\begin{equation}
	\label{eq:bigglauber}
	\hat{S}^{(AB)}(\vec{b}) = \prod_{i=1}^A \prod_{j=1}^B \hat{S}^{(N_i N_j)}(\vec{b}_{ij}),
\end{equation}
where $i$ and $j$ denote the individual nucleons, $\vec{b}$ is the nucleus--nucleus impact parameter and $\vec{b}_{ij}$
is the nucleon--nucleon impact parameter.
We will here consider the simplifying case where
only one projectile (either p or $\gamma^*$, called $n$ below)
collides with a nucleus (A), which reduces
\eqref{eq:bigglauber} considerably:
\begin{equation}
	\hat{T}^{(nA)}(\vec{b}) = 1 - \prod_{i=1}^{A} \hat{S}^{(nN_i)}(\vec{b}_{ni}) = 1 - \prod_{i=1}^A(1 - \hat{T}^{(nN_i)}(\vec{b}_{ni})).
\end{equation}
If no fluctuations in the interaction are included, the projectile-nucleon elastic amplitude can be inserted, and the total and
elastic cross sections can be calculated directly from eqs.~(\ref{Eq:XS1}--\ref{Eq:XS2}).
If fluctuations for projectile and target are included, as calculated for example in the dipole model, the amplitude will depend on
the states of target ($t_i$) and projectile ($p$) respectively. As shown in \sectref{sec:good-walker}, the elastic amplitude
can be calculated as an average over all states. In ref. \cite{Alvioli:2013vk} it was pointed out
that in the evaluation of such an average, the projectile must remain frozen in the same state throughout the passage of the 
target. Similar to \eqref{eq:elprofile} the elastic profile function (at fixed Mandelstam $s$) for a fixed state ($k$) of the projectile scattered on a single target nucleon (all states) becomes:
\begin{align}
	\Gamma_k(\vec{b}) &= \langle \psi_S | \psi_I \rangle = \langle
  \psi_k,\psi_t | \hat{T}(\vec{b}) | \psi_k, \psi_t \rangle = (c_k)^2
  \sum_t |c_t|^2 T_{tk} (\vec{b}) \langle \psi_k, \psi_t | \psi_k,
  \psi_t \rangle \nonu\\
	& = (c_k)^2 \sum_t |c_t|^2 T_{tk}(\vec{b}) \equiv \langle T_{tk}(\vec{b}) \rangle_t,
\end{align}
where previously suppressed indices $k$ and $t$ on $T$ are spelled out
for clarity. For a projectile-nucleus collision, with the projectile
  frozen in the state $k$, the relevant projectile-nucleon ($nN_i$) amplitude becomes:
\begin{equation}
\label{eq:tfluct}
	\langle T_{t_i,k}^{(nN_i)}(\vec{b}_{ni}) \rangle_t \equiv T_k^{(nN_i)}(\vec{b}_{ni}).
\end{equation}
In the short hand notation on the right hand side, the average over the repeated index $t$ is suppressed.
This is the amplitude used to determine which nucleons are ``wounded'' in a collision. If the purpose is to determine which
nucleons participate in the collision either elastically or inelastically, the differential wounded cross section can
be calculated with the normal differential pp total cross section as an
ansatz, $\d\sigma_{\mrm{tot}}/\d^2\vec{b} = 2\langle T \rangle_{p,t}$ from
\eqref{Eq:XS1}. Since the projectile should be frozen in the state $k$, the expression for $T$ from \eqref{eq:tfluct}
is inserted to the differential $\p\p$ total cross section. This just recovers the normal total
projectile-nucleon cross section:
\begin{equation}
	\label{eq:flucttot}
	\frac{\d\sigma_{\mrm{tot}}}{\d^2\vec{b}} = 2 \langle T_k \rangle_p = 2\langle \langle T_{t,k} \rangle_p \rangle_t = 2\langle T \rangle_{p,t}.
\end{equation}
In a Monte Carlo, the number of wounded nucleons can then be generated
by assigning each projectile or nucleon a radius of $\sqrt{\sigma_{\mrm{tot}}/2\pi}$,
where the expression in \eqref{eq:flucttot} has been integrated over
$\d^2\vec{b}$ to give $\sigma_{\mrm{tot}}$. Normally one is not
interested in the number of wounded nucleons including elastic
interaction, but rather those that contribute to particle production (i.e. where there is
a colour exchange). A usual approach is to just use the inelastic cross
section in place of $\sigma_{\mrm{tot}}$ in the Monte Carlo recipe.
This does, however, not account fully for colour fluctuations, as the
inelatic cross section is modified when averaging over target states
with a frozen projectile. Instead of directly using the inelastic
cross section in the Monte Carlo, the modified cross section should 
be used. This cross section was 
dubbed the ``wounded cross section'' in ref. \cite{Bierlich:2016smv},
and can be constructed by generalizing the inelastic cross section,
using  \eqref{eq:tfluct}. The inelastic cross section 
can from eqs.~(\ref{Eq:XS1}--\ref{Eq:XS2}) be directly written down as:
\begin{equation}
	\frac{\d\sigma_{\mrm{inel}}}{\d^2\vec{b}} = 2\langle T(\vec{b}) \rangle_{p,t} - \langle T(\vec{b}) \rangle^2_{p,t}.
\end{equation}
When the frozen projectile is taken into account by inserting $T$ from \eqref{eq:tfluct}, the usual expression is now not recovered, but the average over targets must be made \emph{before} squaring the second term:
\begin{align}
	\label{eq:sigmawounded}
	\frac{\d\sigma_w}{\d^2\vec{b}} &= 2\langle T_k(\vec{b}) \rangle_{p} -
  \langle T_k^2(\vec{b}) \rangle_p 
				=2 \langle T(\vec{b}) \rangle_{t,p} - \langle \langle T(\vec{b}) \rangle^2_t \rangle_p,
\end{align}
with internal indices again suppressed in the last equality. In a Monte
Carlo this can be generated as above, now only by inserting $\sigma_w$
in place of $\sigma_{\mrm{tot}}$. 

Generalizing this procedure to $\gamma^*\A$ collisions 
requires additional considerations. Starting from the elastic 
profile function for $\gamma^*\p$, a contribution from the 
photon fluctuating to a dipole state must be included. Examining only
the hadronic (non--VMD) components of the photon state, gives: 
\begin{align}
|\gamma^*\rangle\sim c_1|\q\qbar\rangle + c_2|\q\qbar\g\rangle
+ \mrm{higher~order~Fock~states}
\end{align}
where quark helicities have been neglected. We keep only the first (leading order) term, as the higher order
Fock states are included in the dipole evolution. Thus with a photon 
wave function given in eqs.~(\ref{Eq:photonWf1}--\ref{Eq:photonWf2}), we obtain:
\begin{equation}
	\label{eq:photon-dipole}
	|\gamma^*\rangle  = \int \d z \int \d^2\vec{r} \left(|\psi_L(z,r)|^2 + |\psi_T(z,r)|^2\right) |\psi_{I}(r,z)\rangle,
\end{equation}
with $|\psi_I\rangle$ a dipole state. The elastic profile is now:
\begin{align}
	\label{eq:prof-gamma-p-1}\Gamma_{\mrm{el}}(\vec{b}) &= \int \d z \int \d^2\vec{r}~\langle
  \psi_{S}(z,\vec{r}) | \hat{T}(\vec{b}) | \psi_{I}(z,\vec{r}) \rangle
  \langle \psi_{I}(z,\vec{r})|\gamma^*\rangle
  \nonu\\
	&= \int \d z \int \d^2 \vec{r}~(|\psi_L(z,\vec{r})|^2 + |\psi_T(z,\vec{r})|^2) \langle T(\vec{b}) \rangle_{p,t}.
\end{align}
The wounded cross section for $\gamma^*$A collisions can now be defined.
The first interaction is calculated using the photon wave function in the
elastic profile function, leading directly to:
\begin{equation}
	\frac{\d\sigma_w}{\d^2\vec{b}} = \int \d z \int \d^2 \vec{r}~(|\psi_L(z,\vec{r})|^2 + |\psi_T(z,\vec{r})|^2) (2 \langle T(\vec{b}) \rangle_{t,p} - \langle \langle T(\vec{b}) \rangle^2_t \rangle_p).
\end{equation}
This first interaction has now turned to photon from a superposition 
of all dipole states into a single, specific dipole (or vector meson). 
\tit{This} is the state that the projectile should be frozen
to throughout the passage of the nucleus: the first interaction chooses
a specific dipole state $|\psi_{I}\rangle_{z,\vec{r}}$ with given $z$ and $\vec{r}$. 
This reduces the elastic profile function for the secondary interactions to the well known \eqref{eq:elprofile}, 
from which a differential wounded cross section has already been calculated (\eqref{eq:sigmawounded}). 

Thus, in a Monte Carlo, the number of
wounded nucleons can be generated with the following method: 
\begin{itemize}
\item First by selecting, for each event, a dipole with $r$ and $z$
corresponding to the wave function weight, $w_{\gamma}$ in \eqref{Eq:WfWeight}
\item Secondly, testing if any nucleons are hit including the
photon wave function normalization proportional to $\alpha_{em}$ (i.e.\
according to \eqref{eq:prof-gamma-p-1})
\item If any nucleons are hit, then subsequently testing all (other)
nucleons, w.r.t.\ the dipole-target weight (i.e.\ \eqref{eq:elprofile})
\end{itemize}

In the following section, colour fluctuations from the introduced dipole model (where $T(\vec{b})$ can be evaluated directly 
from \eqref{eq:unitamp}) are compared to a parameterized approach for
fluctuations in pp collisions and $\gamma^*\p$ collisions,
and finally for $\gamma^*\A$.

\subsection{Colour fluctuations in $\p\p$, $\gamma^*\p$ and $\gamma^*\A$ collisions}

\begin{figure}[t]
\centering
\begin{minipage}{0.45\textwidth}
\includegraphics[width=\textwidth]{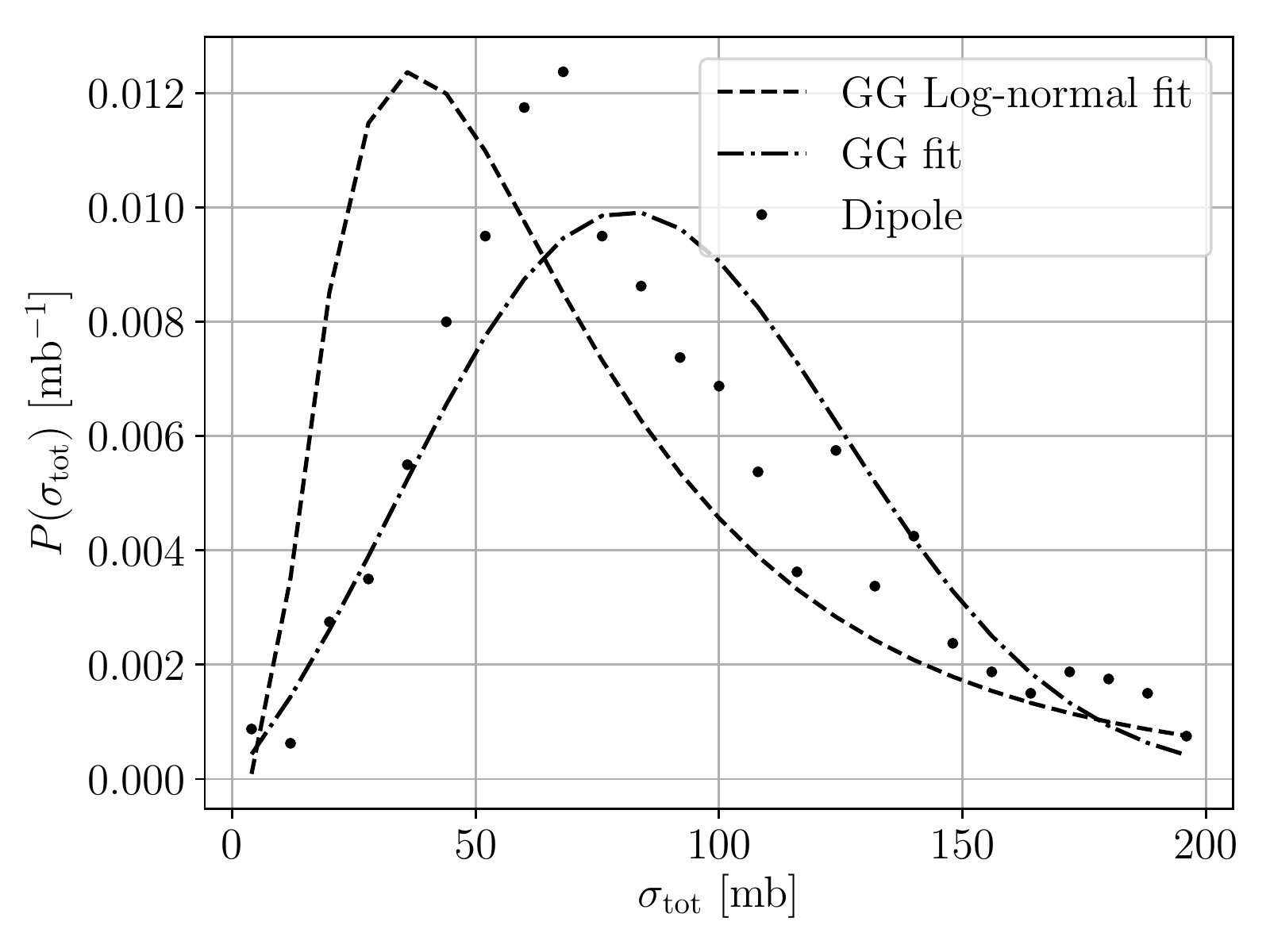}\\
(a)
\end{minipage}
\begin{minipage}{0.45\textwidth}
\includegraphics[width=\textwidth]{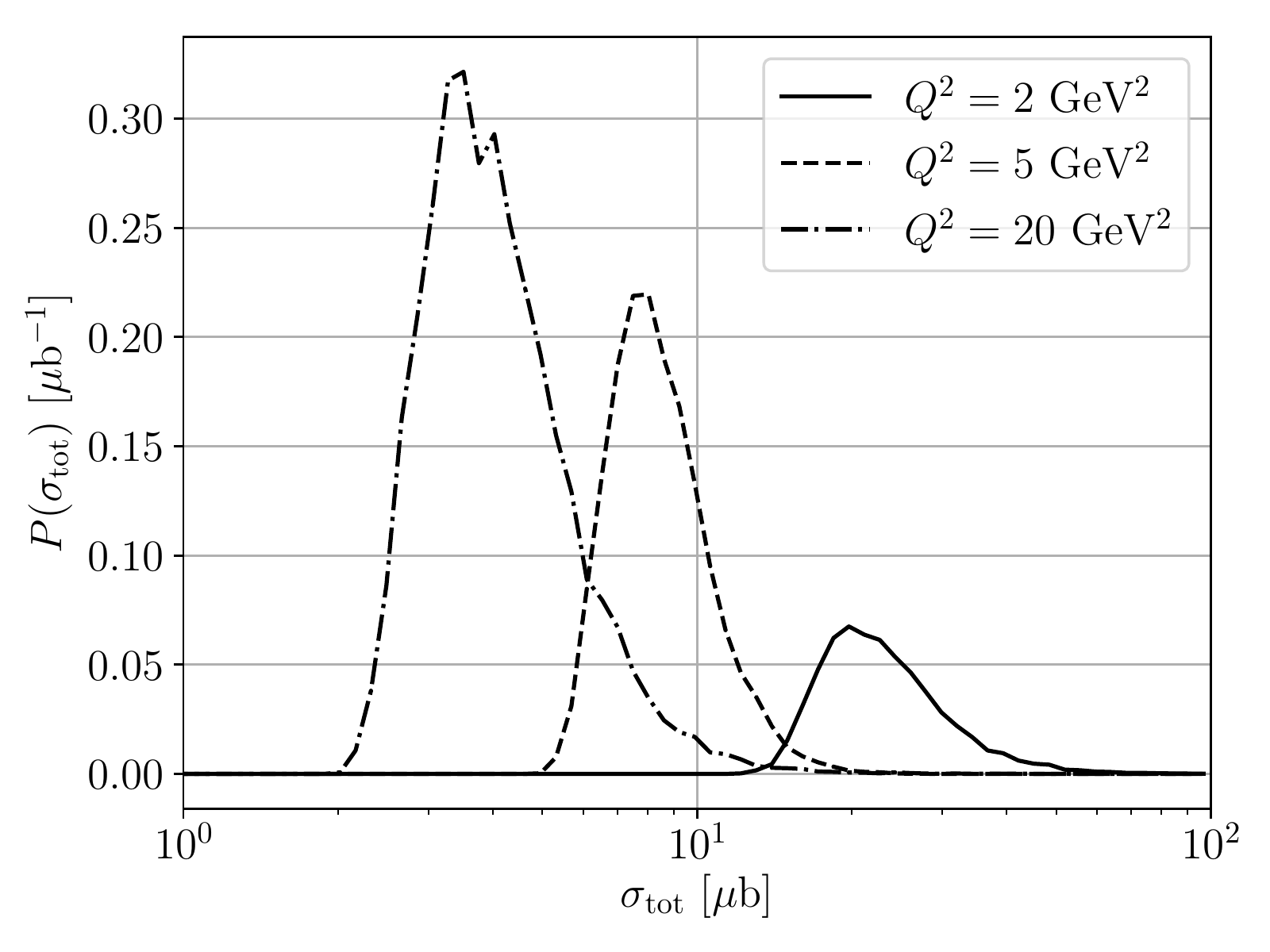}\\
(b)
\end{minipage}
	\caption{\label{Fig:pp-fluctuations}(a) Fluctuating cross section in
  pp at $\sqrt{s} = 5$ TeV, compared to a GG fit (\eqref{eq:GG}) and a
  log-normal fit (\eqref{Eq:AngLogNorm}). (b) Fluctuating cross section
  in $\gamma^*\p$ at $W^2 = 5000$ GeV$^2$ and various $Q^2$, calculated
	with the dipole model (the double peak structure for $Q^2 = 20$ GeV$^2$ is a statistical fluctuation). The cross section is shown on a logarithmic
  horizontal axis, to assess the log-normal approximation (cf.\
  \eqref{Eq:AngLogNorm}).
}
\end{figure}

Fluctuations in the pp cross sections, to estimate the influence of
fluctuations in $\p\A$ collisions, 
are often parametrized using \cite{Heiselberg:1991is,Blaettel:1993ah,Goncalves:2019agu}:
\begin{equation}
	\label{eq:GG}
	P(\sigma) = \rho \frac{\sigma}{\sigma + \sigma_0}\exp\left(-\frac{(\sigma/\sigma_0 - 1)^2}{\Omega^2} \right),
\end{equation}
where $\sigma_0$ and $\Omega$ are parameters, and $\rho$ is a normalization constant. In ref. \cite{Bierlich:2016smv}
is was found that a log-normal distribution (see \eqref{Eq:AngLogNorm} in appendix \ref{sec:angantyr}) describes fluctuations
generated by a dipole approach better. In \figref{Fig:pp-fluctuations} (a) both parametrizations are compared to the fluctuating total cross section in pp at $\sqrt{s} = $ 5 TeV, integrated over $\d^2\vec{b}$.

While the log-normal distribution does better in capturing the skewness
of the distribution, none of the two parametrizations fully describes
the distribution. The problem increases in $\gamma^*\p$ for several
reasons. First of all, any parametrization must include the correct
dependence on DIS kinematics, which changes the average cross section,
cf.\ figure \ref{Fig:pp-fluctuations} (b). Here is shown 
the cross section distributions for three values of $Q^2$ all with $W^2
= 5000$ GeV$^2$ with a logarithmic first axis. This allows for a by-eye
assessment of the validity of a log-normal fit, as a log-normal distribution is Gaussian
with such choice of axes. It is seen directly that
fluctuations in the high-$\sigma$ tails are too large to be described by such a parametrization. 

\begin{figure}[t]
\centering
\includegraphics[width=0.75\textwidth]{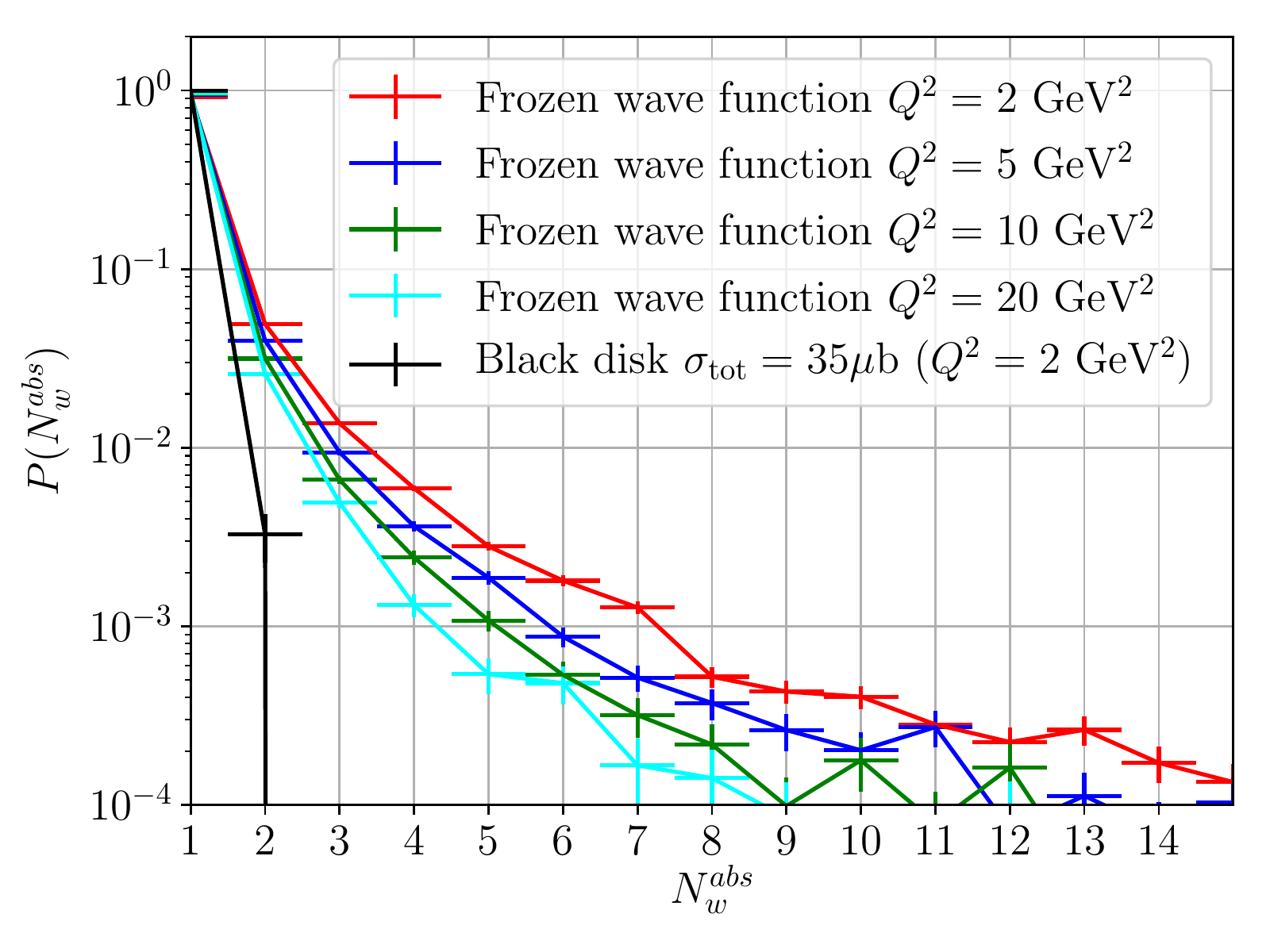}\\
\caption{\label{Fig:ea-glauber} Number of wounded nucleons in a
$\gamma^*$Au collision with $W^2 = 5000$ GeV$^2$ and a range of $Q^2$, 
comparing a treatment with the projectile wave function (denoted wf. in
legend) frozen throughout the passage of the nucleus, to a naive black disk approach. 
}
\end{figure}

Instead of the parametrization approach, the wounded cross section can be 
calculated directly from the dipole evolution. This also allows for simultaneous calculation of both
the part including electromagnetic contributions, and the pure dipole
part (given $z$ and $r$), as introduced in the previous section. This
allows directly for a calculation of the distribution of wounded
nucleons in a $\gamma^*\A$ collision, as shown in figure
\ref{Fig:ea-glauber}. In the figure, the nucleus is taken to be Au-197,
colliding with a virtual photon with $W^2=5000$ GeV$^2$ for a range of
$Q^2$ values, compatible with projected EIC design
\cite{Accardi:2012qut}. Two methods of calculating wounded nucleons are
presented: the full treatment using a frozen wave function, where the
photon wave function has collapsed to a dipole state when probed by the first collision, and
the naive black disk approach, where the photon re-forms after the full
collision and no fluctuations are included. In such a treatment, the cross section for additional collisions has an additional factor $\alpha^2_{\mrm{em}}$ compared to the frozen treatment, from the normalization of the wave function. It is directly visible that 
a full treatment is necessary in order to provide reasonable phenomenological projections for a new collider.

\section{Conclusion and outlook}
One of the main challenges for the understanding of collective effects,
is to grasp how the well-known understanding of flow results from heavy ion 
collisions can be transferred to collision of protons with protons and
heavy nuclei. In this paper we have presented a Monte Carlo implementation
of Mueller's dipole model with several sub-leading corrections,
and with all parameters of the model fixed to total and
semi-inclusive cross sections calculated within the Good-Walker
formalism. This model thus allows for the calculation of proton
substructure without tuning any model parameters to observables
sensitive to said substructure.

The current implementation of the model includes:
\begin{itemize}
	\item BFKL evolution of projectile and target states, be it protons or
  photons, in rapidity and impact parameter space.
	\item Sub-leading corrections in the evolution:
		\begin{enumerate}
			\item Energy-momentum conservation.
			\item Non-eikonal corrections in terms of dipole recoils.
			\item Confinement effects by the introduction of a fictitious gluon mass.
		\end{enumerate}
	\item Projectile--target interactions using the unitarized amplitude, which in a Regge field theory language
	corresponds to multi-Pomeron exchange and Pomeron loops.
	\item Matching to the \PY~MPI model, in order to assign spatial
  vertices to produced partons in $\p\p$ collisions.
	\item Generalization to heavy ion collisions through the \textsc{Angantyr} framework.
\end{itemize}

Besides the implementation of the dipole model, a simpler version has
been provided, based on the geometric properties of the dipole
evolution. This model, denoted the Pascal approximation, allows for easy
insertion of toy-models of sub-leading effects, thus giving a handle on the
importance of such effects. 

We have shown that given simple, but reasonable, assumptions 
of a final--state response (from e.g.\ hydrodynamics or interacting
strings), the eccentricities produced with the implementation provides a 
reasonable description of flow data from the ALICE and CMS experiments.
This includes non-trivial observations such as ratios between $\p\A$ and
$\p\p$ flow coefficients at fixed event multiplicity, normalized
symmetric cumulants in different systems, and ratios between different order 
flow harmonics in $\p\A$ collisions. All are signatures which cannot be 
described in a simpler model, where the spatial structure of MPIs are assumed 
to be distributed according to a rotationally symmetric distribution.
We want to stress that even though we have here chosen flow-type observables
to illustrate the effect of the space-time structure of the initial state
on observations of collective effects, effects linked to enhancement of strangeness 
and baryon production \cite{Biro:1984cf,Bierlich:2014xba,Bierlich:2015rha} and even modifications
of jets in high multiplicity pp collisions \cite{Mangano:2017plv,Bierlich:2019ixq} are expected to be
influenced as well.

Lastly, we have provided the initial steps towards the generation of
fully exclusive final states in electron-ion collisions, by
determining the importance of colour-fluctuations in the collisions with
virtual photons. We have shown that previous parametrizations from
$\p\p$ collisions do not fully capture the colour-fluctuations predicted
by the dipole formulation of BFKL evolution, and thus argue that it is better to calculate the
cross sections directly from the dipole model -- which has not been
possible in the \textsc{Angantyr} model before this work. Secondly, we
stress that the collapse of the photon wave function at first interaction
provides a larger number of wounded nucleons as compared to the black
disk approximation. Each of the wounded nucleons are expected to give
rise to final state activity, thus more complicated final states are
expected with the proper treatment as opposed to the naive
expectations. \\

The implemented dipole model can be improved in several ways, including:
\begin{itemize}
\item Running $\alpha_s$ in the dipole evolution and in the
scattering, which will capture some of the NLO corrections in $\alpha_s$.
\item On longer term, an inclusion of full NLO-BK should be the goal,
though further theoretical development is needed first.
\item Gluon saturation effects in the cascade such
as those included in the CGC formalism. To maintain the current treatment of the effect
of gluon branchings in the cascade (as opposed to CGC), this could be included by
the introduction of a simple swing mechanism, e.g.\ a mock $2\rightarrow1$ 
dipole recombination.
\item Several improvements are expected w.r.t.\ the initial dipole
configuration in protons and photons, as well as in the wave functions of these particles. 
This includes adding the VMD contribution to the photon wave function, 
to be able to study lower $Q^2$ and vector meson production in various processes.
\item New ways of treating MPIs in $\p\p$ collisions by fully merging the dipole
approach with more traditional approaches are foreseen. It is our hope that this
could provide new tools to improve understanding of particle production
mechanisms across collision systems.
\end{itemize}

Detailed understanding of the interplay between the proton geometry and the response of final state interactions in hadronic and heavy ion collisions,
is crucial for the understanding of collectivity and particle production mechanisms. Since 
detailed understanding requires tools which are both accessible and transparent, it is our
hope that the detailed treatment presented here, and the accompanying open Monte Carlo implementation,
can help facilitate this process.

\acknowledgments

We thank G\"osta Gustafson, Leif L\"onnblad and Torbj\"orn Sj\"ostrand
for several discussions during the making of this project and G\"osta, Torbj\"orn 
and Helen Brooks for reading through and commenting on the
manuscript. We also thank the ALICE group at the Niels Bohr Institute, University of
Copenhagen for hospitality during the writing of this work.

This project has received funding in part from the European Research
Council (ERC) under the European Union's Horizon 2020 research
and innovation program (grant agreement No 668679), and in part 
by the Swedish Research Council, contract numbers 2016-05996 and 2017-0034 
as well as the Marie Sk\l odowska-Curie Innovative Training Network MCnetITN3 
(grant agreement 722104).
 
\appendix
\addtocontents{toc}{\protect\setcounter{tocdepth}{1}}
\section{\label{sec:derivation}Appendix: The dipole model}

Below we go through the details of the dipole model not included in
sections \ref{sec:dipole-theory} and \ref{sec:mc-implementation}.

We here work with light cone momenta,
\begin{align}
p_{\pm}=E\pm p_z,
\end{align}
and can thus define the rapidity as
\begin{align}\label{Eq:yDef}
y
=\frac{1}{2}\log\left(\frac{p_+}{p_-}\right)=\log\left(\frac{p_+}{\pT}\right),
\end{align}
with the latter equality valid for massless particles. Hence we can express 
the lightcone momenta in terms of dipole $\pT$ and rapidity,
\begin{align}
p_{\pm}=\pT\exp(\pm y).
\end{align}
The $\pT$ of a dipole can be related to its size through $\pT\sim\hbar/r$.

The dipole-dipole scatterings are defined to occur at rapidity zero.
Thus the evolution of the beams begin at rapidity $y=\pm y_{\mrm{max}}$
and evolve to zero, i.e.\ with negative rapidity steps. For technical reasons, 
the actual evolution is easier to implement with positive steps in
rapidity. Thus the internal rapidity used in the code (and in the next
section) is negated w.r.t.\ the rapidity defined in \eqref{Eq:yDef}:
\begin{align}
y_{\mrm{MC}}=&-y=\log\left(\frac{\pT}{p_+}\right)\Rightarrow\\
p_{\pm}=&\pT\exp(\mp y_{\mrm{MC}}).
\end{align}
where in the forthcoming sections we will skip the subscript $\mrm{MC}$.

\subsection{Mueller's dipole branching}
\begin{figure*}[t]
\centering
\includegraphics[width=0.5\linewidth]{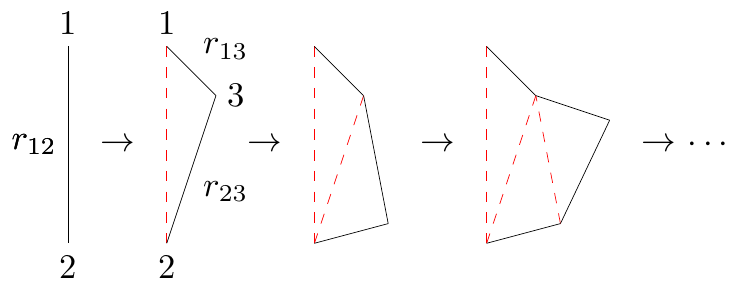}
\caption{\label{Fig:Dipole}
Schematic view of a dipole splitting. The initial dipole is spanned by
partons 1 and 2, that emits a new parton (3), thus creating two new 
dipoles: the dipole spanned by partons 1 and 3 and the dipole spanned 
by partons 2 and 3. This can be succeeded by additional splittings as
indicated by the additional figures following the arrows.
}
\end{figure*}

We begin by examining the dipole splitting function, 
\begin{align}\label{Eq:MuDipSplit}
\frac{\d P}{\d y~\d^2 \vec{r}} =
\frac{N_c\alpha_s}{2\pi^2}\frac{r_{12}^2}{r_{13}^2r_{23}^2},
\end{align}
where $\vec{r}$ is the transverse location of the emitted parton $3$
from the original dipole spanned by partons $1,2$ and $r_{ij}$ the 
length of the dipoles, also shown in figure \ref{Fig:Dipole}. 
In order to turn this into a dipole evolution, a
Sudakov factor, $\Delta(y_{\mrm{min}},y)$, restricting emission between 
$y_{\mrm{min}}$ and $y$, has to be introduced. The full dipole splitting
kernel then reads,
\begin{align}
\frac{\d P}{\d y~\d^2 \vec{r}} =
\frac{N_c\alpha_s}{2\pi^2}\frac{r_{12}^2}{r_{13}^2r_{23}^2}\Delta(y_{\mrm{min}},y),
\end{align}
For event generation to 
proceed, we need to find an overestimate for the above splitting 
probability. The Sudakov factor is included via the veto algorithm, 
and is thus neglected in the expressions below. 

First we sample a transverse location of the emitting dipole.
Assuming partons 1 and 2 located in the $(x,y)$-plane at 
$\vec{r}_1=(0,0)$ and $\vec{r}_2=(1,0)$ with length $r_{12}=1$ fm, 
while the emitted parton is located at $\vec{r}_3=(r_x,r_y)$, we can write the 
splitting probability as,
\begin{align}
\frac{\d P}{\d y~\d r_x~\d r_y} =&
\frac{N_c\alpha_s}{2\pi^2}\frac{r_{12}^2}{(r_x^2+r_y^2)((r_{2,x}-r_x)^2+r_y^2)}\nonu\\
=&\frac{N_c\alpha_s}{2\pi^2}\frac{1}{(r_x^2+r_y^2)((1-r_x)^2+r_y^2)}~,
\end{align}
where in the second step we have inserted the values for $r_{2,x}=1$ fm
and $r_{12}^2=1$ fm$^2$, but suppressed dimensions. These dimensions are
suppressed throughout the section.
This distribution is symmetric around $r_x=1/2$ fm and $r_y=0$ fm, so the
limits of integration can be changed from $r_{x/y}\in]-\infty,\infty[$ to
$r_x\in[-\infty, 1/2]$ and $r_y\in[-\infty, 0]$. \\
The above splitting probability can be overestimated by the function,
\begin{align}
\frac{\d P_1}{\d y~\d r_x~\d r_y} = \frac{N_c\alpha_s}{2\pi^2}\frac{2}{(r_x^2+r_y^2)(r_x^2+r_y^2+\frac{1}{4})}.
\end{align} 
Changing to cylindrical coordinates we obtain,
\begin{align}
\frac{\d P_1}{\d y~r~\d r~\d \phi} =
\frac{N_c\alpha_s}{2\pi^2}\frac{2}{r^2(r^2+\frac{1}{4})},
\end{align} 
which can be integrated from a minimal dipole size, $\rho$. Thus we
obtain,
\begin{align}\label{Eq:rSample}
\frac{\d P_1}{\d y}=
\frac{4N_c\alpha_s}{\pi}\log\left(1+\frac{1}{4\rho^2}\right).
\end{align}

Without energy ordering, the minimal dipole size $\rho$ has to be
fixed to a number larger than zero to avoid the distribution from
blowing up. Here, energy the ordering is introduced by ordering
of positive lightcone momenta \cite{Avsar:2005iz}. Again relating the transverse
momentum of the dipole to its size, gives an expression for $\rho$
related to the kinematics of the parent dipole ($p$),
\begin{align}\label{Eq:pPlusCons}
p_+^3\leq p_+^p \Rightarrow\quad
\pT^3e^{-y}=\frac{1}{\rho}e^{-y}\leq p_+^p\Rightarrow\quad
\rho \geq e^{-y}/p_+^p.
\end{align}
This expression is then used in \eqref{Eq:rSample},
\begin{align}
\frac{\d P_1}{\d y}=
\frac{4N_c\alpha_s}{\pi}\log\left(1+\frac{(\pT^p)^2}{4}e^{2 y}\right).
\end{align}
This overestimate cannot trivially be integrated, so we find yet 
another,
\begin{align}
\frac{\d P_2}{\d y}=&\frac{4N_c\alpha_s}{\pi}\log\left[e^{2
y}\left(1+\frac{(\pT^p)^2}{4}\right)\right]\equiv\frac{4N_c\alpha_s}{\pi}\log\left[e^{2
y}A\right]
\end{align}
which is both integrable and invertible. We take into account the
Sudakov factor by using the Veto algorithm, and thus the rapidity can 
be sampled from this distribution by
\begin{align}
y_i = 1/2\sqrt{\left[\log(A)+
2y_{i-1}\right]^2-\pi\log(R_1)/(N_c\alpha_s)} -
1/2\log(A),
\end{align}
where $R_1$ is a uniformly distributed random number.

From \eqref{Eq:rSample} we can sample both $r$ and $\phi$,
\begin{align}
\phi =& 2\pi R_2,\\
r =&
\sqrt{\frac{1}{4}\frac{\rho^{2R_3}}{(\rho^2+1/4)^{R_3}-\rho^{2R_3}}},
\end{align}
with $R_2,R_3$ two new random numbers. Here we should note that we've
changed the integration limits, such that any $r_x=r\cos(\phi)>1/2$ must
be rejected in the event generation. Half of the remaining events 
should be mirrored to $r_x\rightarrow1-r_x$, and this should be taken 
into account in the overestimate $\d P_1/\d y~\d r_x \d r_y$ as well,
such that
\begin{align}
\frac{\d P_1}{\d y~\d r_x\d r_y} =&
\frac{N_c\alpha_s}{2\pi^2}\frac{2}{(r_x^2+r_y^2)(r_x^2+r_y^2+\frac{1}{4})}\rightarrow\nonu\\
&\frac{N_c\alpha_s}{2\pi^2}\left[\frac{1}{(r_x^2+r_y^2)(r_x^2+r_y^2+\frac{1}{4})}
+ \frac{1}{((1-r_x)^2+r_y^2)((1-r_x)^2+r_y^2+\frac{1}{4})}\right].
\end{align}
The events are weighted to the correct distributions with,
\begin{align}
w_r =&
\frac{(r_x^2+r_y^2+1/4)((1-r_x)^2+r_y^2+1/4)}{((1-r_x)^2+r_y^2)((1-r_x)^2+r_y^2+1/4)
+ (r_x^2+r_y^2)(r_x^2+r_y^2+1/4)},\\
w_y =& \frac{\log(1+\frac{(\pT^p)^2}{4}e^{2y})}{2y+\log(A)},
\end{align}
such that if $w_rw_y<R_4$ the event is rejected and the process is
reiterated.

The evolution of an initial dipole thus goes as follows. Firstly, a
trial emission from the initial dipole is performed according to 
\eqref{Eq:MuDipSplit}. If the rapidity $y_0$ of this emission is 
below the maximally allowed rapidity, then the trial branching is 
accepted, thus two new dipoles are created. Trial emissions are then 
allowed from each of these dipoles using $y_{\mrm{min}}=y_0$ in 
\eqref{Eq:MuDipSplit}. This creates two new emissions with rapidities 
$y_{1,2}$. But here \textit{only} the dipole with the smallest rapidity 
is accepted. Thus after the second iteration we have three dipoles, 
from each of which trial emissions are created and only the emission 
with the smallest rapidity is accepted, thus creating an additional 
dipole. The process is reiterated until no trial emissions are 
produced below the maximally allowed rapidity. The process is visualized
in figure \ref{Fig:Dipole}.

The choice of $\pT$ of the emitted parton is not obvious. Here we 
assign the parton the largest $\pT$ of the system,
\begin{align}
\pT^3 = \frac{1}{\mrm{min}(r_{13}, r_{23})}.
\end{align}

\subsubsection{Ordering of lightcone momenta}
We here rely on approximate energy conservation through ordering of $p_+$. 
This has already been discussed in the above, where we found the 
cutoff for small dipoles in the event generation of $r$, 
\eqref{Eq:pPlusCons}. Thus we have implemented
energy conservation as
\begin{align}
p_+^3 \leq p_+^p,
\end{align}
which implies a rapidity-dependent cutoff for smaller dipoles.
 
Momentum conservation is introduced through the ordering
of $p_-$, 
\begin{align}
p_-^3 \geq \mrm{max}(p_-^1, p_-^2),
\end{align}
where it should be noted that this requirement is applied \textit{after}
the recoils have been taken into account. This choice also sets an upper 
bound for the dipole size through
\begin{align}
p_-^3 \geq p_-^p \Rightarrow\quad \pT e^{y_3} = \frac{1}{r} e^{y_3} \geq
p_-^p \Rightarrow\quad r \leq \frac{e^{y_3}}{p_-}
\end{align}

\subsubsection{Recoil effects}
The recoil of the emitted parton is shared equally between the partons
spanning the emitting dipole. Energy conservation requires that the
energy of the emitter after the emission of a new dipole equals the
energy of the emitter before the collision minus the recoil,
\begin{align}\label{Eq:pPlusRecoil}
p_+^{\mrm{after}}=p_+^{\mrm{before}}-p_+^{\mrm{recoil}}.
\end{align}

The recoil cannot be determined from first principles thus
have to make an ansatz. The choice here is also from \cite{Avsar:2005iz},
\begin{align}
p_+^{1,\mrm{recoil}}=&\frac{r_{23}}{r_{13}+r_{23}}p_+^3\nonu\\
p_+^{2,\mrm{recoil}}=&\frac{r_{13}}{r_{13}+r_{23}}p_+^3
\end{align}
thus the recoil on parton $1$ depends on the length of the dipole
spanned by partons 2, 3 and vice versa. Energy conservation is 
satisfied in the event generation by always requiring that 
$p_+^{i,\mrm{recoil}}\leq p_+^{i,\mrm{before}}$.

The recoil will also affect the $\pT$ of the emitter. Here the choice is 
\begin{align}
\pT^{i, \mrm{after}}=\mrm{max}\left(\pT^i, \frac{1}{r_{i3}}\right),
\end{align}
where $i=1,2$ are the initial partons and $3$ is the emitted parton.

Changing both the $\pT$ and $p_+$ of the emitter thus also requires us
to change the rapidity of the emitter for consistency, 
\begin{align}
y^{i,\mrm{after}}=\log\left(\frac{\pT^{i,\mrm{after}}}{p_+^{i,\mrm{after}}}\right).
\end{align}
Note here that the rapidity of the parent after the recoil will always
be larger than the rapidity of the parent dipole \textit{before} the
recoil. This is because $p_+$ after the recoil is always smaller than
$p_+$ before the recoil, while the $\pT$ is after the recoil is always
larger than or equal to the $\pT$ before the recoil. Because of this, we must
require that the rapidity of the emitters after the recoil is smaller
than the rapidity of the emitted gluon,
$y^{1,\mrm{after}},y^{2,\mrm{after}}\leq y^3$.

\subsubsection{Effects of confinement}
Here it should be noted that the modified Bessel functions behave as
$K_1(x)\sim1/x$ for small arguments, while falling off exponentially
at large arguments, $K_1(x)\sim\sqrt{\pi/x}\exp(-x)$.
Thus the confined distribution is overestimated by the unconfined
distribution, and the introduction of confinement only
adds an additional weight $f(\mrm{confined})/f(\mrm{unconfined})$ that vetoes
events with large dipole sizes.

\subsection{Initial states}

The initial dipole configuration depends on the particle. Here we present
two types: a proton sampler and a photon sampler. The difference here
lies both in the number of initial dipoles (three for protons, one for
photons) and in the wave function of the particle itself.

\subsubsection{Protons}
The initial state proton is not known from QCD, but instead has to be
described by some phenomenological model. At rest, it consists primarily
of three valence quarks, which we can view as endpoints of the initial
dipoles. The configuration of these dipoles, however, is not known, thus
we here work with a single scenario: An equilateral triangle.

We allow the dipole length to be distributed according to a Gaussian of
mean $r_{0}$ and width $\sigma_r$, such that the length of the
initial dipoles is given as:
\begin{align}
r=r_{0}+r_{\mrm{w}} R_g
\end{align}
with $R_g$ a Gaussian random number. The center of the triangle is fixed at origo. 

\subsubsection{Photons}

The wave function used in this work is presented in eqs.~(\ref{Eq:photonWf1}--\ref{Eq:photonWf2}). 
The full cross section for $\gamma^*\p$ is then given as
\begin{align}
\label{eq:gp-xsec2}
\sigma^{\gamma^*\p}(s)=\int_0^1\d
z\int_0^{r_{\mrm{max}}}r\d
r\int_0^{2\pi}\d\phi\left(|\psi_L(z,r)|^2+|\psi_T(z,r)|^2\right)\sigma(z,\vec{r}),
\end{align}
with $\sigma(z,\vec{r})$ the dipole-dipole scattering cross sections
given in equations~(\ref{Eq:XS1}-\ref{Eq:XS3}). The dipole-dipole scattering cross section goes
roughly as the square of the size of the largest dipole, 
$\sigma(z,\vec{r})\sim r^2$, thus an overestimate of the $\gamma^*\p$
cross section can be found by sampling the parameters from the
following distributions,
we obtain  
\begin{align}
z=&R_1,\\
\phi=&2\pi R_2\\
r=&r_{\mrm{max}}R_3,\\
\sigma_{\gamma^*\p}^O=&\frac{2\pi r_{\mrm{max}}}{N}\sum_{i=1}^Nr_i^3\left(|\psi_L(z_i,r_i)|^2+|\psi_T(z_i,r_i)|^2\right)
\end{align}
The maximal value obtained in the sum is kept to accept or reject the
integrand in the algorithm, where first and $z_i,r_i$ are chosen and
then accepted w.r.t.\
\begin{align}\label{Eq:WfWeight}
w_{\gamma}=\frac{r_i^3\left(|\psi_L(z_i,r_)|^2+|\psi_T(z_i,r_i)|^2\right)}{(\mrm{max.}\quad
\mrm{value})}.
\end{align}
If this weight is less than a new random number, $w_{\gamma}<R_4$, the event is rejected.
If kept, the event is given a weight $w=\sigma_{\gamma^*\p}^O/r_i^2$ to
take into account the overestimation of the dipole-dipole scattering
cross section.

\section{\label{sec:angantyr}Appendix: The \textsc{Angantyr} model for heavy ion collisions}

The \textsc{Angantyr} model for heavy ion collisions is based on the following
four components:
\begin{itemize}
\item Firstly, the position of the nucleons inside the nuclei has to be
determined.
\item Secondly, the number of interacting nucleons and binary $NN$
collisions has to be calculated within the Glauber-Gribov
(GG) formalism.
\item Thirdly, the contribution to the final state of each interacting
nucleon has to be determined. Here \textsc{Angantyr} uses the wounded nucleon
model by Bia\l as, Bleszy\'{n}ski and Czy\.z \cite{Bialas:1976ed}.
\item Lastly, any hard partonic subcollision has to be modeled, thus
introducing the concepts of primary and secondary absorptive
interactions.
\end{itemize}

Each of the four components will here be shortly reviewed. For the full
explanation, see \cite{Bierlich:2016smv,Bierlich:2018xfw}. 

The nucleon distribution is generated using a
Woods-Saxon potential:
\begin{align}
\rho(r) =& \frac{\rho_0\left(1+wr^2/R^2\right)}{1+\exp\left((r-R)/a\right)}
\end{align}
with $\rho(r)$ the radial density of the nucleons, $R$ the radius of the
nucleus, $a$ the skin width and $w$ the Fermi parameter, introducing a
varying density but set to zero in \textsc{Angantyr}. The $A$ nucleons are thus
generated randomly according to
$P(\vec{r}_i)=\rho(\vec{r}_i)\d^3\vec{r}_i$, assuming isospin
invariance, such that $\p=\n$. \textsc{Angantyr} uses the hard core assumption,
such that a new position for a nucleus is tried if the distance to its
neighbours falls below twice the hard-core radius $R_h$. Once the
nuclear distributions are set up, the impact parameter of the collision
is sampled using a Gaussian distribution. This information is then
passed the GG framework, which determines the fluctuations of the
target and projectiles.

The fluctuations arise because of fluctuations in the proton
wave function. Because the wave function enters in the cross section
calculations and because it is assumed that the projectile state is 
frozen during its interaction with the target, these fluctuation 
are then translated into fluctuating cross sections. In the dipole 
model, the probabilistic nature of the dipole evolution gives rise 
to different dipole configurations before the collisions, thus giving
rise to different dipole-dipole interactions and hence integrated cross
sections. \textsc{Angantyr} uses a probability distribution for the
cross section in $\p\A$ extracted from \DP:
\begin{align}
\label{Eq:AngLogNorm}P_{\mrm{tot}}\left(\log\sigma\right)=&\frac{1}{\Omega\sqrt{2\pi}}\exp\left[-\frac{\log^2\left(\sigma/\sigma_0\right)}{2\Omega^2}\right],\\
\label{Eq:AngEl}\langle T(\vec{b},\sigma)\rangle =&T_0\Theta\left(\sqrt{\frac{\sigma}{2\pi T_0}}- b\right),
\end{align}
with $\sigma=\int\d^2\vec{b}\langle 2T(\vec{b})\rangle$ and \eqref{Eq:AngEl}
describing a slightly modified version of the elastic scattering
amplitude. The parameters $\sigma_0,\Omega,T_0$ are tuned to data. For
$\A\A$ the fluctuations are instead determined by a Gamma function, 
\begin{align}
P(r)=&\frac{r^{k-1}e^{-r/r_0}}{\Gamma(k)r_0^k},\\
T(\vec{b},r_p,r_t)=&T_0(r_p+r_t)\Theta\left(\sqrt{\frac{(r_p+r_t)^2}{2T_0}} - b\right),\\
T_0(r_p+r_t)=&\left(1-\exp\left[-\frac{\pi(r_p+r_t)^2}{\sigma_t}\right]\right)^{\alpha},
\end{align}
where $P(r)$ determines fluctuations in the nucleon radius $r_p$ and
$r_k$. $T(\vec{b},r_p,r_t)$ again describes a slightly modified elastic
amplitude with an opacity $T_0$ depending on the radii of both the
target and projectile. Here, the parameters $\sigma_t,\alpha,k,r_0$ are
tuned to data. The number of wounded target nucleons in $\p\A$
collisions is then determined by
\begin{align}\label{Eq:wtpA}
\frac{\d\sigma_{\mrm{Wt}}}{\d^2\vec{b}}=&1-\langle\langle S_{pt}\rangle_t^2\rangle_p,
\end{align}
with $S_{pt}$ the S-matrix for a given target ($t$) and projectile ($p$)
state. Subscript on the brackets determines averages over one side only.
In $\A\A$ collisions \textsc{Angantyr} distinguishes between
absorptively and diffractively wounded nucleons, with the former
dominating given by,
\begin{align}
\frac{\d\sigma_{\mrm{abs}}}{\d^2\vec{b}}=&1-\langle S_{pt}^2\rangle_{pt},
\end{align}
and the latter determined by generating the auxiliary states $p',t'$ for
both target and projectile, and from these determining the number of wounded 
target states with either $t$ or $t'$ from \eqref{Eq:wtpA}, i.e.\ using either $S_{pt'}$
or $S_{pt}$ in the derivation. Non-negative probabilities are ensured by
shuffling when to use $t,t'$.

Once the number of wounded target and projectile states has been
determined, the wounded nucleon model is used to create final-state partons,
\begin{align}
\frac{\d N_{\mrm{ch}}}{\d\eta}=&w_pF(\eta)+w_tF(-\eta),
\end{align}
with the functions $F(\pm\eta)$ determined from the MPI framework of
\PY. Each nucleon in the target (and projectile) is allowed to interact
several times, similar to an ordinary $\p\p$ collision containing
several MPIs. Thus the pairs of projectile-target nucleons are ordered
w.r.t.\ their impact-parameter $b_{\mu\nu}$. The list is iterated over
several times in order to determine which pairs give rise to a primary
absorptive scattering, and which are secondary. Once a pair has been
selected, the MPI framework of \PY~is used to generate an event, and the
pair is marked as having interacted in a primary interaction. If the pair 
is again chosen to interact, it will be marked as a secondary
interaction. After the determination of the absorptive interactions, the
diffractive ones are chosen by iterating the list several times, thus
creating primary and secondary diffractive interactions. An already
wounded nucleon cannot be further excited, but an unwounded nucleon can
participate in several diffractive interactions, until itself becomes
wounded.

After the determination of the absorptive and diffractively primary and
secondary interactions, each of the events are passed to \PY~and the
parton-level events are stacked on top of each other. The \PY~description 
of single-diffractive events are modified to look like
non-diffractive ones, to describe the secondary absorptive events, while
diffractive primary and secondary events remain unmodified. We are thus
left with a large set of parton-level events that can be passed to the
hadronisation framework of \PY~and further analysed.

\section{\label{sec:eccAppendix}Appendix: Additional eccentricity figures}

\begin{figure}[t]
\begin{minipage}[c]{0.475\linewidth}
\centering
\includegraphics[width=\linewidth]{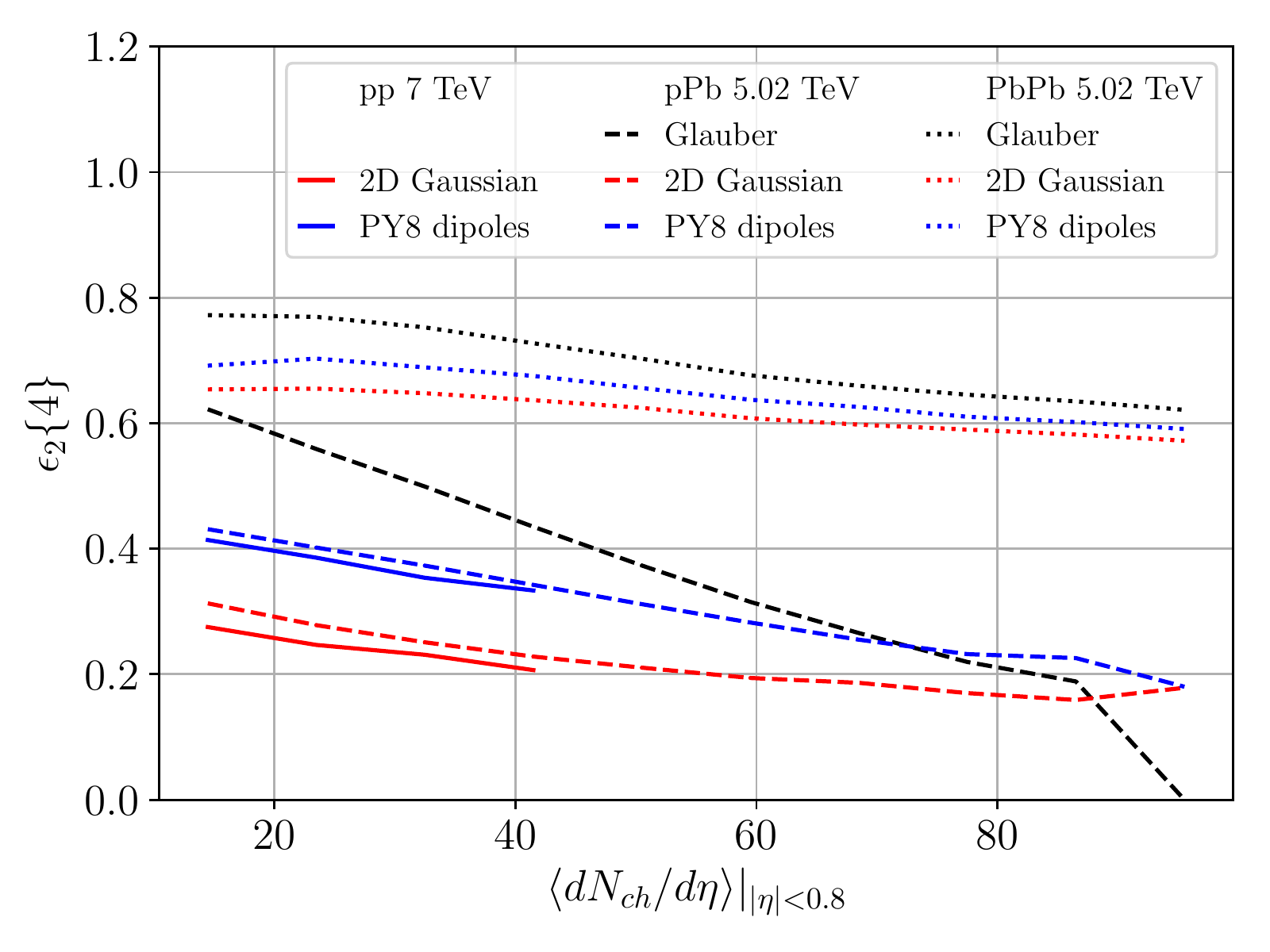}\\
(a)
\end{minipage}
\hfill
\begin{minipage}[c]{0.475\linewidth}
\centering
\includegraphics[width=\linewidth]{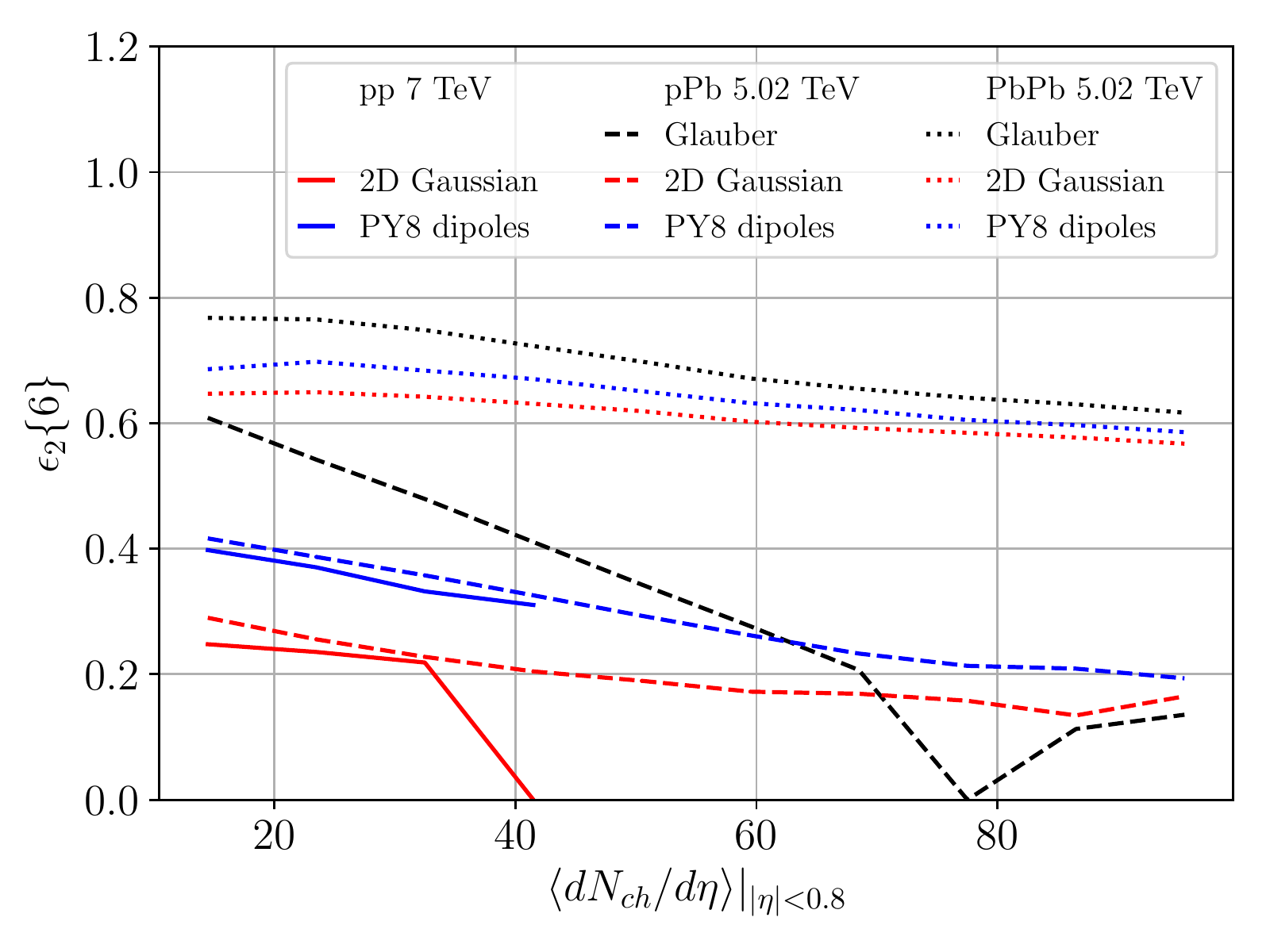}\\
(b)
\end{minipage}
\hfill
\begin{minipage}[c]{0.475\linewidth}
\centering
\includegraphics[width=\linewidth]{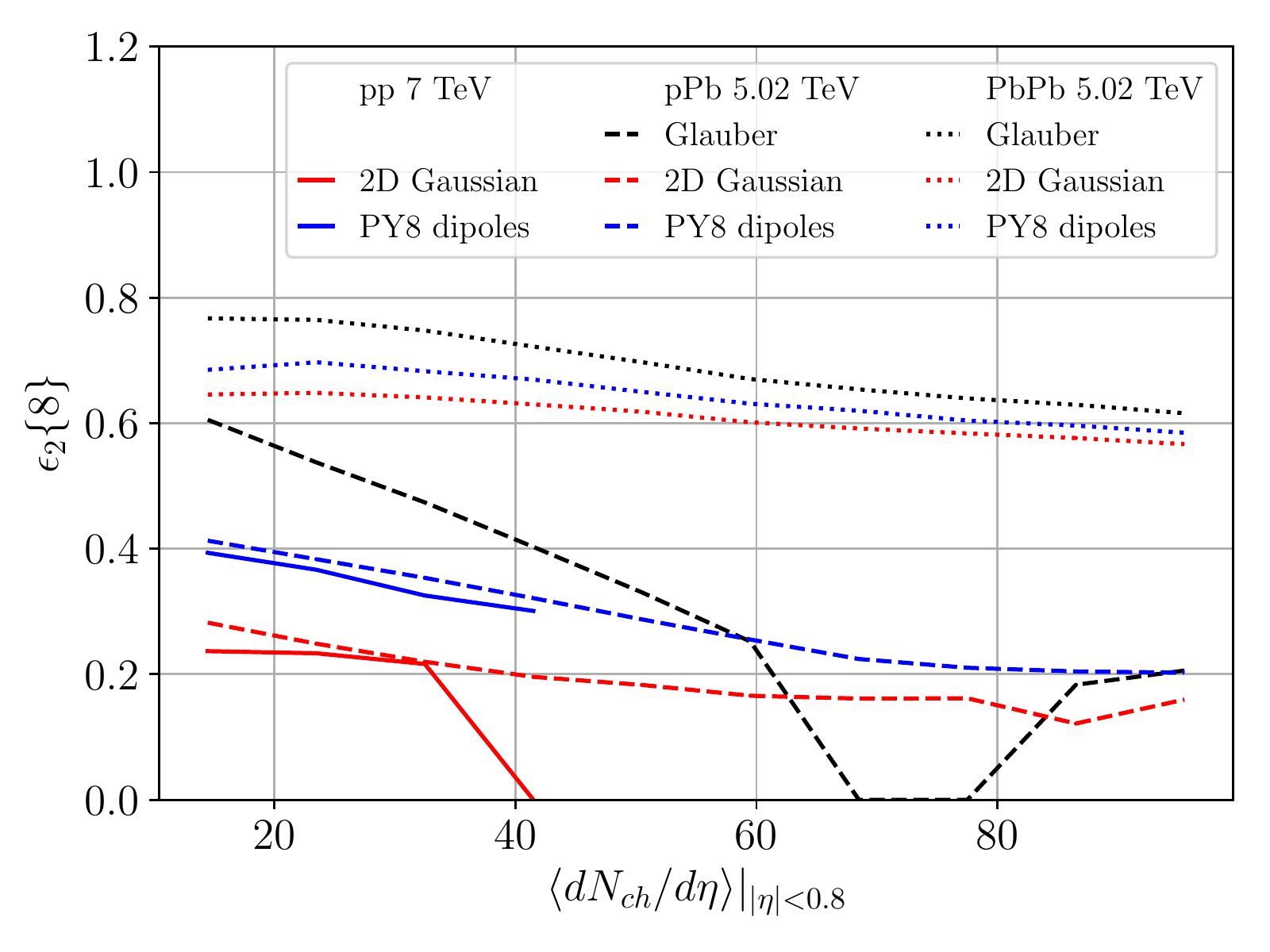}\\
(c)
\end{minipage}
\hfill
\begin{minipage}[c]{0.475\linewidth}
\centering
\includegraphics[width=\linewidth]{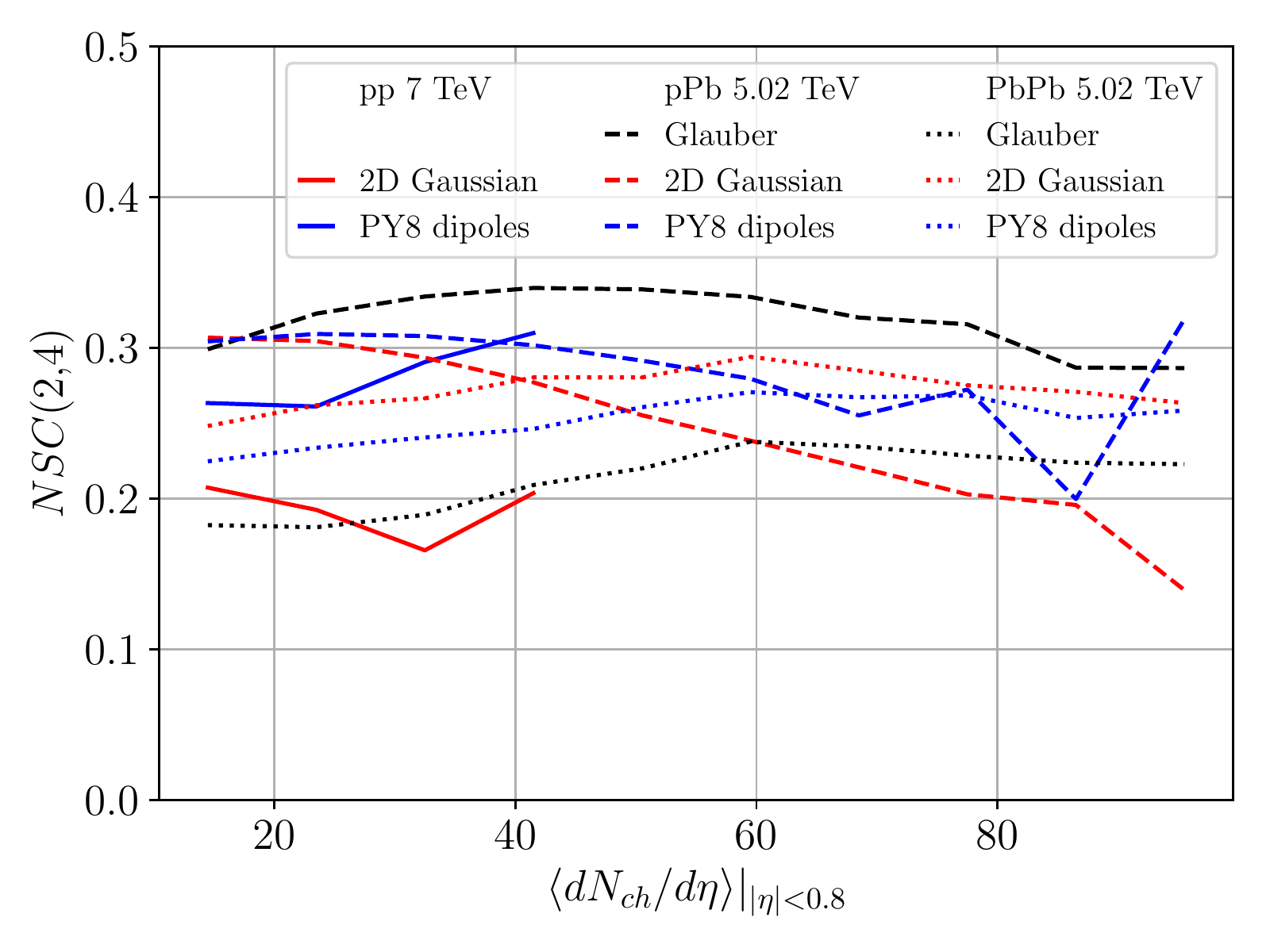}\\
(d)
\end{minipage}
\caption{\label{Fig:ecc-higher} (a-c) The eccentricity $\epsilon_2$ with
higher order cumulants $\{4,6,8\}$. (d) The normalised symmetric cumulant $NSC(4,2)$
as a function of average multiplicity for $\p\p,\p\A,\A\A$ systems.
}
\end{figure}

In this section, we show additional eccentricity figures not presented
in the main body of the text. Figure \ref{Fig:ecc-higher} (a-c) shows $\epsilon_2$
using higher-order cumulants in the evaluation. It is evident that the
eccentricities are the same regardless of the number of particles used
in the calculation, except for the effects from lack of statistics in
the high-multiplicity tail for both the $\p\p$ and $\p\Pb$ figures.
Figure \ref{Fig:ecc-higher} (d) shows the normalized symmetric cumulant
$NSC(4,2)$. This cumulant is positive in the entire multiplicity range,
consistent with measurements in ALICE. Here, it is evident that
discrimination between models would be possible in both $\p\p$ and
$\p\Pb$ collisions, as opposed to $NSC(3,2)$ where discriminatory power
was not evident in $\p\p$ collisions.

\begin{figure}[t]
\begin{minipage}[c]{0.475\linewidth}
\centering
\includegraphics[width=\linewidth]{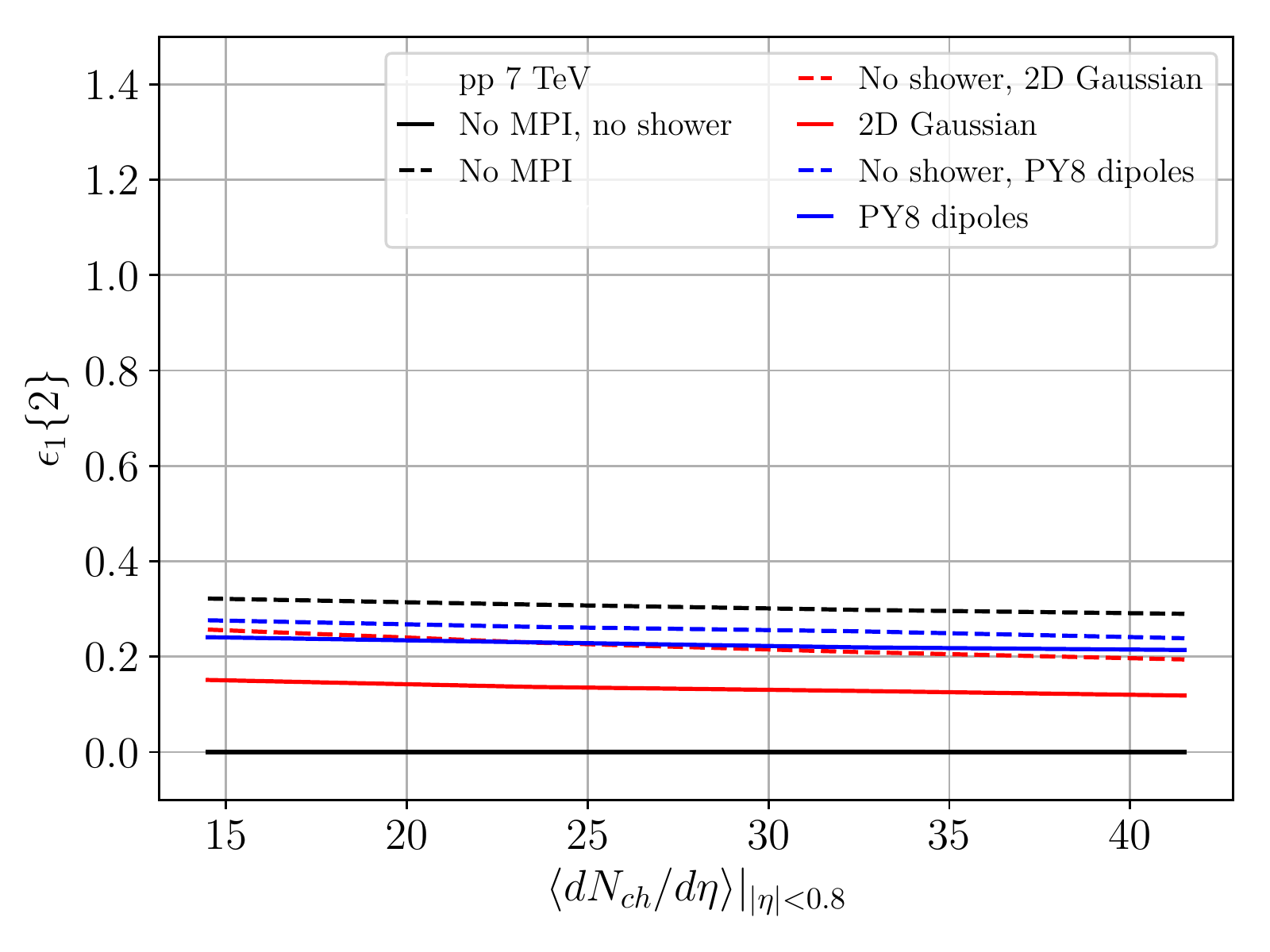}\\
(a)
\end{minipage}
\hfill
\begin{minipage}[c]{0.475\linewidth}
\centering
\includegraphics[width=\linewidth]{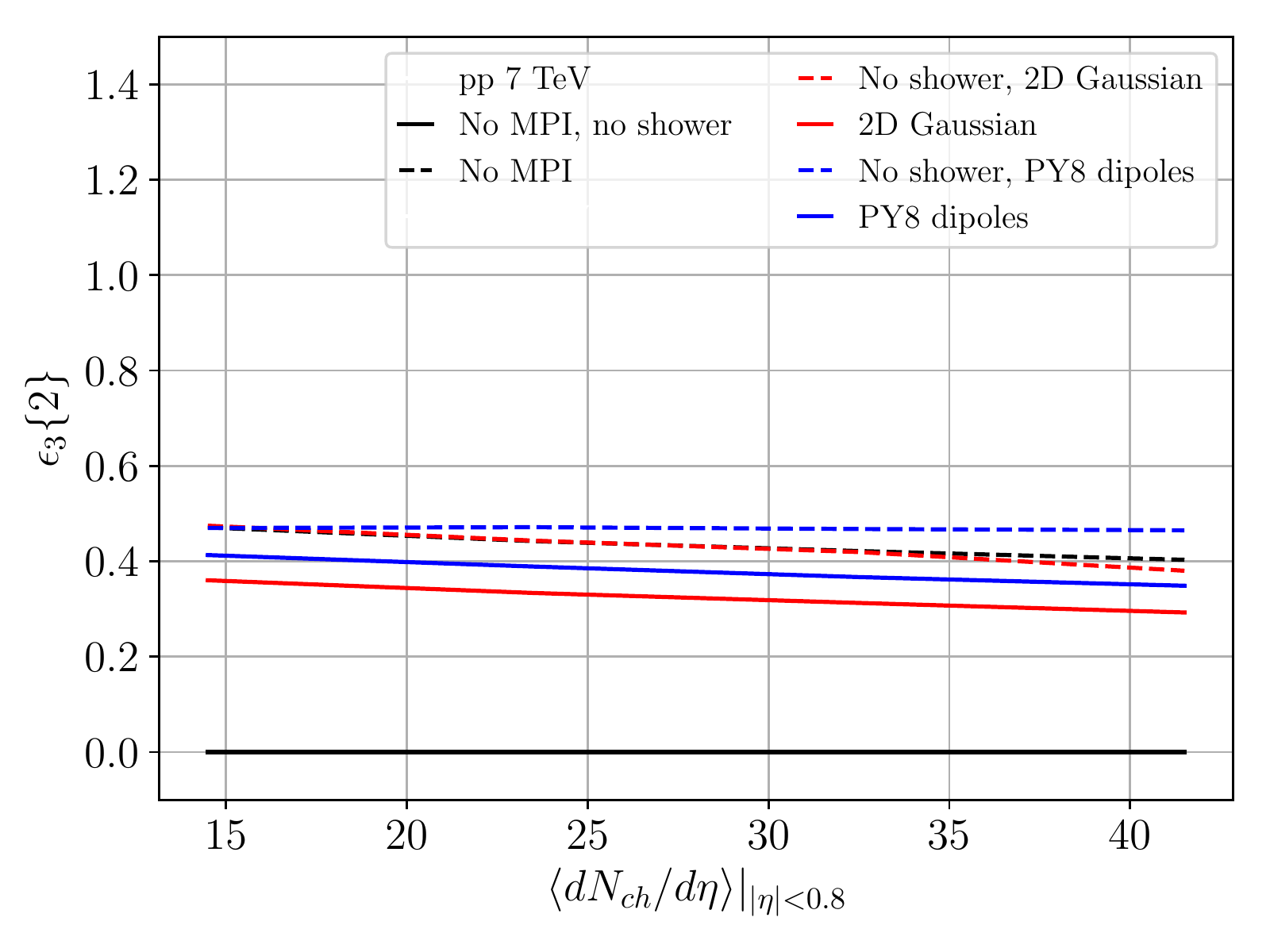}\\
(b)
\end{minipage}
\caption{\label{Fig:ecc-zero2} $\epsilon_{1,3}\{2\}$ shown for different
MPI vertex assignments with and without the shower smearing.
}
\end{figure}

For completeness, we also show the eccentricities $\epsilon_{1,3}$
obtained in $\p\p$ collisions with and without shower smearing in 
figure \ref{Fig:ecc-zero2}. Both are shown to give an estimate of the 
effects on the size of the additional terms in the Fourier expansion 
of the flow coefficients.

\bibliographystyle{JHEP}
\bibliography{lutp1932}

\end{document}